\documentclass[%
 aip,
 amsmath,amssymb,
 reprint,%
]{revtex4-1}
\usepackage{graphicx}
\usepackage{subfigure}
\usepackage{dcolumn}
\usepackage{bm}
\usepackage{color}
\usepackage[T1]{fontenc}
\usepackage{mathptmx}
\usepackage{multirow}
\usepackage{array}
\usepackage{float}
\usepackage{etoolbox}
\usepackage{booktabs}
\usepackage{url}
\usepackage{CJKutf8}
\usepackage{pifont}
\usepackage{epstopdf, epsfig}
\usepackage{graphicx}
\usepackage{longtable}
\usepackage[
    colorlinks=true,
    linkcolor=blue,
    urlcolor=blue,
    citecolor=blue
]{hyperref}

\usepackage{titlesec}
\titleclass{\subsubsubsection}{straight}[\subsubsection]
\newcounter{subsubsubsection}[subsubsection]
\renewcommand\thesubsubsubsection{\thesubsubsection.\arabic{subsubsubsection}}
\titleformat{\subsubsubsection}
  {\normalfont\normalsize\bfseries}{\thesubsubsubsection}{1em}{}
\titlespacing*{\subsubsubsection}
  {0pt}{1.5ex plus .1ex minus .2ex}{1ex plus .2ex}

\begin{document}
\preprint{AIP/123-QED}

\title{Kinetic study of compressible Rayleigh-Taylor instability with time-varying acceleration}

\author{Huilin Lai \begin{CJK*}{UTF8}{gbsn} (赖惠林) \end{CJK*}}
\affiliation{School of Mathematics and Statistics \& Key Laboratory of Analytical Mathematics and Applications (Ministry of Education) \& Fujian Key Laboratory of Analytical Mathematics and Applications (FJKLAMA) \& Center for Applied Mathematics of Fujian Province (FJNU), Fujian Normal University, 350117 Fuzhou, P. R. China.}
\author{Chuandong Lin \begin{CJK*}{UTF8}{gbsn} (林传栋) \end{CJK*}}
\thanks{Corresponding author: linchd3@mail.sysu.edu.cn}
\affiliation{Sino-French Institute of Nuclear Engineering and Technology, Sun Yat-sen University, Zhuhai 519082, P. R.China.}
\affiliation{Key Laboratory for Thermal Science and Power Engineering of Ministry of Education, Department of Energy and Power Engineering, Tsinghua University, Beijing 100084, P. R. China.}
\author{Hao Xu \begin{CJK*}{UTF8}{gbsn} (徐浩) \end{CJK*}}
\affiliation{School of Mathematics and Statistics \& Key Laboratory of Analytical Mathematics and Applications (Ministry of Education) \& Fujian Key Laboratory of Analytical Mathematics and Applications (FJKLAMA) \& Center for Applied Mathematics of Fujian Province (FJNU), Fujian Normal University, 350117 Fuzhou, P. R. China.}
\author{Hailong Liu \begin{CJK*}{UTF8}{gbsn} (刘海龙) \end{CJK*}}
\affiliation{Laoshan Laboratory, Qingdao 266237, P. R. China}
\author{Demei Li \begin{CJK*}{UTF8}{gbsn} (李德梅)\end{CJK*}}
\thanks{Corresponding author: dmli079@fjnu.edu.cn}
\affiliation{School of Mathematics and Statistics \& Key Laboratory of Analytical Mathematics and Applications (Ministry of Education) \& Fujian Key Laboratory of Analytical Mathematics and Applications (FJKLAMA) \& Center for Applied Mathematics of Fujian Province (FJNU), Fujian Normal University, 350117 Fuzhou, P. R. China.}
\author{Bailing Chen \begin{CJK*}{UTF8}{gbsn} (陈百灵) \end{CJK*}}
\affiliation{School of Mathematics and Statistics \& Key Laboratory of Analytical Mathematics and Applications (Ministry of Education) \& Fujian Key Laboratory of Analytical Mathematics and Applications (FJKLAMA) \& Center for Applied Mathematics of Fujian Province (FJNU), Fujian Normal University, 350117 Fuzhou, P. R. China.}

\date{\today}

\begin{abstract}	
Rayleigh-Taylor (RT) instability commonly arises in compressible systems with time-dependent acceleration in practical applications. To capture the complex dynamics of such systems, a two-component discrete Boltzmann method is developed to systematically investigate the compressible RT instability driven by variable acceleration. Specifically, the effects of different acceleration periods, amplitudes, and phases are systematically analyzed. The simulation results are interpreted from three key perspectives: the density gradient, which characterizes the spatial variation in density; the thermodynamic non-equilibrium strength, which quantifies the system's deviation from local thermodynamic equilibrium; and the fraction of non-equilibrium regions, which captures the spatial distribution of non-equilibrium behaviors. Notably, the fluid system exhibits rich and diverse dynamic patterns resulting from the interplay of multiple competing physical mechanisms, including time-dependent acceleration, RT instability, diffusion, and dissipation effects. These findings provide deeper insights into the evolution and regulation of compressible RT instability under complex driving conditions.
\end{abstract}

\maketitle

\section{\label{sec:level1} Introduction}

Rayleigh-Taylor (RT) instability occurs at the interface between two fluids when a denser (heavier) fluid is supported or accelerated by a less dense (lighter) one\cite{Rayleigh1882,Taylor1950}. The mixing driven by RT instability plays a pivotal role in a wide range of natural and engineering phenomena, including corona formation\cite{hoogenboom2006rayleigh}, inertial confinement fusion\cite{rahimi2013reduction}, supernova explosions \cite{ribeyre2004compressible, abarzhi2019supernova}, the formation of underground salt domes \cite{selig1966families}, and the evolution of volcanic islands \cite{ghosh2020cold}. Acceleration is a critical factor influencing the development of RT instability. While constant acceleration governs relatively steady processes such as the formation of salt domes and volcanic islands, variable acceleration plays a crucial role in more dynamic and complex phenomena, including inertial confinement fusion and supernova explosions. Therefore, a comprehensive understanding of the effects of time-dependent acceleration on RT instability is of significant practical and theoretical importance. 

Research on RT instability generally falls into three main categories: experimental studies\cite{morgan2020experiments,lherm2022rayleigh}, theoretical analyses\cite{meshkov2019group,matsumoto2017linear,abarzhi2020scale}, and numerical simulations\cite{liu2018surface,shin2022numerical,luo2021effects}. Experimental studies provide intuitive and convincing insights, but they are often time-consuming, costly, and may involve safety risks. Theoretical analyses, though relatively straightforward, are typically constrained by simplifying assumptions, which limit their applicability to complex real-world scenarios. In contrast, numerical simulations have gained increasing attention with the advancement of computational science and technology. They offer several advantages, including reduced cost, shorter research cycles, improved safety, and the ability to generate detailed and comprehensive data. Common numerical approaches include direct numerical simulation (DNS)\cite{liang2019direct}, moving particle semi-implicit (MPS) methods\cite{guo2018numerical}, implicit large eddy simulation (ILES)\cite{youngs2017rayleigh}, and smooth particle hydrodynamics (SPH)\cite{rahmat2014numerical}, among others. Over the past few decades, these numerical methods have been widely employed to investigate various aspects of RT instability. For example, Hamzehloo \textit{et al.} utilized DNS to study the effects of different combinations of Atwood number, Reynolds number, surface tension, and initial perturbation amplitude on RT instability\cite{hamzehloo2021direct}. Song \textit{et al.} adopted ILES to examine RT instability in the presence of a density gradient layer\cite{song2021numerical}. Shadloo \textit{et al.} applied SPH to explore incompressible RT instability with surface tension effects\cite{shadloo2013simulation}.

The RT instability, particularly under conditions of variable acceleration, has been extensively studied over the past few decades\cite{aslangil2016numerical, aslangil2022effects, boffetta2019suppression, ramaprabhu2013rayleigh, ramaprabhu2016evolution, hu2020evolution, livescu2021rayleigh}. Aslangil \textit{et al.} investigated the dynamics of RT instability driven by single or double acceleration inversions and observed that the mixed fluid layer ceases to grow following acceleration inversion\cite{aslangil2022effects}. Boffetta \textit{et al.} examined the effects of time-periodic acceleration on RT instability, discovering that such acceleration inhibits RT-induced turbulent mixing \cite{boffetta2019suppression}. Ramaprabhu \textit{et al.} analyzed the RT instability under acceleration profiles described by $g(t) \sim t^n$ with $n \ge 0$, along with acceleration histories inspired by linear electric motor experiments\cite{ramaprabhu2016evolution}. Hu \textit{et al.} conducted numerical and theoretical studies on the evolution of RT instability under conditions of discontinuous interface acceleration caused by radiation, and showed that this scenario is equivalent to the classical RT instability with effective acceleration\cite{hu2020evolution}.
Livescu \textit{et al.} investigated the evolution of RT instability when gravity is suddenly set to zero or reversed\cite{livescu2021rayleigh}. Banerjee \textit{et al.} studied the ablative RT instability under variable acceleration, revealing that the curvature and asymptotic growth rate of the bubble tip tend to saturate at finite values\cite{banerjee2023ablative}.

Existing research has primarily relied on macroscopic models, such as the Euler and Navier-Stokes (NS) equations, which are based on the assumption of equilibrium or near-equilibrium conditions. While these models are effective in capturing large-scale hydrodynamic behaviors, they often fall short in describing the intricate thermodynamic non-equilibrium (TNE) phenomena within fluid systems. To overcome this limitation and explore the underlying non-equilibrium mechanisms driving the evolution of RT instability, we employ the discrete Boltzmann method (DBM), a coarse-grained physical model developed from the lattice Boltzmann method (LBM)\cite{Wang2020, wei2018novel, wei2022small, chai2008lattice, chai2020}. In a seminal review, Xu \textit{et al.} introduced the concept of DBM and emphasized that the non-conserved moments of $(f_i - f_i^{\mathrm{eq}})$ can be used to quantitatively measure the deviation from thermodynamic equilibrium and to characterize the corresponding non-equilibrium effects \cite{xu2012lattice,Xu2024}. DBM incorporates additional physical constraints that enable more accurate detection of non-equilibrium states and facilitate the extraction of detailed information\cite{gan2018discrete, gan2022discrete}. Its primary aim is to effectively capture TNE behaviors. This forms the foundation of the current DBM modeling strategy. DBM is particularly suitable for investigating TNE behaviors that are often neglected in conventional macroscopic fluid models and cannot be directly addressed by molecular dynamics simulations due to limitations in spatial and temporal resolution.

The DBM has been successfully applied to investigate various complex physical systems, including shock waves\cite{lin2020hydrodynamic,liu2023discrete,zhang2023specific}, multiphase flows\cite{gan2015discrete,gan2022discrete,gan2012physical,wang2023high,sun2024droplet,sun2022thermodynamic}, reactive flows\cite{lin2018mrt,su2022nonequilibrium,yan2013lattice,zhang2016kinetic,ji2022three,lin2019discrete,ji2021three}, and hydrodynamic instabilities\cite{shan2023,song2024plasma,yang2023influence,gan2019nonequilibrium,Li2022FOP,Lai2016,Chen2022discrete,li2022rayleigh,Chen2022,Ye2020,Chen2021,Li2018,Chen2016Prandtl}. In the context of RT instability research, Lai \textit{et al.} employed DBM to simulate RT instability in compressible fluids, investigating the interaction between hydrodynamic non-equilibrium (HNE) and TNE effects \cite{Lai2016}. Chen \textit{et al.} developed a DBM model that incorporates intermolecular interactions to study the impact of interfacial tension, viscosity, and heat conduction on 2D single-mode RT instability\cite{Chen2022discrete}. Li \textit{et al.} utilized DBM with tracers to explore the effects of viscosity, constant acceleration, compressibility, and Atwood number on RT instability under multi-mode perturbations\cite{li2022rayleigh}. Furthermore, Chen \textit{et al.} conducted numerical studies on compressible RT instability with random multimode initial perturbations at continuous interfaces, revealing the physical mechanisms underlying the evolution of non-equilibrium intensity during the RT process\cite{Chen2022}. Ye \textit{et al.} applied DBM to investigate the influence of the Knudsen number on compressible RT instability, finding that an increase in Knudsen number inhibits RT instability while enhancing TNE effects\cite{Ye2020}. Additionally, Chen \textit{et al.} analyzed the effect of specific heat ratio on compressible RT instability, focusing on key physical quantities such as temperature gradients and the proportion of the non-equilibrium region\cite{Chen2021}. Li \textit{et al.} studied the compressible RT instability under multi-mode initial perturbations using DBM, emphasizing the TNE effects in the evolution of RT instability\cite{Li2018}. Chen \textit{et al.} also investigated the impact of viscosity, heat conduction, and Prandtl number on RT instability using the multi-relaxation time DBM\cite{Chen2016Prandtl}. Lai \textit{et al.} further explored the RT instability under varying accelerations using DBM, finding that higher acceleration results in a faster increase in non-equilibrium strength during the early stages, followed by a slower decrease in the later stages\cite{lai2023influences}. These studies have significantly advanced our understanding of the complex TNE behaviors in macroscopic fluid flows.

The previously mentioned DBMs have been primarily applied to single-component fluids, limiting their ability to provide detailed insights into the flow field, such as the specific flow velocity, temperature, and pressure of each chemical species. Recent advancements in two- and multi-component DBMs for fluid systems have significantly progressed the study of fluid instabilities\cite{lin2017discrete,lin2019kinetic,Zhang2020Two, chen2023effects,Lin2024CTP}. Lin \textit{et al.} introduced a two-component DBM to investigate the invariants of tensors associated with non-equilibrium effects in compressible RT instability involving two chemical species\cite{lin2017discrete}. Zhang \textit{et al.} developed a DBM based on the ellipsoidal statistical Bhatnagar-Gross-Krook model to study the impact of Prandtl number effects on Kelvin-Helmholtz (KH) instability\cite{Zhang2020Two}. Lin \textit{et al.} further expanded their work by introducing a multiple-relaxation-time DBM for multi-component mixtures, incorporating non-equilibrium effects, and exploring the influence of thermal conductivity on KH instability\cite{Lin2021PRE}. Chen \textit{et al.} applied a two-component DBM to examine the effects of interface inclination on compressible RT instability, finding that larger inclination angles accelerate the system's evolution\cite{chen2023effects}. Lin \textit{et al.} also proposed a multi-relaxation-time DBM with a split collision approach for both subsonic and supersonic compressible reacting flows, where each chemical species is represented by its own discrete distribution functions\cite{Lin2024CTP}.

In this paper, the evolution of compressible RT instability under time-varying acceleration is numerically simulated using a two-component DBM. The variations of key physical quantities during the RT process are analyzed from both macroscopic and mesoscopic perspectives. The structure of the paper is organized as follows: Section \ref{II} provides a brief overview of the two-component DBM. In Section \ref{III}, numerical simulations of the compressible RT instability with time-varying acceleration are presented. Section \ref{IV} concludes the paper with a summary of the findings.

\section{Two-component DBM}\label{II}

The discrete Boltzmann equation for two-component fluids takes the following form \cite{lin2017discrete}:
\begin{equation}\label{e1}
	\frac{{\partial {f_i^{\sigma}}}}{{\partial \textit{\textbf{t}}}} + {v_{i\alpha}^{\sigma}}\dfrac{{\partial
			{f_i^{\sigma}}}}{{\partial{r_{\alpha}}}}+\sum\limits_{\alpha}\frac{m^{\sigma}}{T^{\sigma}}a_{\alpha}(u_{\alpha}^{\sigma}
	-{v_{i\alpha}^{\sigma}}){f_i^{{\sigma}eq}}
	=-\dfrac{1}{\tau^{\sigma}}({f_i^{\sigma}} - {f_i^{{\sigma}eq}})
	\tt{,}
\end{equation}
where the superscript ${\sigma}$ denotes the fluid species, $r_{\alpha}$ the spatial coordinate, $\tau^{\sigma}$ the
relaxation time, ${m^{\sigma}}$ the particle mass, $u_{\alpha}^{\sigma}$ the flow velocity, and $T^\sigma$ the
temperature, $v_{i\alpha}^{\sigma}$ the discrete velocity, $f_i^{{\sigma}}$ ($f_i^{{\sigma}eq}$) the discrete
(equilibrium) distribution function, and $i =1, 2, \cdots, N$, with $N$ the total number of discrete velocities.

\begin{figure}[H]
	\center\includegraphics*%
	[bbllx=30pt,bblly=0pt,bburx=390pt,bbury=380pt,width=0.5\textwidth]{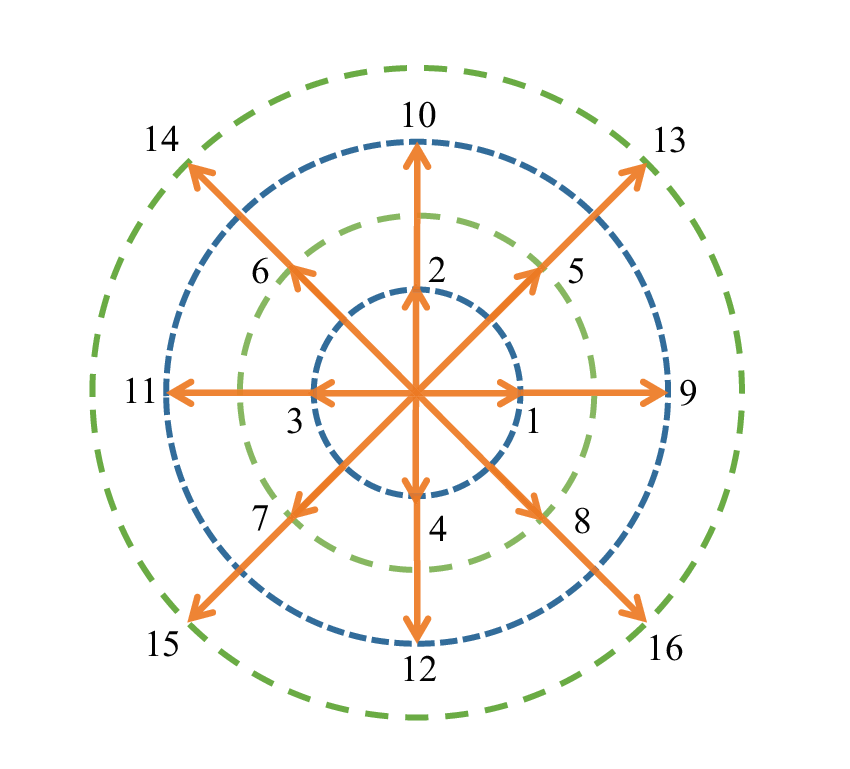}
	\caption{The sketch of D2V16 discrete velocity model.}\label{FIG01}
\end{figure}

As displayed in Fig. \ref{FIG01}, the discrete velocity model D2V16 is selected in this work, with the expressions
taking the following form:
\begin{equation}\label{e2}
	\textbf{v}_i=\left\{
	\begin{aligned}
		&v_a\Big[\cos\frac{(i-1)\pi}{2},\sin\frac{(i-1)\pi}{2}\Big],i=1,\cdots,4,\\
		&v_b\Big[\cos\frac{(2i-1)\pi}{4},\sin\frac{(2i-1)\pi}{4}\Big],i=5,\cdots,8,\\
		&v_c\Big[\cos\frac{(i-9)\pi}{2},\sin\frac{(i-9)\pi}{2}\Big],i=9,\cdots,12,\\
		&v_d\Big[\cos\frac{(2i-9)\pi}{4},\sin\frac{(2i-9)\pi}{4}\Big],i=13,\cdots,16,\\
	\end{aligned}
	\right.
\end{equation}
and
\begin{equation}\label{e3}
	\eta_{i}=\left\{
	\begin{aligned}
		&\eta_{a}, 1 \le {i} \le 4 \text{,}\\
		&\eta_{b}, 5 \le {i} \le 8 \text{,}\\
		&\eta_{c}, 9 \le {i} \le 12 \text{,}\\
		&\eta_{d}, 13\le {i} \le 16 \text{,}\\
	\end{aligned}
	\right.
\end{equation}
where $v_a$, $v_b$, $v_c$, $v_d$ and $\eta_a$, $\eta_b$, $\eta_c$, $\eta_d$ are tunable parameters. Specifically, $\eta_{i} = \eta_{0}$ when $i=1, \cdots, 4$; otherwise, $\eta_i=0$.

The individual particle number density, mass density, and flow velocity of each component ${\sigma}$ are defined as
follows:
\begin{equation}\label{e4}
	n^{\sigma}=\sum\limits_{i}f_i^{\sigma}
	\tt{,}
\end{equation}
\begin{equation}\label{e5}
	\rho^{\sigma}=m^{\sigma}n^{\sigma}
	\tt{,}
\end{equation}
\begin{equation}\label{e6}
	\textbf{u}^{\sigma}=\dfrac{1}{n^{\sigma}}\sum\limits_{i}f_i^{\sigma}\textbf{v}_i
	\tt{.}
\end{equation}

The mixing particle number density, mass density, and the macroscopic velocity of the system are expressed by
\begin{equation}\label{e7}
	n=\sum\limits_{\sigma}n^{\sigma}
	\tt{,}
\end{equation}
\begin{equation}\label{e8}
	\rho=\sum\limits_{\sigma}{\rho^{\sigma}}
	\tt{,}
\end{equation}
\begin{equation}\label{e9}
	\textbf{u}=\dfrac{1}{\rho}\sum\limits_{\sigma}\rho^{\sigma}{\textbf{u}^{\sigma}}
	\tt{.}
\end{equation}

The internal energy per unit volume of the component ${\sigma}$ and the internal energy per unit volume of the system
are
\begin{equation}\label{e10}
	E^{\sigma}={\dfrac{1}{2}}{m^{\sigma}}\sum\limits_{i}f_i^{\sigma}\left(|\textbf{v}_i-\textbf{u}|^2+{\eta_i}^2\right)
	\tt{,}
\end{equation}
and
\begin{equation}\label{e11}
	E=\sum\limits_{\sigma}{E^{\sigma}}
	\tt{,}
\end{equation}
respectively.

The individual and mixing temperatures are respectively
\begin{equation}\label{e12}
	T^{\sigma}=\dfrac{2}{D+I^{\sigma}}\dfrac{E^{\sigma}}{n^{\sigma}}
	\tt{,}
\end{equation}
and
\begin{equation}\label{e13}
	T=\dfrac{2E}{\sum\limits_{\sigma}n^{\sigma}(D+I^{\sigma})}
	\tt{,}
\end{equation}
where $D=2$ is the space dimension, $I^{\sigma}$ represents the number of extra degrees of freedom, and ${\eta_i}$ is
used to describe the internal energy of extra degrees of freedom.

The Chapman-Enskog (CE) multi-scale analysis indicates that the DBM is consistent with the NS equations in the
hydrodynamic limit \cite{Lin2016CNF}. To achieve this aim, $f_i^{{\sigma}eq}$ should satisfy the following seven moment
relations:
\begin{equation}\label{e14}
	\iint{{f}^{{\sigma}eq}}{d\textbf{v}}d{\eta} =\sum\limits_{i}f_{i}^{{\sigma}eq}
	\tt{,}
\end{equation}
\begin{equation}\label{e15}
	\iint{{f}^{{\sigma}eq}}{v_{\alpha}}{d\textbf{v}}d{\eta} =\sum\limits_{i}f_{i}^{{\sigma}eq}
	{v_{i\alpha}^{\sigma}}
	\tt{,}
\end{equation}
\begin{equation}\label{e16}
	\iint{{f}^{{\sigma}eq}}(v^2+{\eta}^2){d\textbf{v}}d{\eta} =\sum\limits_{i}f_{i}^{{\sigma}eq}
	(v_{i}^{{\sigma}2}+{\eta}_{i}^{{\sigma}2})
	\tt{,}
\end{equation}
\begin{equation}\label{e17}
	\iint{{f}^{{\sigma}eq}}{v_{\alpha}}{v_{\beta}}{d\textbf{v}}d{\eta}
	=\sum\limits_{i}f_{i}^{{\sigma}eq}{v_{i{\alpha}}^{\sigma}}{v_{i{\beta}}^{\sigma}}
	\tt{,}
\end{equation}
\begin{equation}\label{e18}
	\iint{{f}^{{\sigma}eq}}(v^2+{\eta}^2){v_{\alpha}}{d\textbf{v}}d{\eta}
	=\sum\limits_{i}f_{i}^{{\sigma}eq}(v_{i}^{{\sigma}2}+{\eta}_{i}^{{\sigma}2})
	{v_{i{\alpha}}^{\sigma}}
	\tt{,}
\end{equation}
\begin{equation}\label{e19}
	\iint{{f}^{{\sigma}eq}}{v_{\alpha}}{v_{\beta}}{v_{\chi}}{d\textbf{v}}d{\eta}
	=\sum\limits_{i}f_{i}^{{\sigma}eq}{v_{i{\alpha}}^{\sigma}}{v_{i{\beta}}^{\sigma}}
	{v_{i{\chi}}^{\sigma}}
	\tt{,}
\end{equation}
\begin{equation}\label{e20}
	\iint{{f}^{{\sigma}eq}}(v^2+{\eta}^2){v_{\alpha}}{v_{\beta}}{d\textbf{v}}d{\eta}
	=\sum\limits_{i}f_{i}^{{\sigma}eq}(v_{i}^{{\sigma}2}+{\eta}_{i}^{{\sigma}2})
	{v_{i{\alpha}}^{\sigma}}{v_{i{\beta}}^{\sigma}}
	\tt{.}
\end{equation}

In formulas \eqref{e14}-\eqref{e20}, the integral is extended over the phase space $(\textbf{v},\eta)$, and
$f^{{\sigma}eq}$ represents the equilibrium distribution function expressed by
\begin{equation}\label{e21}
	\begin{split}
		{f^{{\sigma}eq}} = & \, n^{\sigma} \left( \dfrac{m^{\sigma}}{2\pi k T^{\sigma}} \right)^{D/2}
		\left( \dfrac{m^{\sigma}}{2\pi I^{\sigma} k T^{\sigma}} \right)^{1/2} \\
		& \times \exp\left[ -\dfrac{m^{\sigma}|{\textbf{v}}-{\textbf{u}}|^2}{2kT^{\sigma}} 
		- \dfrac{m^{\sigma} \eta^2}{2I^{\sigma}kT^{\sigma}} \right] \tt{,}
	\end{split}
\end{equation}
where ${n^{\sigma}}$ is the particle number density, ${\textbf{u}}$ the mixture velocity, ${\textbf{v}}$ the velocity of
particle translational motion, $k=1$ the Boltzmann constant, and ${\eta}$ is a parameter utilized to describe the
rotational and/or vibrational energies.

By applying a linear transformation between the velocity and moment spaces, the seven moments in Eqs. (\ref{e14}) -
(\ref{e20}) can be  expressed in the matrix form as follows:
\begin{equation}\label{e22}
	{\textbf{M}}\,{\textbf{f}}^{{\sigma}eq}={\hat{\textbf{f}}}^{{\sigma}eq}
	\tt{,}
\end{equation}
where ${\textbf{f}}^{{\sigma}eq}$ and ${\hat{\textbf{f}}}^{{\sigma}eq}$ are a set of the particle discrete equilibrium
distribution functions in velocity and moment spaces, respectively. The transformation matrix $\textbf{M}$ comprises
components defined by the discrete parameters $v_{i\alpha}^{\sigma}$ and ${\eta}_{i}^{{\sigma}}$. If the matrix
$\textbf{M}$ is invertible, the above formula can be reformulated as:
\begin{equation}\label{e23}
	{\textbf{f}}^{{\sigma}eq}={\textbf{M}}^{-1}\,{\hat{\textbf{f}}}^{{\sigma}eq}
	\tt{.}
\end{equation}

With respect to the seven kinetic moment relations mentioned above, the first three equations in \eqref{e14} -
\eqref{e16} are referred to as conserved moment relations, where the equilibrium distribution function
$f_i^{{\sigma}eq}$ can be substituted with the distribution function $f_i^{\sigma}$. However, for the remaining four
relationships in Eqs. \eqref{e17} - \eqref{e20}, substituting $f_i^{\sigma}$ for $f_i^{{\sigma}eq}$ may result in a
discrepancy between the two sides in a non-equilibrium system. This discrepancy reflects the system's deviation from the
local equilibrium state in moment space, and it can be used to describe the TNE effects. Accordingly, the nonequilibrium
quantity is defined as follows:
\begin{equation}\label{e24}
	{\boldsymbol\Delta_{m,n}^{{\sigma}*}}={m^{\sigma}}[{\textbf{M}_{m,n}^*}({f_i}^{\sigma})-
	{\textbf{M}_{m,n}^*}({{f_i}^{{\sigma}eq}})]
	\tt{,}
\end{equation}
where ${\boldsymbol\Delta_{m,n}^{{\sigma}*}}$ describes the thermal fluctuation characteristics of microscopic
particles. The subscript ``$m,n$'' signifies the reduction of the $m$-order tensor to the $n$-order tensor. Physically,
${\boldsymbol\Delta}_{2}^{{\sigma}*}$ stands for the non-organized momentum flux,
${\boldsymbol\Delta}_{3,1}^{{\sigma}*}$ and ${\boldsymbol\Delta}_{3}^{{\sigma}*}$ denote the non-organized energy, and
${\boldsymbol\Delta}_{4,2}^{{\sigma}*}$ represents the flux of non-organized energy flux. The term
${\textbf{M}_{m,n}^*}$ refers to the kinetic central moments of the velocity and equilibrium distribution functions,
defined using the relative velocity $\textbf{{v}}_{i}^{{\sigma}*}={{\textbf{v}}_{i}^{\sigma}-{\textbf{u}}}$. The details
are presented below:
\begin{equation}\label{e25}
	\left\{
	\begin{array}{ll}
		\textbf{M}_{2}^*({f_i}^{{\sigma}})=\sum\limits_i {f_i}^{{\sigma}} \textbf{v}_{i}^{{\sigma}*}\textbf{v}_{i}^{{\sigma}*},
		\\[10pt]
		\textbf{M}_{3}^*({f_i}^{{\sigma}})=\sum\limits_i {f_i}^{{\sigma}} \textbf{v}_{i}^{{\sigma}*}
		\textbf{v}_{i}^{{\sigma}*}\textbf{v}_{i}^{{\sigma}*},\\[10pt]
		\textbf{M}_{3,1}^*({f_i}^{{\sigma}})=\sum\limits_i {f_i}^{{\sigma}}(\textbf{v}_{i}^{{\sigma}*} \cdot
		\textbf{v}_{i}^{{\sigma}*}+{\eta}_{i}^{{\sigma}2})\textbf{v}_{i}^{{\sigma}*},\\[10pt]
		\textbf{M}_{4,2}^*({f_i}^{{\sigma}})=\sum\limits_i {f_i}^{{\sigma}}(\textbf{v}_{i}^{{\sigma}*}\cdot
		\textbf{v}_{i}^{{\sigma}*}+{\eta}_{i}^{{\sigma}2})\textbf{v}_{i}^{{\sigma}*}\textbf{v}_{i}^{{\sigma}*}
		\tt{,}
	\end{array}
	\right.
\end{equation}
and
\begin{equation}\label{e26}
	\left\{
	\begin{array}{ll}
		\textbf{M}_{2}^*(f_i^{{\sigma}eq})=\sum\limits_i f_i^{{\sigma}eq} \textbf{v}_{i}^{{\sigma}*}\textbf{v}_{i}^{{\sigma}*},
		\\[10pt]
		\textbf{M}_{3}^*(f_i^{{\sigma}eq})=\sum\limits_i f_i^{{\sigma}eq} \textbf{v}_{i}^{{\sigma}*}
		\textbf{v}_{i}^{{\sigma}*}\textbf{v}_{i}^{{\sigma}*},\\[10pt]
		\textbf{M}_{3,1}^*(f_i^{{\sigma}eq})=\sum\limits_i f_i^{{\sigma}eq}(\textbf{v}_{i}^{{\sigma}*} \cdot
		\textbf{v}_{i}^{{\sigma}*}+{\eta}_{i}^{{\sigma}2})\textbf{v}_{i}^{{\sigma}*},\\[10pt]
		\textbf{M}_{4,2}^{{\sigma}*}(f_i^{{\sigma}eq})=\sum\limits_i f_i^{{\sigma}eq}(\textbf{v}_{i}^{{\sigma}*}\cdot
		\textbf{v}_{i}^{{\sigma}*}+{\eta}_{i}^{{\sigma}2})\textbf{v}_{i}^{{\sigma}*}\textbf{v}_{i}^{{\sigma}*}
		\tt{.}
	\end{array}
	\right.
\end{equation}

Based on the above-defined nonequilibrium quantity, the following nonequilibrium quantities are introduce to measure the
global TNE effect of the system:
\begin{equation}\label{e27}
	{|{\boldsymbol\Delta}_{2}^{{\sigma}*}|}={|{\boldsymbol\Delta}_{2xx}^{{\sigma}*}|}+{|{\boldsymbol\Delta}_{2xy}^{{\sigma}*}|}+
	{|{\boldsymbol\Delta}_{2yy}^{{\sigma}*}|}
	\tt{,}
\end{equation}
\begin{equation}\label{e28}
	{|{\boldsymbol\Delta}_{3,1}^{{\sigma}*}|}={|{\boldsymbol\Delta}_{3,1x}^{{\sigma}*}|}+{|{\boldsymbol\Delta}_{3,1y}^{{\sigma}*}|}
	\tt{,}
\end{equation}
\begin{equation}\label{e29}
	{|{\boldsymbol\Delta}_{3}^{{\sigma}*}|}={|{\boldsymbol\Delta}_{3xxx}^{{\sigma}*}|}+{|{\boldsymbol\Delta}_{3xxy}^{{\sigma}*}|}+
	{|{\boldsymbol\Delta}_{3xyy}^{{\sigma}*}|}+{|{\boldsymbol\Delta}_{3yyy}^{{\sigma}*}|}
	\tt{,}
\end{equation}
\begin{equation}\label{e30}
	{|{\boldsymbol\Delta}_{4,2}^{{\sigma}*}|}={|{\boldsymbol\Delta}_{4,2xx}^{{\sigma}*}|}+{|{\boldsymbol\Delta}_{4,2xy}^{{\sigma}*}|}+
	{|{\boldsymbol\Delta}_{4,2yy}^{{\sigma}*}|}
	\tt{.}
\end{equation}

The total TNE quantity is obtained by summing the above quantities, which describes the degree of the system's deviation
from its equilibrium state:
\begin{equation}\label{e31}
	{|{\boldsymbol\Delta}^{{\sigma}*}|}={|{\boldsymbol\Delta}_{2}^{{\sigma}*}|}+{|{\boldsymbol\Delta}_{3,1}^{{\sigma}*}|}+
	{|{\boldsymbol\Delta}_{4,2}^{{\sigma}*}|}+{|{\boldsymbol\Delta}_{3}^{{\sigma}*}|}
	\tt{.}
\end{equation}

Moreover, the following TNE strength function is defined to describe the global average TNE in the whole fluid system:
\begin{equation}\label{e32}
	{\overline{D}^{\sigma}}=\dfrac{1}{L_xL_y}\int_{0}^{L_x}\int_{0}^{L_y}|{\boldsymbol\Delta}^{{\sigma}*}|d{x}d{y}
	\tt{,}
\end{equation}
where ${\overline{D}^{\sigma}}$ is the global average TNE strength, and $L_{\alpha}$ denotes the boundary length, with
${\alpha}=x$ or $y$.

\section{Numerical Simulations}\label{III}

\begin{figure}[!ht]
	\begin{center}
		\includegraphics[width=0.3\textwidth]{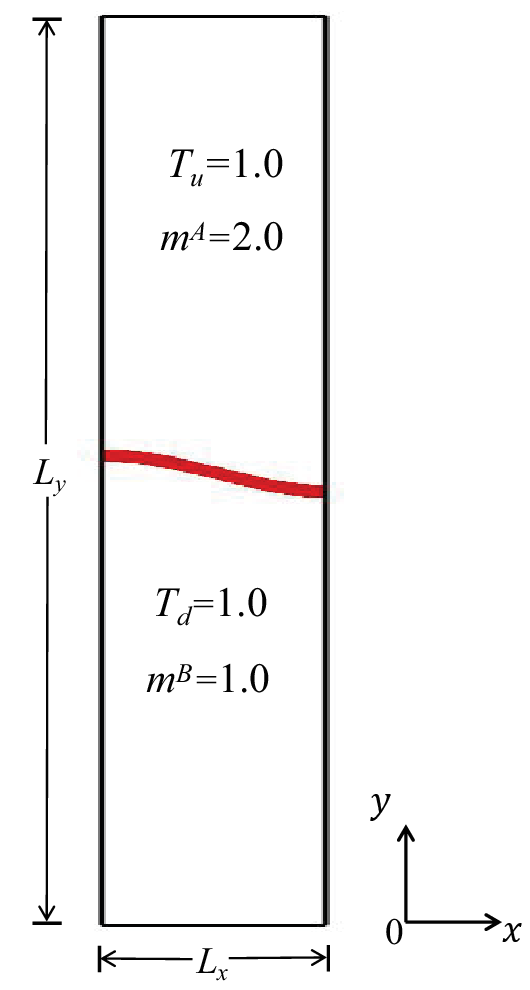}
	\end{center}
	\caption{The initial configuration for the compressible RT instability.}\label{FIG02}
\end{figure}

In this section, we employ the two-component DBM to explore the impact of the time-varying acceleration on the
compressible RT instability. The initial configuration, depicted in Fig. \ref{FIG02}, consists of a domain with a width
$L_x=0.025$ and a length $L_y=0.2$. The two-dimensional computational domain is initially divided into two distinct
regions, separated by a perturbed interface located at the midpoint of the flow field. The upper region is occupied by
component $A$, characterized by a particle mass $m^{\sigma} = 2.0$ and a temperature $T_{u} = 1.0 $. In contrast,
component $B$ fills the lower region, with a particle mass $ m^{\sigma} = 1.0$ and a temperature $ T_{d} = 0.1 $. The
system is subjected to a gravitational field with a time-varying acceleration $\textbf{a} = (0, a_y)$, where $a_y $ is
defined as:
\begin{equation}\label{e33}
	a_y=a_0+A_0\sin(\omega t+\Phi)
	\tt{,}
\end{equation}
where $a_0 = -1$ indicates the initial acceleration, $\omega=2 \pi/T_0$ denotes the frequency, $T_0$ represents the
period, $A_0$ is the amplitude, and $\Phi$ signifies the phase of the time-varying acceleration. Under the condition of
static equilibrium, ${\nabla}{p}={\rho}{\textbf{a}}$, the initial concentrations are set as,
\begin{equation}\label{e34}
	\begin{split}
		\left\{\begin{array}{l}
			{{n}^{A}}=\dfrac{p_{m}}{T_{u}}{\exp}\left[
			\dfrac{m^{A}{g}}{T_{u}}({y_{m}}-{y})\right],\ {{n}^{B}}=0,\ {y}>{y_{m}},\\\\
			{{n}^{B}}=\dfrac{p_{m}}{T_{d}}{\exp}\left[
			\dfrac{m^{B}{g}}{T_{d}}({y_{m}}-{y})\right],\ {{n}^{A}}=0,\ {y}<{y_{m}},\\
		\end{array}\right.\\
	\end{split}
\end{equation}
where $p_{m}=4.0$ represents the pressure at the material interface and ${y_m}=L_y/2+A \cos(\pi x/L_x)$ denotes the
interface location, with a perturbation amplitude of $A=L_y/50$. Moreover, the hyperbolic tangent function ${\tanh}$ is
used to smooth the transition layer of temperature across the material interface as follows,
\begin{equation}\label{e35}
	{T}=\dfrac{{T_u}+{T_d}}{2}+\dfrac{{T_u}-{T_d}}{2}{\tanh}\dfrac{{y}-{y_m}}{W}
	\tt{,}
\end{equation}
where the interfacial transition layer width is set to ${W}={{L_y}/200}$.

In addition, the relaxation time is $\tau = 4 \times 10^{-5}$, the discrete parameters are $(v_a$, $v_b$, $v_c$, $v_d$,
${{\eta}_0}$)=($5.5$, $2.5$, $0.7$, $0.9$, $5.3)$. The mirror-reflection boundary conditions are applied in the $x$ and
$y$ directions. A grid convergence test is conducted to validate the simulation results, seen more details in Appendix
\ref{A}. As a result, to ensure computational efficiency with numerical accuracy, the grid number $200 \times 1600$ is
chosen for the following simulations.

\subsection{Effect of the period of time-varying acceleration on RT instability}

\begin{figure}[!ht]
	\centering
	\includegraphics[scale=0.35]{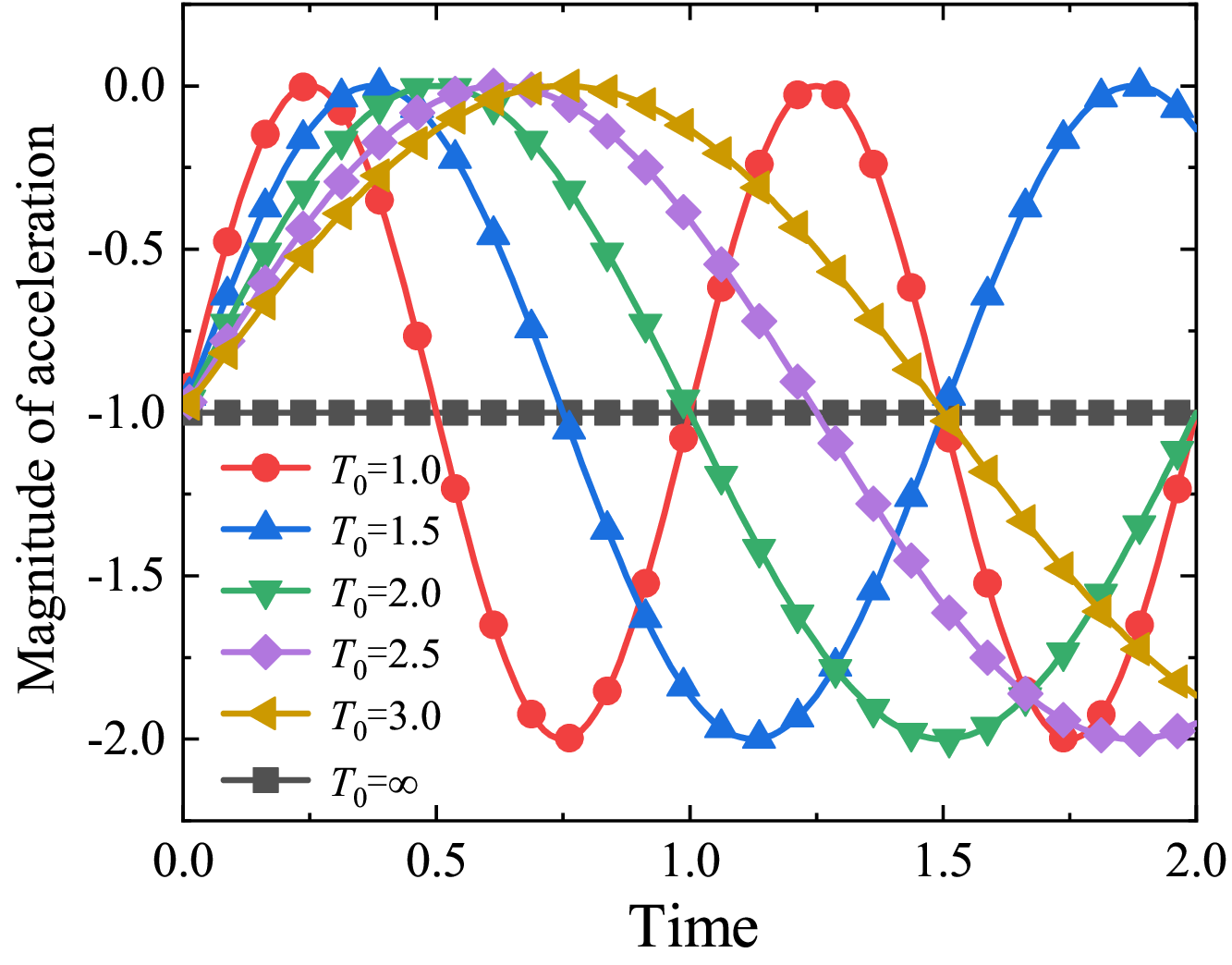}
	\caption{The evolution of acceleration for various periods of the time-varying acceleration.}\label{FIG03}
\end{figure}

The change in period adjusts the frequency of acceleration, thereby affecting the vibrational characteristics and
development rate of the interface disturbances. Therefore, studying the impact of period variation on fluid systems is
of significant importance. In this section, the effect of the period of time-varying acceleration $T_0$ on the evolution
of RT system is explored. To isolate the effect of $T_0$, the amplitude and phase are fixed at $A_0=1$ and $\Phi=0$,
respectively. The chosen values for $T_0$ are $1.0$, $1.5$, $2.0$, $2.5$, $3.0$, and $\infty$. Notably, $ T_0 = \infty $
corresponds to the case of a constant acceleration $a_0$. Figure \ref{FIG03} illustrates the acceleration evolution for
these different periods of the time-varying acceleration.

It is well-established that the physical gradient is intricately connected to the TNE effect. To elucidate the
nonequilibrium mechanisms underlying the evolution of RT instability, we first focus on analyzing the density gradient.
The global average density gradient in the $x$ direction is given by
\begin{equation}\label{e37}
	|{\overline{{\nabla_{x}}{\rho}}}| = \iint_{\Omega} |{{\nabla_{x}}{\rho}}| dxdy/{(L_xL_y)}
	\tt{,}
\end{equation}
the global average density gradient in the $y$ direction is expressed as
\begin{equation}\label{e38}
	|{\overline{{\nabla_{y}}{\rho}}}| = \iint_{\Omega} |{{\nabla_{y}}{\rho}}| dxdy/{(L_xL_y)}
	\tt{,}
\end{equation}
and the global average density gradient is defined as
\begin{equation}\label{e39}
	|{\overline{{\nabla}{\rho}}}| = \iint_{\Omega} |{{\nabla}{\rho}}| dxdy/{(L_xL_y)}
	\tt{,}
\end{equation}
where $\Omega\in[0,L_x]\times[0,L_y]$.

\begin{figure}[htbp]
	\begin{center}
		\includegraphics[width=0.5\textwidth]{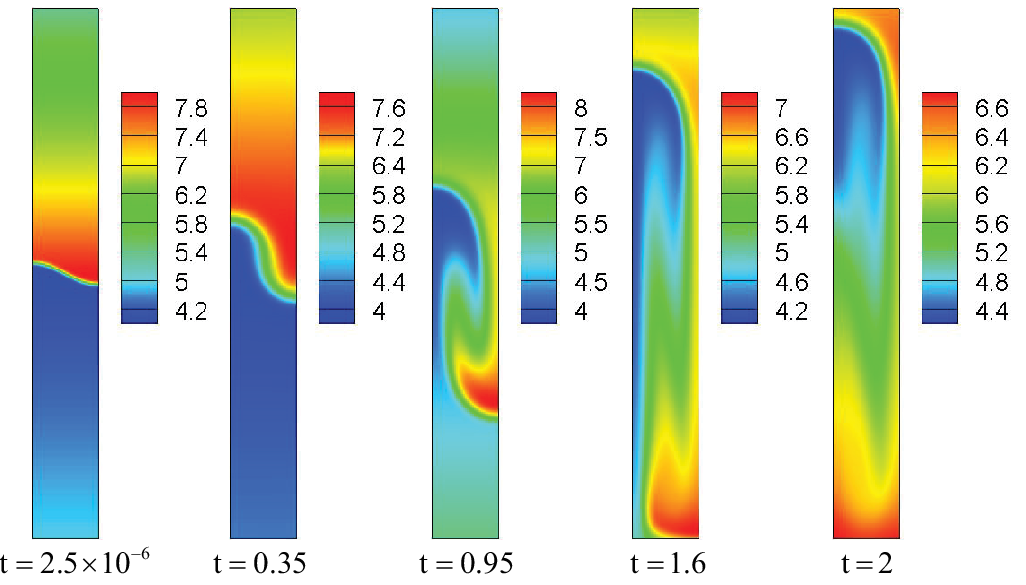}
	\end{center}
	\caption{Density contours in the evolution of the RT instability for the case of $T_0=2.0$.}\label{FIG04}
\end{figure}

To provide a clearer understanding, Fig. \ref{FIG04} displays the density contours during the evolution of the RT
instability for the case of $T_0=2.0$ at time instants $t = 2.5 \times 10^{-6}$, $0.35$, $0.95$, $1.6$, and $2$,
respectively. It is evident that, due to the dissipation and diffusion effects, the transition layer broadens and
smooths. Meanwhile, the interface elongates and deforms, gradually forming characteristic spike and bubble structures.
As time goes on, the mixing between the two components intensifies further, and the spike (bubble) structure continues
to extend downward (upward).
\begin{figure*}[htbp]
	\begin{center}
		\includegraphics[width=0.8\textwidth]{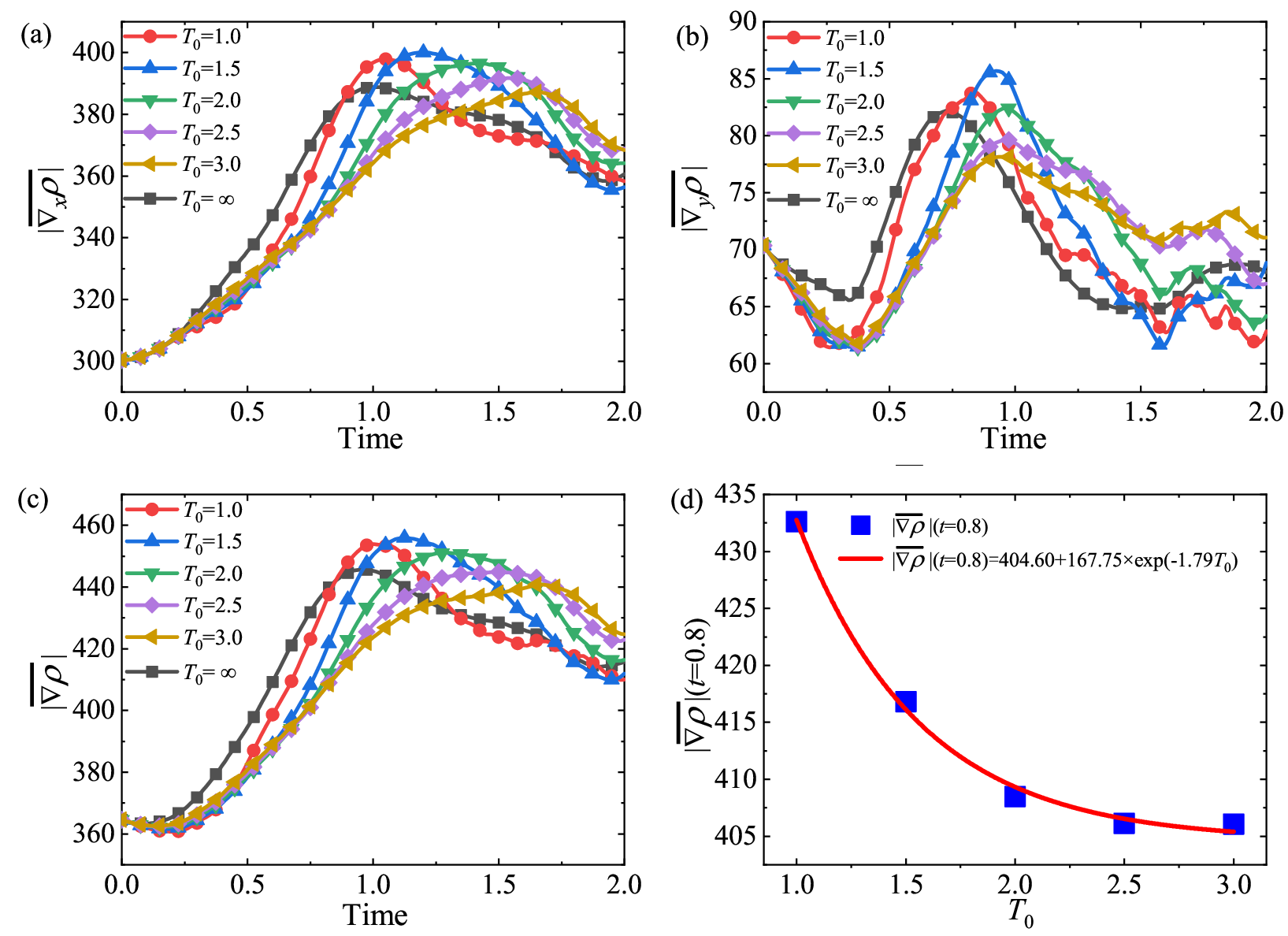}
	\end{center}
	\caption{Evolution of the global average density gradient with various periods of time-varying acceleration: (a)
		$|{\overline{{\nabla_{x}}{\rho}}}|$, (b) $|{\overline{{\nabla_{y}}{\rho}}}|$, (c) $|{\overline{{\nabla}{\rho}}}|$. (d)
		The relationship between the value of the global average density gradient at $t=0.8$ and the period of time-varying
		acceleration.}\label{FIG05}
\end{figure*}

Figure \ref{FIG05} (a) depicts the evolution of the global average density gradient in the $x$ direction
$|{\overline{{\nabla_{x}}{\rho}}}|$ under different periods of time-varying acceleration $T_0$. The value of
$|{\overline{{\nabla_{x}}{\rho}}}|$ initially increases and then decreases, which results from the competitive mechanism
of the elongation of the fluid interface and the interpenetration of the two components. In the first stage, with the
slow formation of spike and bubble structures near the interface, the elongation of the fluid interface plays a dominant
role, resulting in an increase in $|{\overline{{\nabla_{x}}{\rho}}}|$. In the second stage, the two components mixed
with a deeper degree, the transition layer gradually widens, and the vortex structures gradually dissipate. During this
process, the effects of dissipation and diffusion, leading to a reduction in $|{\overline{{\nabla_{x}}{\rho}}}|$. In
addition, compared to the case of constant acceleration, the period of time-varying acceleration $T_0$ suppresses the
evolution of the RT system in the early stage but promotes it in the later stage.

Figure \ref{FIG05} (b) illustrates the evolution of the global average density gradient in the $y $ direction $
|{\overline{{\nabla_{y}}{\rho}}}|$ under various periods of time-varying acceleration $T_0$. Generally,
$|{\overline{{\nabla_{y}}{\rho}}}|$ decreases first, then increases, and finally declines. Take $T_0=1.0$ as an example,
from $t=0.0$ to $0.35$, $|{\overline{{\nabla_{y}}{\rho}}}|$ initially decreases. At the beginning, two disturbance waves
emerge at the interface. As time progresses, the disturbance waves propagate around and dissipate gradually, reducing
the local physical quantity gradient. Subsequently, from $t=0.35$ to $t=0.95$, $|{\overline{{\nabla_{y}}{\rho}}}|$
increases rapidly. In this process, as the two components interpenetrate, the fluid interface elongates and twists
vertically, causing the density in the $y$ direction to become inhomogeneous. Therefore, the
$|{\overline{{\nabla_{y}}{\rho}}}|$ increases rapidly. Finally, for $t > 0.95$, as the mixing of the two components
becomes nearly complete, the vortex structures gradually dissipate, and the physical gradient in the $ y $ direction
smooths out, resulting in a gradual decrease in $|{\overline{{\nabla_{y}}{\rho}}}|$.

Figure \ref{FIG05} (c) presents the evolution of the global average density gradient $|{\overline{{\nabla}{\rho}}}|$
under different periods of time-varying acceleration $T_0$. In fact, $|{\overline{{\nabla}{\rho}}}|$ is determined by
its components in the $x$ and $y$ directions. Therefore, the physical mechanisms of the evolution of the global average
density gradient $|{\overline{{\nabla}{\rho}}}|$ can be elucidated through a comprehensive analysis of
$|{\overline{{\nabla_{x}}{\rho}}}|$ and $|{\overline{{\nabla_{y}}{\rho}}}|$. Additionally, to understand the nonlinear
characteristics at the early stage of RT instability evolution, the fitting relationship between
$|{\overline{{\nabla}{\rho}}}|$ and $T_0$ is depicted in Fig. \ref{FIG05} (d) at a time $t=0.8$. The specific fitting
function is given by $|{\overline{{\nabla}{\rho}}}|_{t=0.8}=404.60+167.75\times\exp(-1.79T_0)$. Clearly,
$|{\overline{{\nabla}{\rho}}}|$ decreases exponentially with increasing $T_0$. Physically, a smaller period of
time-varying acceleration induces more pronounced changes in the physical field, resulting in sharper density gradients
$|{\overline{{\nabla}{\rho}}}|$ in the RT system.
\begin{figure*}[htbp]
	\begin{center}
		\includegraphics[width=0.7\textwidth]{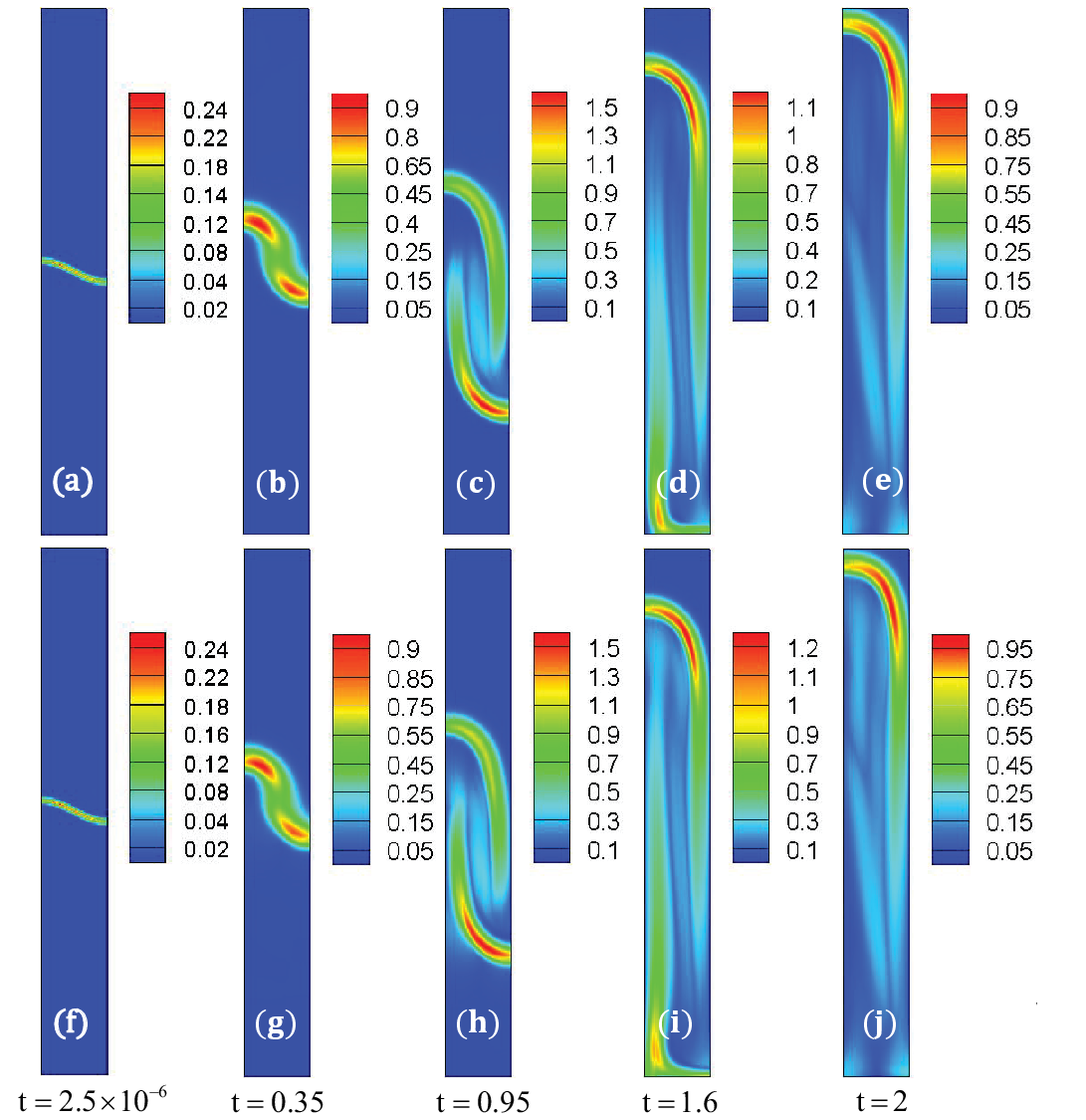}
	\end{center}
		\caption{Spatial distributions of the TNE strength in the case of $T_0=2.0$. The first group of subgraphs (a)-(e)
		correspond to the component $\sigma=A$, and the second group of subgraphs (f)-(j) correspond to the component
		$\sigma=B$.}\label{FIG06}
\end{figure*}

To provide a more intuitive understanding of the TNE behaviors in RT instability, Fig. \ref{FIG06} displays the contours
of the global average TNE strength for the case of $T_0=2.0$ at six characteristic time instants $t = 2.5 \times
10^{-6}$, $0.35$, $0.95$, $1.6$, and $2$, respectively. It can be seen that the non-equilibrium strength near the
interface is highest during the early stage, primarily due to the sharp physical gradient at the interface. As time
progresses, the transition layer gradually expands, with the higher-density region developing downward to form spike
structures and the lower-density region extending upward to form bubble structures. Throughout the process, the
non-equilibrium intensity remains consistently high around the spike and bubble structures. In the later stage, the two
components fully mix, the vortex structures gradually blur due to the diffusion and dissipation, leading the system
towards equilibrium.
\begin{figure*}[htbp]
	\begin{center}
		\includegraphics[width=0.85\textwidth]{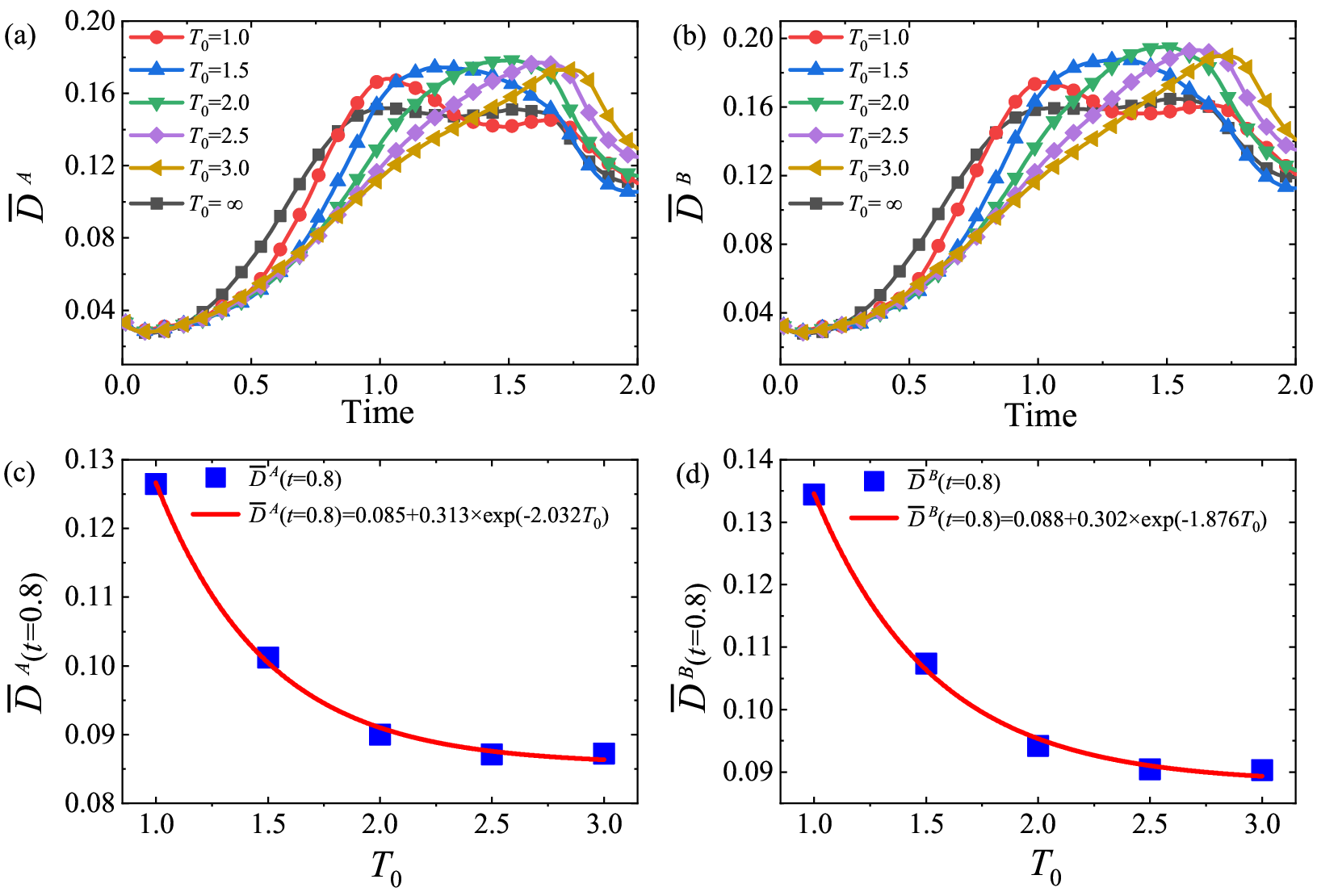}
	\end{center}
\caption{The global average TNE strength of components $A$ (a) and $B$ (b) with various periods of time-varying
	acceleration. The fitting relationship of the value of ${\overline{D}^{A}}$ (c) and ${\overline{D}^{B}}$ (d) at $t=0.8$
	and the period of time-varying acceleration $T_0$.}\label{FIG07}
\end{figure*}

Furthermore, Figs. \ref{FIG07} (a) and (b) illustrate the evolution of the global average TNE strength
${\overline{D}^{\sigma}}$ for the two components, $\sigma=A$ and $\sigma=B$, respectively. It is evident that
${\overline{D}^{\sigma}}$ initially shows a slight decline, then rises, and eventually decreases. Physically, in the
early phase, two disturbance waves emerge around the material interface, and then propagate outward with attenuation of
energy and reduction in physical quantity gradients, leading to a drop in local TNE intensity. Subsequently, as the
fluids interpenetrate, vortex structures form, increasing the complexity of the fluid's structure and the contact area
between the two components, thereby enhancing local TNE strength. In the later phase, diffusion and dissipation weaken
the physical gradient, resulting in a reduction in local TNE strength.

Additionally, Figs. \ref{FIG07} (c) and (d) show the relationships between the global average TNE strength
${\overline{D}^{\sigma}}$ at $t=0.8$ and the period of time-varying acceleration $T_0$. The fitting relationships for
the two components  $\sigma=A$ and $\sigma=B$ are given by:
${\overline{D}^{A}}_{t=0.8}=0.085+0.313\times\exp(-2.032T_0)$ and
${\overline{D}^{B}}_{t=0.8}=0.088+0.302\times\exp(-1.876T_0)$, respectively. It is evident that
${\overline{D}^{\sigma}}$ decrease exponentially with $T_0$ increasing. Moreover, a comparison between the cases of
time-varying and constant accelerations reveals that the periods of those time-varying accelerations suppress the TNE
effects of the system in the early stage but enhance the TNE effects in the later stage.
\begin{figure*}[htbp]
	\begin{center}
		\includegraphics[width=0.85\textwidth]{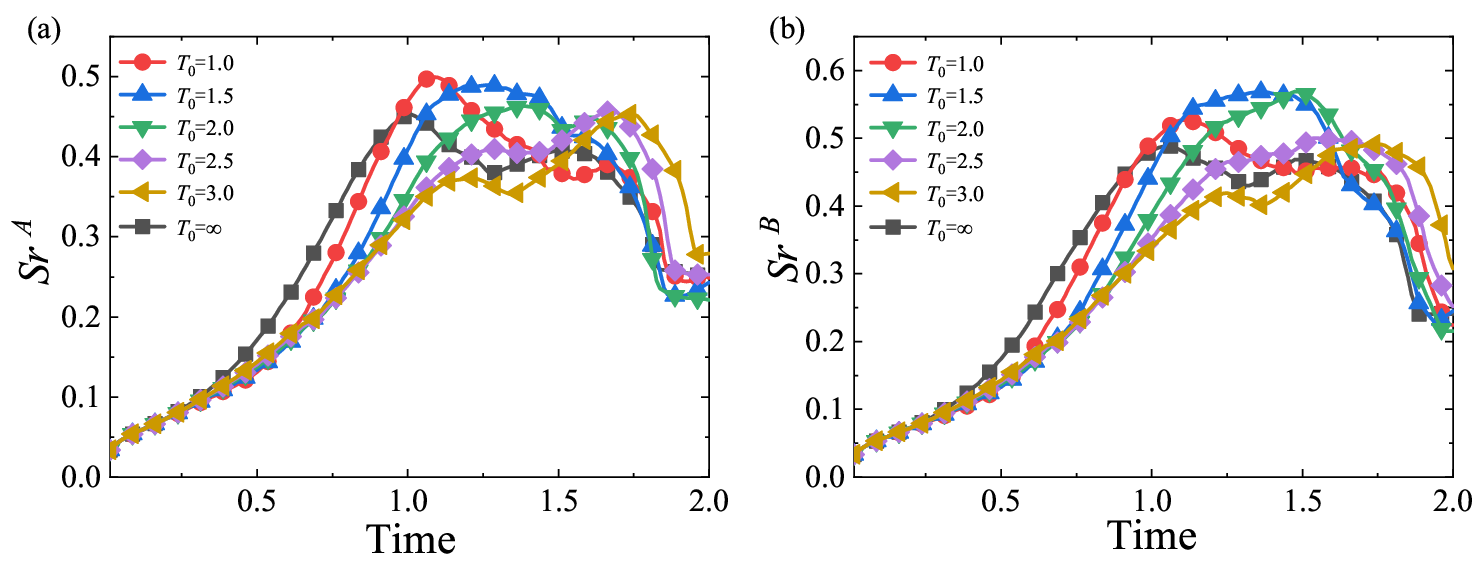}
	\end{center}
		\caption{Evolution of the proportion of nonequilibrium region of component $A$ (a) and $B$ (b) under various periods of
		time-varying acceleration. }\label{FIG08}
\end{figure*}

To further analyze the global average TNE intensity of the system, Fig. \ref{FIG08} illustrates the evolution of the
proportion of non-equilibrium regions $Sr^{\sigma}$ for the two components $\sigma=A$ and $\sigma=B$. It is shown that,
for all cases, the $Sr^{\sigma}$ initially increases and then decreases. And in the early phase, the curves nearly
overlap, while in the later phase the time for $Sr^{\sigma}$ to reach its peak becomes longer as the period of
time-varying acceleration increases (except the special case of $T_0 = \infty $). Physically, the time-varying effects
of acceleration have not yet manifested in the early stages, leading to the overlapping curves in the initial phase.
During the ascending phase, the perturbation interface continuously stretches and becomes dominant, leading to an
increase in the contact area between the two components. As a result, the non-equilibrium region expands and
$Sr^{\sigma}$ rises. In the descending phase, the dominant mechanism shifts to the thorough mixing of the two
components, causing the spike and bubble structures in the fluid system to dissipate due to diffusion, reducing the
non-equilibrium area and causing $Sr^{\sigma}$ to decrease. Furthermore, Fig. \ref{FIG03} shows that as the period
increases, the rate at which acceleration changes from $-1$ to $0$ slows down. This slows the weakening of the pressure
difference, the descent of the heavy fluid, the ascent of the light fluid, and the stretching of the interface.
Consequently, the rate of increase in the non-equilibrium area also slows down, and the time to reach the peak is
therefore extended.

\subsection{Effect of the amplitude of time-varying acceleration on RT instability}

\begin{figure}[!ht]
	\centering
	\includegraphics[scale=0.350]{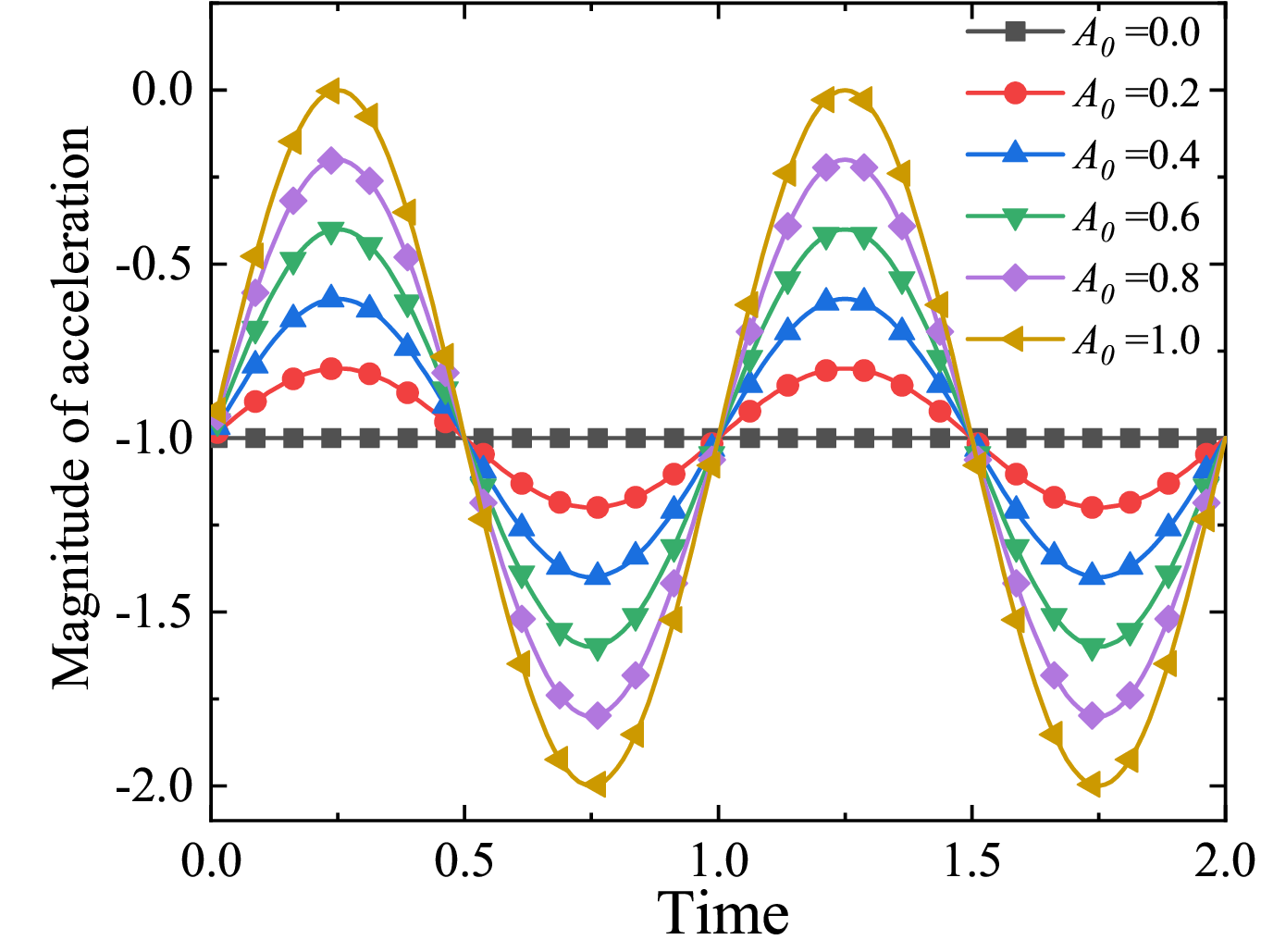}
	\caption{The evolution of the acceleration with different $A_0$.}\label{FIG09}
\end{figure}

The amplitude of the time-varying acceleration is related to the range of acceleration fluctuations. Different
amplitudes lead to varying acceleration changes experienced by the interface, which in turn affects the intensity of the
disturbances acting on the interface. In this section, the effect of the amplitude of time-varying acceleration $A_0$ on
RT instability is examined. In the numerical simulation, the period $T_0$ and phase $\Phi$ are fixed at $1.0$ and $0.0$,
respectively, while the amplitude $A_0$ is varied from $0.0$ to $1.0$ in increments of $0.2$. Figure \ref{FIG09} shows
the evolution of acceleration for different values of time-varying amplitude. It is important to note that when $A_0 =
0$, the acceleration remains constant $-1.0$.
\begin{figure*}[htbp]
	\begin{center}
		\includegraphics[width=0.85\textwidth]{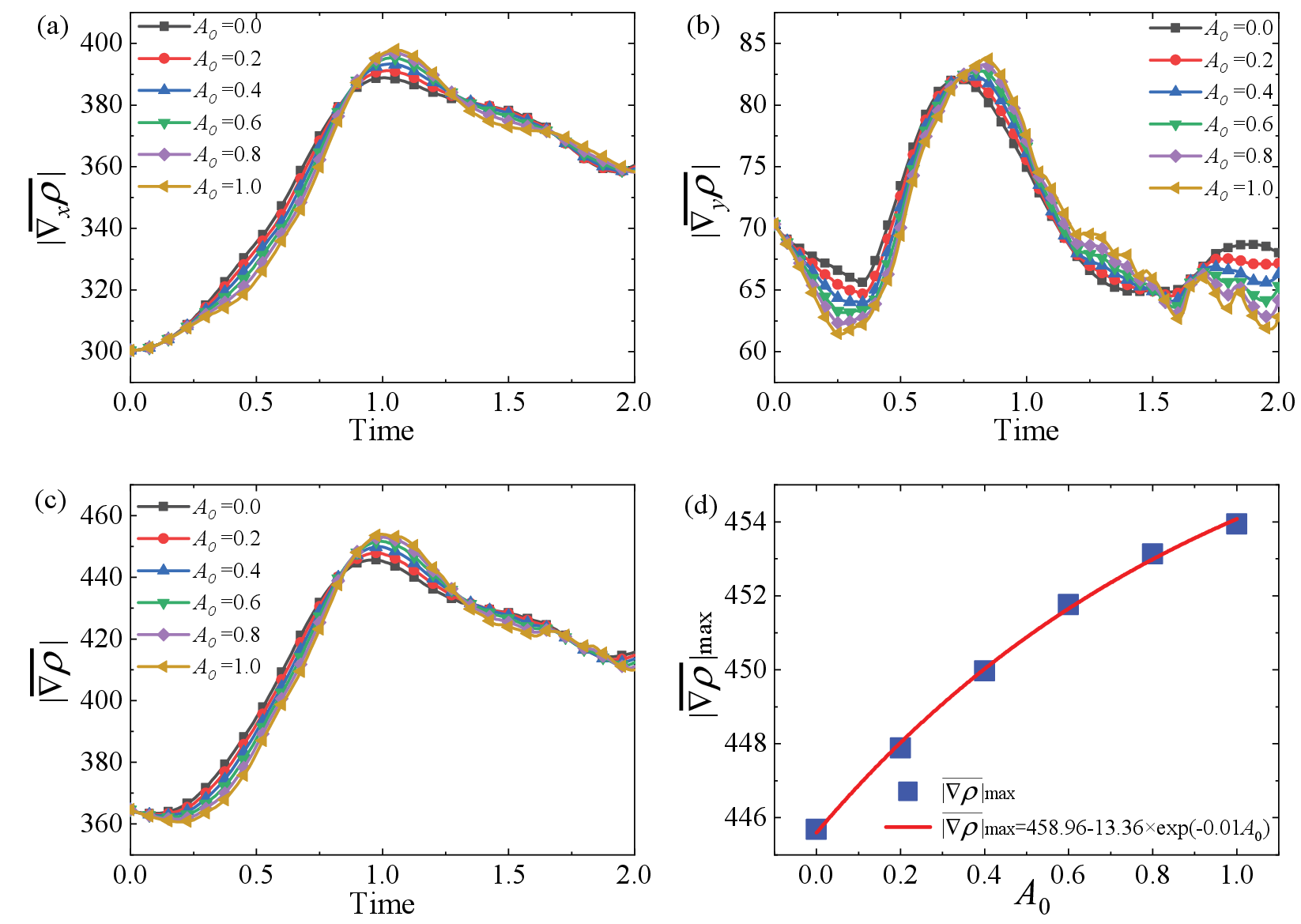}
	\end{center}
	\caption{Evolution of the average density gradient with various amplitudes of time-varying acceleration: (a)
		$|{\overline{{\nabla_{x}}{\rho}}}|$, (b) $|{\overline{{\nabla_{y}}{\rho}}}|$, and (c) $|{\overline{{\nabla}{\rho}}}|$.
		(d) The relationship between the maximum of $|{\overline{{\nabla}{\rho}}}|$ and the amplitude of time-varying
		acceleration.}\label{FIG10}
\end{figure*}

Figure \ref{FIG10} (a) illustrates the evolution of the average density gradient in the $x$ direction
$|{\overline{{\nabla_{x}}{\rho}}}|$. It can be seen that $|{\overline{{\nabla_{x}}{\rho}}}|$ first increases and then
decreases. Physically, $|{\overline{{\nabla_{x}}{\rho}}}|$ reflects the inhomogeneity of the density field in the $x$
direction and is primarily manifested in the amplitude of the disturbed interface. In the early stage, the influence of
acceleration on the density gradient in the $x$ direction is relatively weak. Therefore, the disturbed interface in each
case has similar change in amplitude, and the curves are close to each other.
As time progresses, the disturbed interface is stretched during the evolution of the fluid system, shear stress
facilitates the gradual formation of vortex structures within the fluid, and the impact of acceleration gradually
increases. This process enhances fluid nonlinearity, leading to an increase in the density gradient. In the later stage,
the dissipative and diffusive effects of the system make the interface become blurred, and the vortex structures
gradually disappear, leading to a smoothing of the physical gradient and a decrease in the density gradient.

Figure \ref{FIG10} (b) depicts the evolution of the average density gradient in the $y$ direction
$|{\overline{{\nabla_{y}}{\rho}}}|$. It can be observed that $|{\overline{\nabla_{y}}\rho}|$ initially decreases, then
rises, and finally declines with oscillations. In the initial decline phase, as $ A_0 $ increases,
$|{\overline{{\nabla_{y}}{\rho}}}|$ decreases more rapidly and significantly. As the absolute value of the acceleration
decreases, the external force on the fluids weakens gradually, the fluids rise due to the pressure difference in the $y$
direction, the density field tends to be uniform, and the value of the density gradient falls.
Additionally, as the amplitude $ A_0 $ increases, the absolute value of the time-varying acceleration reduces faster,
leading to a weakening in the changes of density stratification. Subsequently, the interface is stretched along the $y$
direction and gradually curls, forming spike and bubble structures. This results in an enhanced density variation in the
$y$ direction, leading to a rapid increase in $|{\overline{{\nabla_{y}}{\rho}}}|$. When $t > 0.85$, the mixing of the
two components reaches saturation, vortex structures gradually dissipate due to diffusion, and the density field becomes
smoother, resulting in a decrease in $|\overline{\nabla_{y} \rho}|$.

Figure \ref{FIG10} (c) represents the evolution of the average density gradient $|{\overline{{\nabla}{\rho}}}|$. For all
cases, $|{\overline{{\nabla}{\rho}}}|$ decreases first, then increases,and finally decreases with slight oscillations.
Actually, the evolution of the $|{\overline{{\nabla}{\rho}}}|$ can be inferred by analyzing its components
$|{\overline{{\nabla_{x}}{\rho}}}|$ and $|{\overline{{\nabla_{y}}{\rho}}}|$. In addition, Fig. \ref{FIG10} (d) shows the
the maximum of $|{\overline{{\nabla}{\rho}}}|_{max}$ versus the amplitude of time-varying acceleration $A_0$. The
fitting relation is $|{\overline{{\nabla}{\rho}}}|_{max}=458.96-13.36\times\exp(-0.01A_0)$. Obviously, the
$|{\overline{{\nabla}{\rho}}}|_{max}$ increases exponentially as the $A_0$ increases.
\begin{figure*}[htbp]
	\begin{center}
		\includegraphics[width=0.85\textwidth]{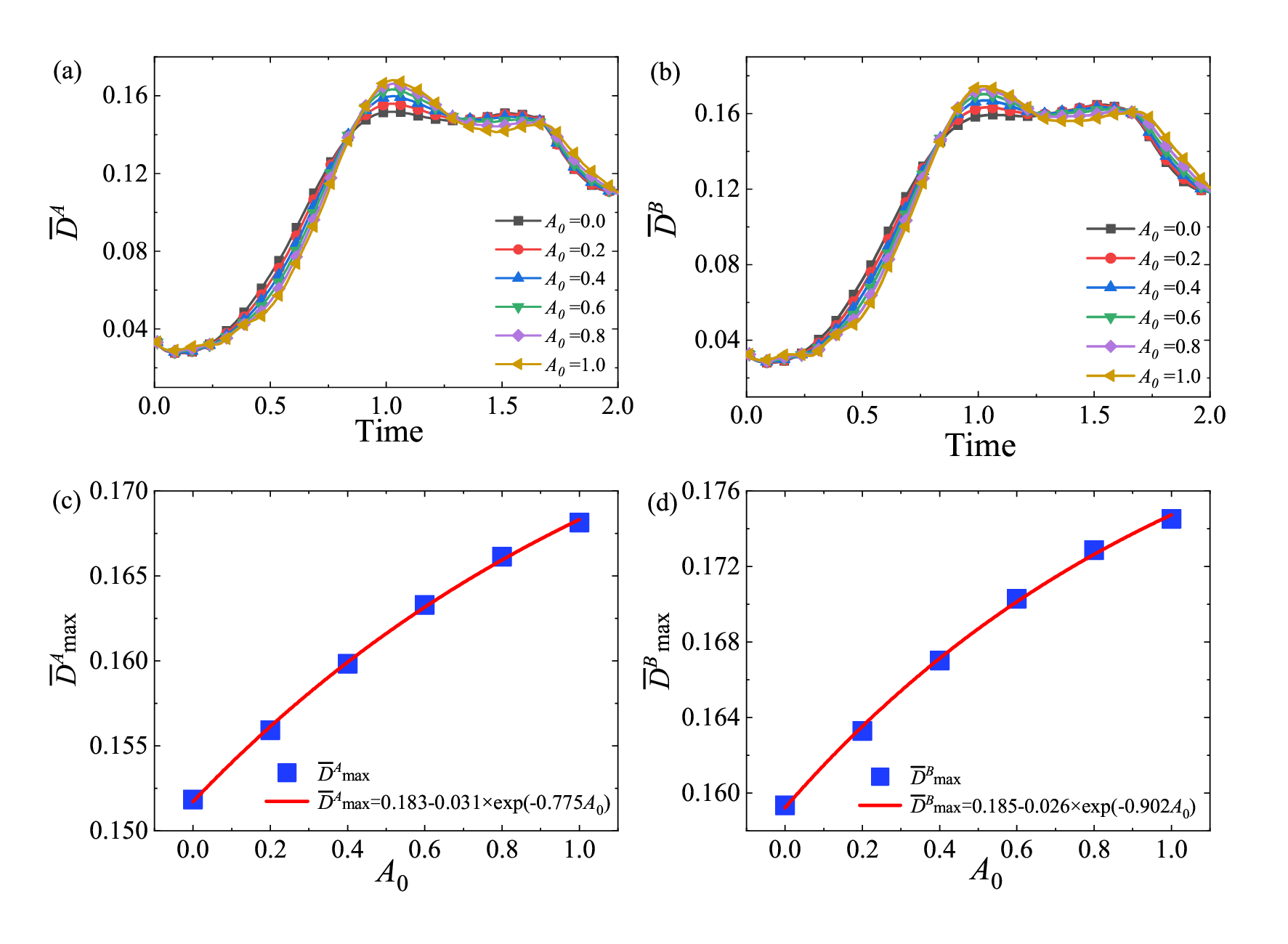}
	\end{center}
	\caption{The average TNE strength of components $A$ (a) and $B$ (b) with various amplitudes of time-varying
	acceleration. The fitting curve of the maximum of the average TNE strength for component $A$ (c) and $B$ (d) with
	various amplitudes of time-varying acceleration.}\label{FIG11}
\end{figure*}

Furthermore, Figs. \ref{FIG11} (a) and (b) display the evolution of the average TNE strength ${\overline{D}^{\sigma}}$
under different amplitudes of time-varying acceleration $A_0$. It is evident that ${\overline{D}^{\sigma}}$ initially
decreases, then rises, and finally declines with oscillations. Take $A_0=1.0$ as an example. From $t = 0.0$ to $t =
0.125$, the interface becomes smoother under the effect of thermal diffusion, and the local physical gradient weakens,
resulting in a slight decrease in ${\overline{D}^{\sigma}}$. Subsequently, from $t = 0.125$ to $1.0$, the formation of
spike and bubble structures increases the contact area between the two components, enhancing the TNE effect and causing
${\overline{D}^{\sigma}}$ to rise rapidly. Finally, the mixing degree of the two components approaches saturation, and
the spike and bubble structures gradually disappear during the diffusion process of the two components. This leads to a
weakening of the local physical gradient, resulting in a decrease in ${\overline{D}^{\sigma}}$. It should be mentioned
that the oscillations result from the periodic changing acceleration, which leads to the spacial nonuniform distribution
of the physical field and the emergence of thermodynamic nonequilibrium effects.

Additionally, Figs. \ref{FIG11} (c) and (d) illustrate the fitting curve between the amplitude of time-varying
acceleration $A_0$ and the maximum of the average TNE strength ${\overline{D}^{\sigma}_{max}}$ for the two components
$A$ and $B$. These relationships are represented by exponential functions:
${\overline{D}^{A}_{max}}=0.183-0.031\times\exp(-0.775A_0)$ and
${\overline{D}^{B}_{max}}=0.185-0.026\times\exp(-0.902A_0)$, respectively. Physically, as $A_0$ increases, the changes
in acceleration become more pronounced, leading to a more complex fluid system and a larger nonequilibrium region. As a
result, the TNE effect is enhanced by the acceleration with a larger amplitude.
\begin{figure*}[htbp]
	\begin{center}
		\includegraphics[width=0.8\textwidth]{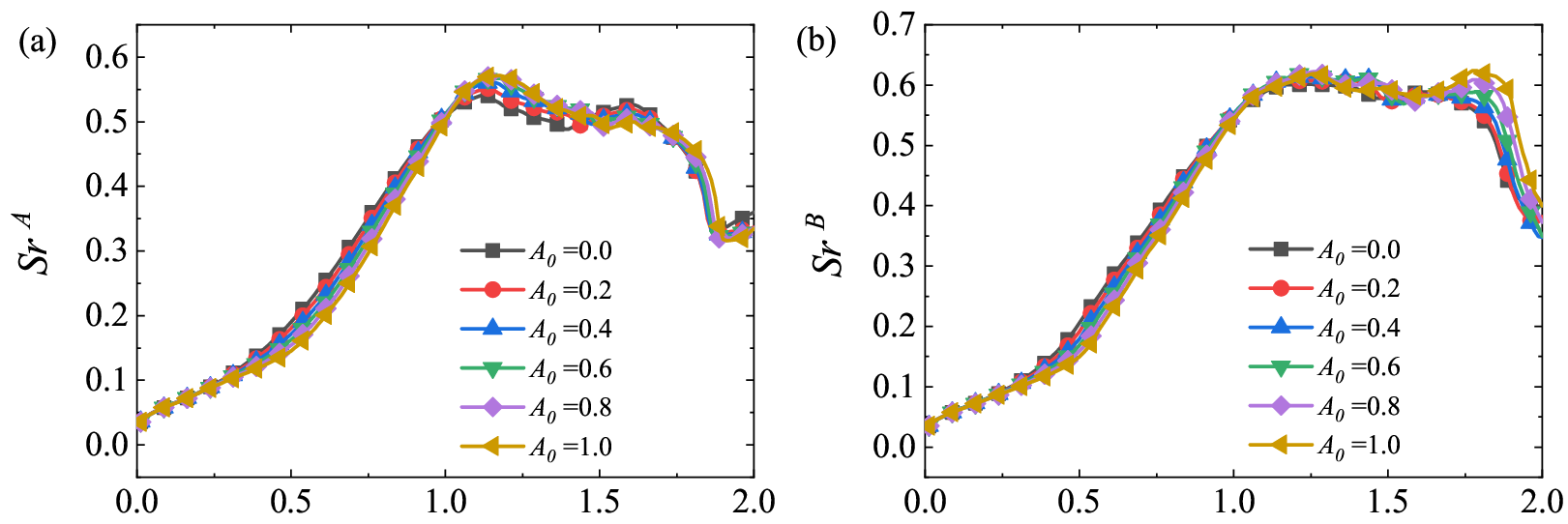}
	\end{center}
	\caption{Evolution of the proportion of nonequilibrium region of component $A$ (a) and $B$ (b) under various amplitudes
		of time-varying acceleration. }\label{FIG12}
\end{figure*}

Figures \ref{FIG12} (a) and (b) show the evolution of the proportion of non-equilibrium region ${Sr^{\sigma}}$ with
various amplitudes of time-varying acceleration $A_0$. It can be found that ${Sr^{\sigma}}$ increases first and then
decreases with oscillations. Physically, ${Sr^{\sigma}}$ increases with the expanding contact area between the two
components in the early stage and decreases as the TNE strength of the system diminishes in the later stage.

\subsection{Effect of the phase of time-varying acceleration on RT instability}

\begin{figure}[!ht]
	\centering
	\includegraphics[scale=0.350]{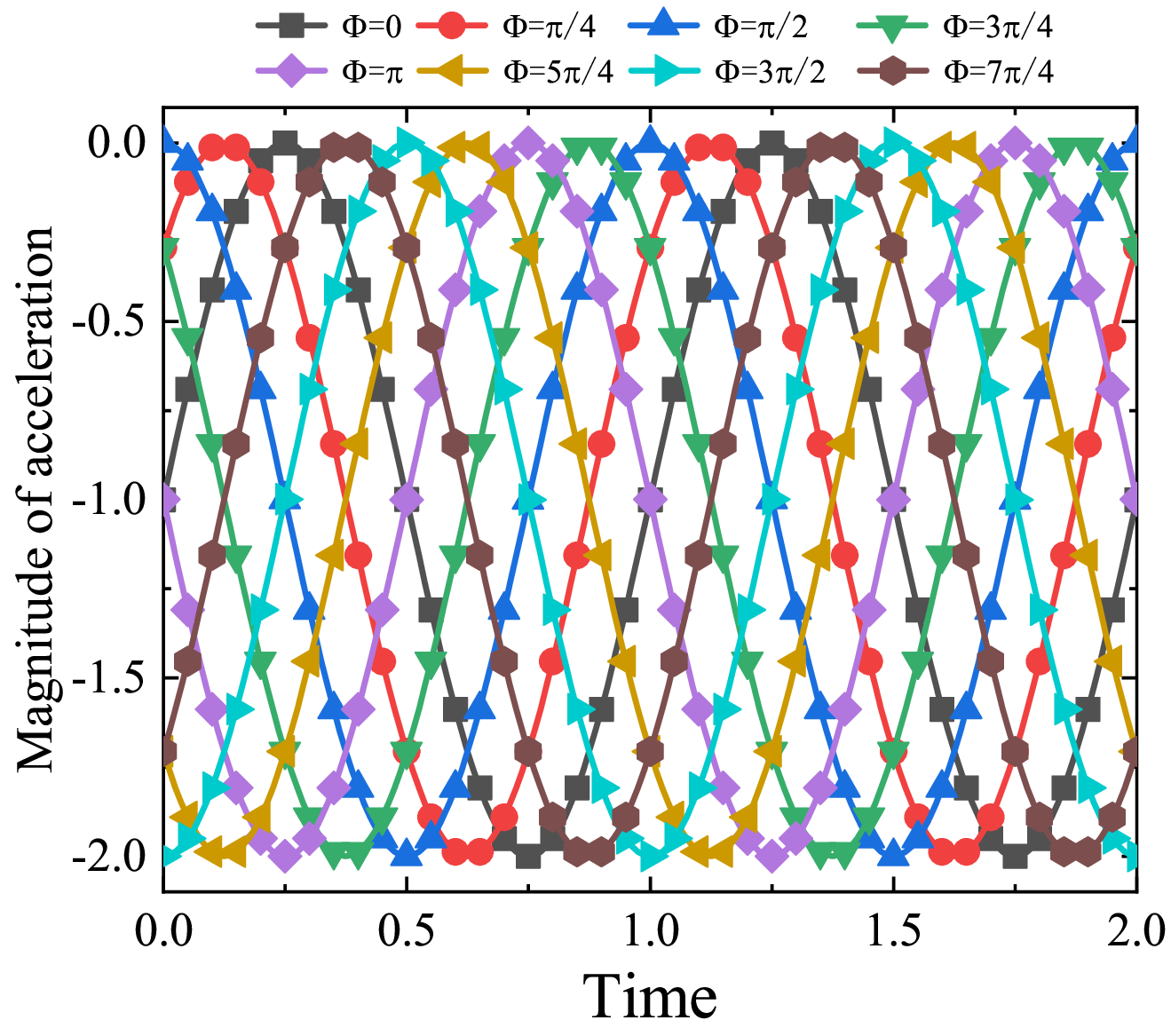}
	\caption{The evolution of the acceleration with different phases.}\label{FIG13}
\end{figure}

\begin{figure*}[!ht]
	\begin{center}
		\includegraphics[width=0.8\textwidth]{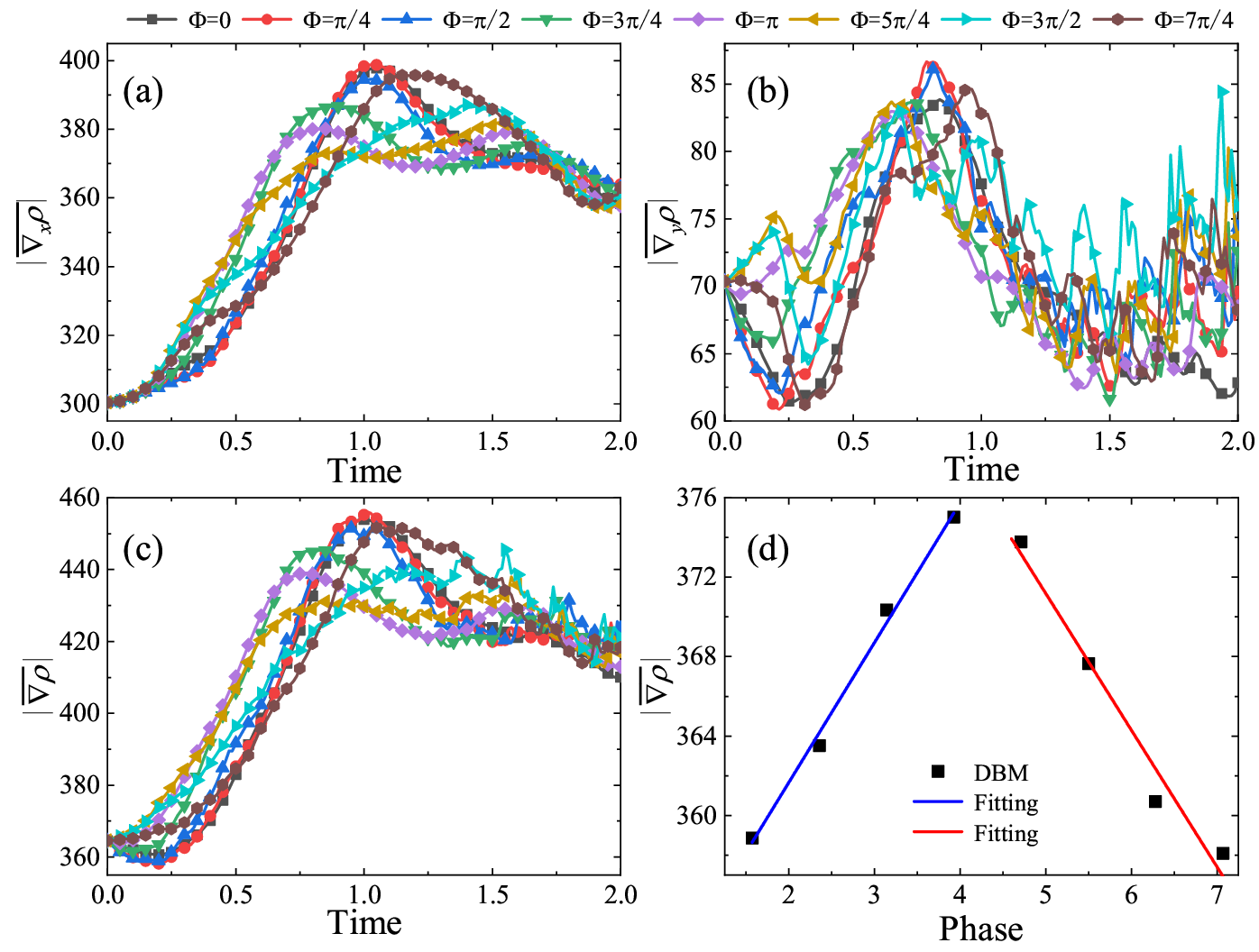}
	\end{center}
\caption{Evolution of the average density gradient with various phases of time-varying acceleration: (a)
	$|{\overline{{\nabla_{x}}{\rho}}}|$, (b) $|{\overline{{\nabla_{y}}{\rho}}}|$, and (c) $|{\overline{{\nabla}{\rho}}}|$.
	(d) The relationship between the $|{\overline{{\nabla}{\rho}}}|$ at $t=0.2$ and the phase of time-varying
	acceleration.}\label{FIG14}
\end{figure*}

\begin{figure*}[!ht]
	\begin{center}
		\includegraphics[width=0.8\textwidth]{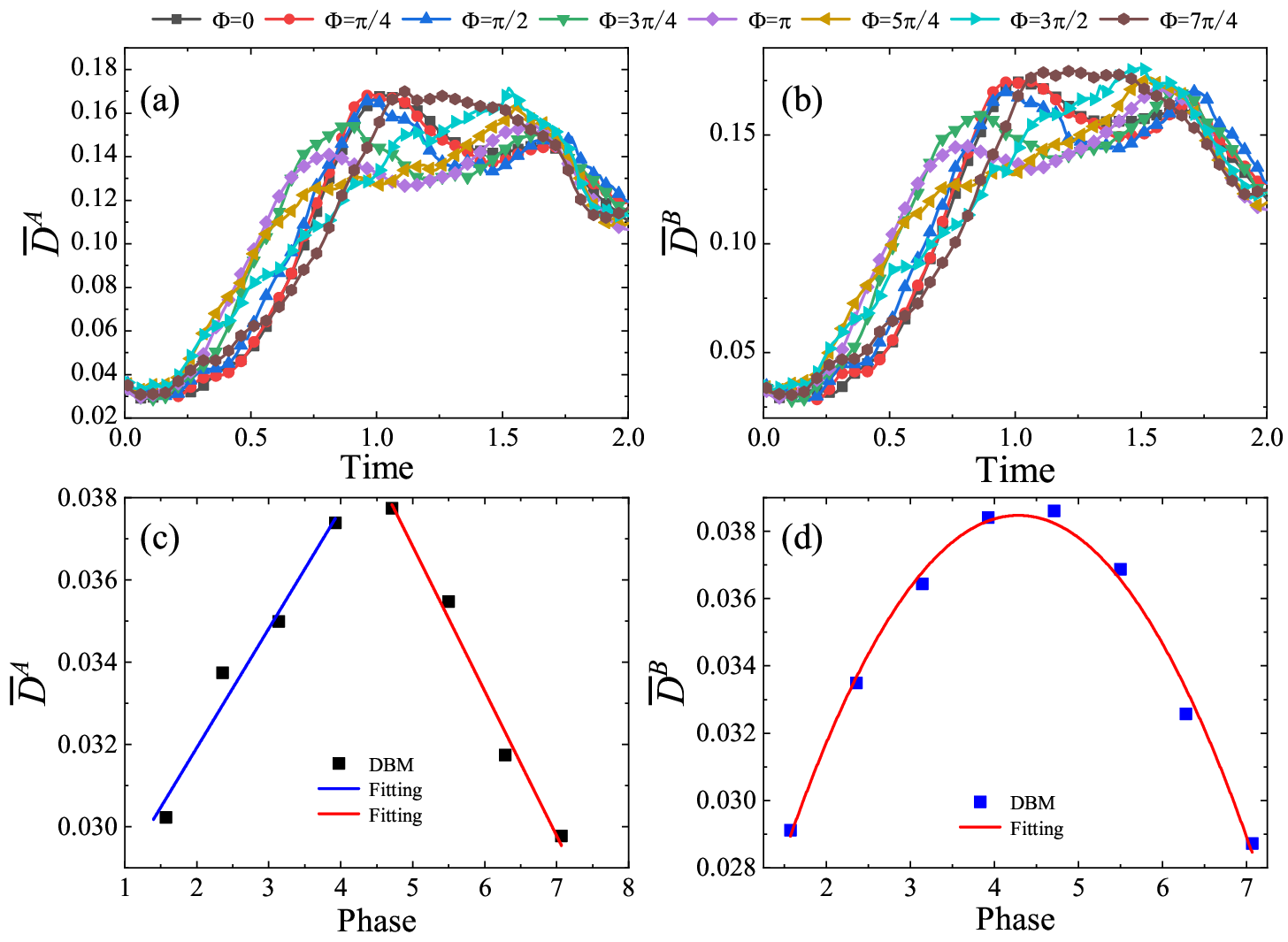}
	\end{center}
		\caption{Evolution of the average TNE strength for components $A$ (a) and $B$ (b). Relationship between the phase of
		time-varying acceleration and the average TNE strength at $t= 0.2$ for components $A$ (c) and $B$ (d).}\label{FIG15}
\end{figure*}

The phase of time-varying acceleration affects the interface perturbation as well. In this part, let us study the
influence of the phase $\Phi$ on the compressible RT instability. Figure \ref{FIG13} depicts the evolution of the
acceleration with eight different phases of time-varying acceleration: $\Phi=0$, $\pi/4$, $\pi/2$, $3\pi/4$, $\pi$,
$5\pi/4$, $3\pi/2$, and $7\pi/4$, respectively. The period is fixed at $T_0 = 1$ and the amplitude is chosen as $A_0=1$.
From a mathematical perspective, $\Phi = 0$ is equivalent to $\Phi = 2\pi$, and $\Phi = \pi/4$ is also equivalent to
$\Phi = 9\pi/4$, and so on. Based on the following analysis in Fig. \ref{FIG14} (d) and Figs. \ref{FIG15} (c) and (d),
it is found that dividing the phase into two intervals: $ \pi/2 \leq \Phi \leq 5\pi/4$ and $ 3\pi/2 \leq \Phi \leq
9\pi/4$, reveals inherent regularities, which facilitates a clearer understanding of the physical mechanisms.

Figure \ref{FIG14} (a) illustrates the evolution of the average density gradient in the $x$ direction
$|{\overline{{\nabla_{x}}{\rho}}}|$. It can be observed that $|{\overline{{\nabla_{x}}{\rho}}}|$ initially increases and
subsequently decreases. At the initial stage, the effect of acceleration on the density gradient in the $x$ direction is
not remarkable. As a result, the amplitude of the disturbed interface exhibits similar changes across all cases, causing
the curves to overlap. In the later stage, significant differences in $|{\overline{{\nabla_{x}}{\rho}}}|$ arise due to
the phase differences of the acceleration. On the contrary, Fig. \ref{FIG14} (b) shows that the effect of the
acceleration on the $y$-direction becomes evident early on. As the phase difference changes, the overall variation
becomes irregular and is accompanied by oscillatory phenomena.

Furthermore, Fig. \ref{FIG14}(c) demonstrates that the global average density gradient $|\overline{\nabla \rho}|$ first
decreases, then increases, and  subsequently decreases again with oscillations when $\Phi$=$0$, $\pi/4$, $\pi/2$,
$3\pi/4$. In contrast, $|\overline{\nabla \rho}|$ initially increases, then decreases when $\Phi$ = $\pi$, $5\pi/4$,
$3\pi/2$, $7\pi/4$. These behaviors arise from several competing physical mechanisms:
(i) The diffusion effect at the material interface leads to a reduction in the density gradient.
(ii) When the initial acceleration is less than $a_0 = -1.0 $, the fluid interface experiences an increasing pressure
difference in the $y $-direction, causing the interface to move downward. This enhances fluid mixing and accelerates the
evolution, increasing the density gradient. Conversely, when the initial acceleration is greater than $a_0 = -1.0$, the
pressure difference decreases, causing the interface to move upward, leading to a more uniform density.
(iii) As the system evolves, the two components begin to penetrate each other, stretching the interface in the vertical
direction, and forming spike and bubble structures. The contact area between the two media increases, making the
physical field more complex.
(iv) After the two components are fully mixed, the diffusion causes the spikes and bubbles to gradually dissipate,
leading to a reduction in the macroscopic gradient of physical quantities.

The above four mechanisms interact and influence the development of the global average density gradient. As a result,
for initial acceleration equal to $a_0 = -1.0$, the first mechanism dominates in the initial phase. When the initial
acceleration is greater than $a_0 = -1.0 $, the first and second mechanisms dominate, causing the density gradient to
decrease in the early stage.
If the initial acceleration is less than $a_0 = -1.0 $, the second mechanism becomes dominant, leading to an increase in
the density gradient initially. During the subsequent increasing phase, the third mechanism takes the lead. In the final
decreasing phase, the fourth mechanism dominate, causing the density gradient to decrease. Additionally, due to the
periodic variation of the time-varying acceleration, the external forces acting on the system also fluctuate, leading to
oscillations in the density gradient during the later stages of the RT evolution.

In Fig. \ref{FIG14} (d), we performed the fitting of $|\overline{\nabla \rho}|$ at $t = 0.2$ for two phase ranges:
$\pi/2 \leq \Phi \leq 5\pi/4$ and $3\pi/2 \leq \Phi \leq 9\pi/4$, respectively. The black squares represent the
numerical results. The first blue line corresponds to the fit for the first range, while the red line represents the fit
for the second range. The fitting functions are $|\overline{\nabla \rho}|_{t=0.2} = 347.58 + 7.04 \Phi$ and
$|\overline{\nabla \rho}|_{t=0.2} = 405.59 - 6.88 \Phi$, respectively. It can be seen that $|\overline{\nabla
	\rho}|_{t=0.2}$ increases linearly with phase within the first range, whereas the trend is reversed in the second range.

Figures. \ref{FIG15} (a) and (b) display the evolution of the average TNE strength ${\overline{D}^{\sigma}}$ for various
acceleration phases $\Phi$. Obviously, for all cases, ${\overline{D}^{\sigma}}$ experiences a slight decrease, then
rises, and finally oscillating declines. It should be noted that the curves of ${\overline{D}^{\sigma}}$ depart from
each other for different acceleration phase $\Phi$, and the differences become large in the later stage. Physically, the
acceleration affects the physical fields and the effect of time-varying acceleration on the interface disturbances
gradually strengthens.
Additionally, Figs. \ref{FIG15} (c) and (d) illustrate the fitting curve between the phase of time-varying acceleration
$A_0$ and the average TNE strength ${\overline{D}^{\sigma}}$ for the two components $A$ and $B$ at t=0.2. For the
component $A$, we performed a fitting of ${\overline{D}^{A}}$ for two phase ranges: $\pi/2 \leq \Phi \leq 5\pi/4$ and
$3\pi/2 \leq \Phi \leq 9\pi/4$, respectively. The black squares represent the numerical results. The first blue line
corresponds to the fit for the first range, while the red line represents the fit for the second range. The fitting
functions are ${\overline{D}^{A}}_{t=0.2} $=$ 2.61\times 10^{-2}+2.9 \times10^{-3} \Phi$ and ${\overline{D}^{A}}_{t=0.2}
= 5.44 \times 10^{-2}- 3.52 \times 10^{-3} \Phi$. For the component $B$, interestingly, the fitting function is a
quadratic function given by: $ {\overline{D}^{B}}_{t=0.2} = 1.47 \times 10^{-2} + 1.1 \times 10^{-2} \Phi - 1.29 \times
10^{-2} \Phi^2$ within the whole range $\pi/2 \le \Phi \le 9\pi/4$.
\begin{figure*}[!ht]
	\begin{center}
		\includegraphics[width=0.8\textwidth]{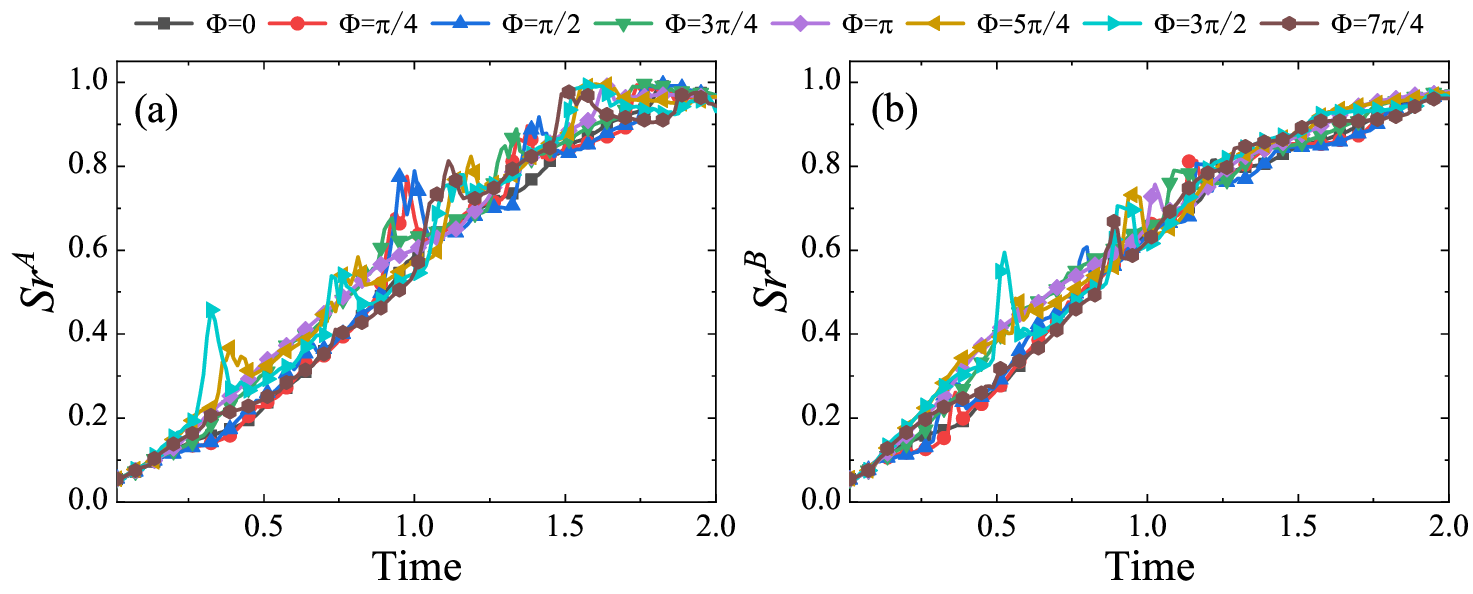}
	\end{center}
\caption{Evolution of the proportion of non-equilibrium region of component $A$ (a) and $B$ (b) under various phases of
	time-varying acceleration.}\label{FIG16}
\end{figure*}

Figures \ref{FIG16} (a) and (b) displays the evolution of the proportion of non-equilibrium regions $Sr^{\sigma}$ for
the two components $A$ and $B$, respectively. It is evident that $Sr^{\sigma}$ increases over time overall, with some
differences between different phases. Before approximately $t = 0.1 $, the differences are small, as the acceleration
effects have not yet manifested. In the later stage, the differences in the $Sr^{\sigma}$ gradually increase, though
they remain relatively small. Additionally, at $t = 2$, the $Sr^{\sigma}$ approaches 1, indicating that the whole system
has evolved to a non-equilibrium state.

\section{Conclusions}\label{IV}
In this paper, a two-component discrete Boltzmann model (DBM) is utilized to study the compressible Rayleigh-Taylor (RT) process under the time-varying acceleration.
The period, amplitude, and phase of time-varying acceleration are investigated in detail. The analysis centers on three
key aspects: the average density gradient $|{\overline{{\nabla}{\rho}}}|$, the average hydrodynamic non-equilibrium (TNE) strength
${\overline{D}^{\sigma}}$, and the proportion of non-equilibrium regions $Sr^{\sigma}$. In fact, the average density
gradient serves as a traditional TNE quantity, characterizing the spatial variation of density
within the system. The average TNE strength is a TNE quantity that reflects the deviation of
the distribution function from its equilibrium counterpart, offering insights into the degree of departure from
TNE state. The proportion of non-equilibrium regions describes the TNE state
from a geometric perspective, highlighting the spatial distribution of non-equilibrium behavior across the system.

For various periods and amplitudes, the changes in these quantities exhibit similar trends. Specifically,
$|{\overline{{\nabla}{\rho}}}|$ and ${\overline{D}^{\sigma}}$ initially decrease, then increase, and finally decrease
again with oscillations. $Sr^{\sigma}$ shows a trend of increasing first and then decreasing. In addition, shorter
periods lead to earlier peaks in these quantities, while larger amplitudes result in lower values of $|\overline{\nabla
	\rho}|$ but higher values of ${\overline{D}^{\sigma}}$ in the initial stage. The maximum values of these physical
quantities increase exponentially with increasing amplitude in the later stage.

For various phases, the changes in these quantities become relatively complex. In the range $0 \leq \Phi \leq 3\pi/4$,
$|\overline{\nabla \rho}|$ first decreases, then increases, and decreases again, accompanied by oscillations. In the
range $\pi/4 \leq \Phi \leq 7\pi/4$, it shows a trend of initially increasing and then decreasing.
${\overline{D}^{\sigma}}$ generally displays an initial increase followed by a decrease. $Sr^{\sigma}$ demonstrates an
overall increasing trend, and at $t = 2$, the proportion approaches 1 for all phases, indicating that the system is
nearing a non-equilibrium state.

Physically, the influence of time-varying acceleration on RT instability can be summarized into four mechanisms:
(i) When the initial acceleration is less than $a_0 = -1.0$, the pressure difference increases, causing the interface to
move downward, enhancing mixing and increasing gradients. Conversely, when the initial acceleration is greater than $a_0
= -1.0$, the pressure difference decreases, causing the interface to move upward and physical gradients to decrease;
(ii) The stretching of the interface increases the contact area, forming spikes and bubbles, which enhance the local
non-equilibrium strength; (iii) The diffusion effect smooths the interface, reducing density gradients; (iv) Dissipation
effect leads to the disappearance of vortices, reducing flow velocity. These findings contribute to a deeper understanding of RT instability, particularly in the context of time-varying accelerations, which is important for various applications in fluid dynamics and instability studies.

\section*{Acknowledgements}
This work was supported by the National Natural Science Foundation of China (under Grant No. U2242214), the
Guangdong Basic and Applied Basic Research Foundation (under Grant No. 2024A1515010927), the China Scholarship Council
(No. 202306380288), Humanities and Social Science Foundation of the Ministry of Education in China (under Grant No.
24YJCZH163), Fujian Provincial Units Special Funds for Education and Research (No.2022639), and Fundamental Research
Funds for the Central Universities, Sun Yat-sen University (under Grant No. 24qnpy044). This work was partly supported
by the Open Research Fund of Key Laboratory of Analytical Mathematics and Applications (Fujian Normal University),
Ministry of Education, P. R. China (under Grant No. JAM2405).

\appendix
\section{Grid Convergence Test}\label{A}

\begin{figure}[!ht]
	\centering
	\includegraphics[scale=0.3]{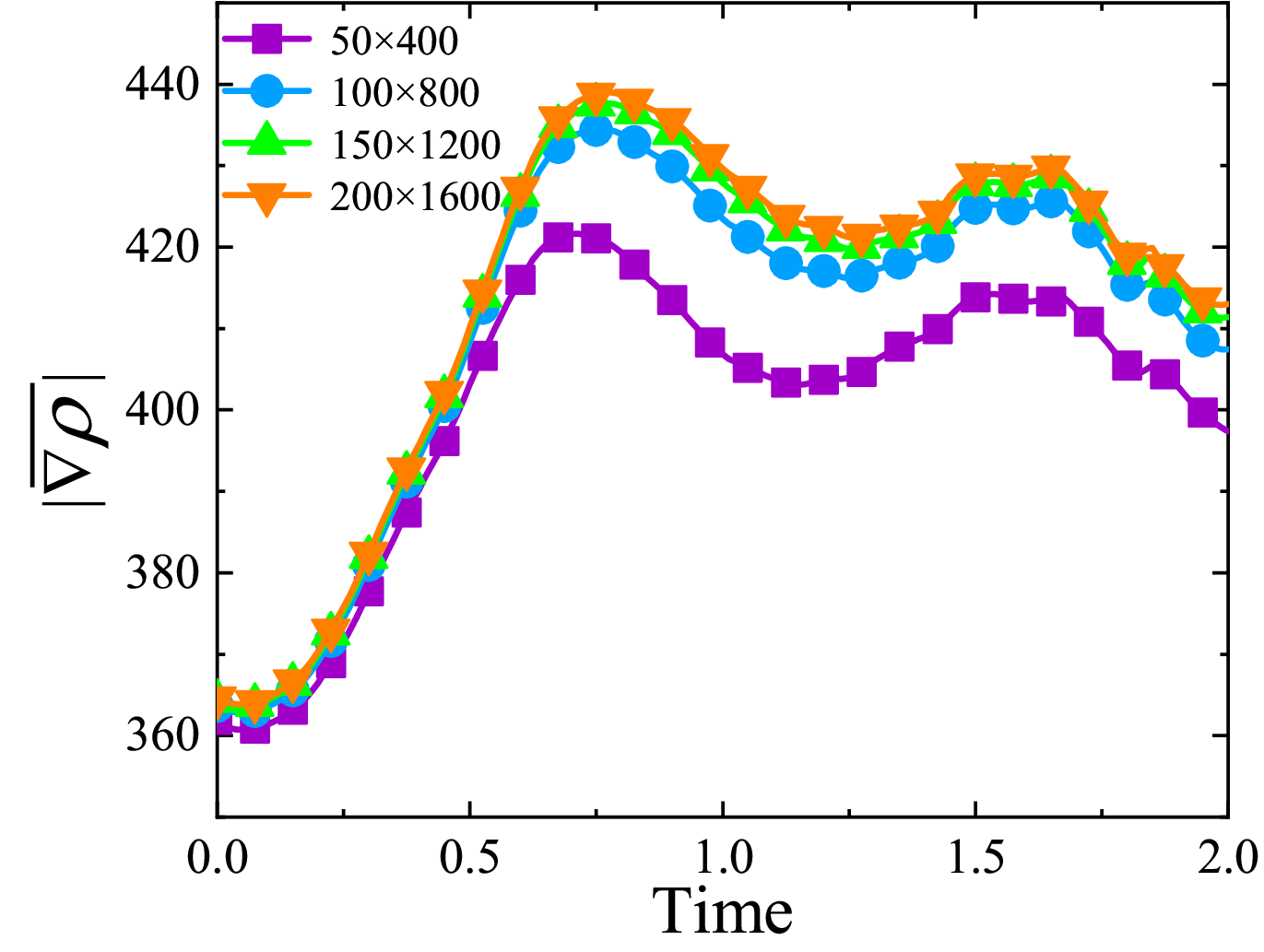}
	\caption{Grid independence test of the RT instability: the evolution of the average density gradient with various mesh
		grids.
	}\label{FIG17}
\end{figure}
We perform a grid independence test to guarantee an accurate and efficient simulation of the RT instability. Figure
\ref{FIG17} depicts the average density $|\overline{\nabla \rho}|$ in the RT process under a fixed time step $\Delta t =
2.5 \times 10^{-6}$ and four different mesh grids ${N_x}{\times}{N_y} = 50{\times}400$, $100{\times}800$,
$150{\times}1200$, and $200{\times}1600$, respectively. As the mesh grid size increases, the numerical errors
progressively decrease. The differences between simulations using $150\times1200$ and $200\times1600$ grids are minimal.
Therefore, to balance accuracy and efficiency, we have chosen to use a $200\times1600$  grid for this simulation.

\section*{Data Availability}
The data that support the findings of this study are available from the corresponding author upon reasonable request.

\section*{References}
\bibliography{DDBMtoVRT}

\begin{thebibliography}{76}%
\makeatletter
\providecommand \@ifxundefined [1]{%
 \@ifx{#1\undefined}
}%
\providecommand \@ifnum [1]{%
 \ifnum #1\expandafter \@firstoftwo
 \else \expandafter \@secondoftwo
 \fi
}%
\providecommand \@ifx [1]{%
 \ifx #1\expandafter \@firstoftwo
 \else \expandafter \@secondoftwo
 \fi
}%
\providecommand \natexlab [1]{#1}%
\providecommand \enquote  [1]{``#1''}%
\providecommand \bibnamefont  [1]{#1}%
\providecommand \bibfnamefont [1]{#1}%
\providecommand \citenamefont [1]{#1}%
\providecommand \href@noop [0]{\@secondoftwo}%
\providecommand \href [0]{\begingroup \@sanitize@url \@href}%
\providecommand \@href[1]{\@@startlink{#1}\@@href}%
\providecommand \@@href[1]{\endgroup#1\@@endlink}%
\providecommand \@sanitize@url [0]{\catcode `\\12\catcode `\$12\catcode
  `\&12\catcode `\#12\catcode `\^12\catcode `\_12\catcode `\%12\relax}%
\providecommand \@@startlink[1]{}%
\providecommand \@@endlink[0]{}%
\providecommand \url  [0]{\begingroup\@sanitize@url \@url }%
\providecommand \@url [1]{\endgroup\@href {#1}{\urlprefix }}%
\providecommand \urlprefix  [0]{URL }%
\providecommand \Eprint [0]{\href }%
\providecommand \doibase [0]{http://dx.doi.org/}%
\providecommand \selectlanguage [0]{\@gobble}%
\providecommand \bibinfo  [0]{\@secondoftwo}%
\providecommand \bibfield  [0]{\@secondoftwo}%
\providecommand \translation [1]{[#1]}%
\providecommand \BibitemOpen [0]{}%
\providecommand \bibitemStop [0]{}%
\providecommand \bibitemNoStop [0]{.\EOS\space}%
\providecommand \EOS [0]{\spacefactor3000\relax}%
\providecommand \BibitemShut  [1]{\csname bibitem#1\endcsname}%
\let\auto@bib@innerbib\@empty
\bibitem [{\citenamefont {Rayleigh}(1882)}]{Rayleigh1882}%
  \BibitemOpen
  \bibfield  {author} {\bibinfo {author} {\bibfnamefont {L.}~\bibnamefont
  {Rayleigh}},\ }\bibfield  {title} {\enquote {\bibinfo {title} {{Investigation
  of the character of the equilibrium of an incompressible heavy fluid of
  variable density}},}\ }\href {\doibase 10.1112/plms/s1-14.1.170} {\bibfield
  {journal} {\bibinfo  {journal} {Proc. London Math. Soc.}\ }\textbf {\bibinfo
  {volume} {1}},\ \bibinfo {pages} {170--177} (\bibinfo {year}
  {1882})}\BibitemShut {NoStop}%
\bibitem [{\citenamefont {Taylor}(1950)}]{Taylor1950}%
  \BibitemOpen
  \bibfield  {author} {\bibinfo {author} {\bibfnamefont {G.~I.}\ \bibnamefont
  {Taylor}},\ }\bibfield  {title} {\enquote {\bibinfo {title} {{The instability
  of liquid surfaces when accelerated in a direction perpendicular to their
  plane}},}\ }\href {\doibase 10.1098/rspa.1950.0052} {\bibfield  {journal}
  {\bibinfo  {journal} {Proc. R. Soc. London A}\ }\textbf {\bibinfo {volume}
  {201}},\ \bibinfo {pages} {192--196} (\bibinfo {year} {1950})}\BibitemShut
  {NoStop}%
\bibitem [{\citenamefont {Hoogenboom}\ and\ \citenamefont
  {Houseman}(2006)}]{hoogenboom2006rayleigh}%
  \BibitemOpen
  \bibfield  {author} {\bibinfo {author} {\bibfnamefont {T.}~\bibnamefont
  {Hoogenboom}}\ and\ \bibinfo {author} {\bibfnamefont {G.~A.}\ \bibnamefont
  {Houseman}},\ }\bibfield  {title} {\enquote {\bibinfo {title}
  {{Rayleigh-Taylor instability as a mechanism for corona formation on
  Venus}},}\ }\href {\doibase https://doi.org/10.1016/j.icarus.2005.11.001}
  {\bibfield  {journal} {\bibinfo  {journal} {Icarus}\ }\textbf {\bibinfo
  {volume} {180}},\ \bibinfo {pages} {292--307} (\bibinfo {year}
  {2006})}\BibitemShut {NoStop}%
\bibitem [{\citenamefont {Shamami}\ and\ \citenamefont
  {Ghasemizad}(2013)}]{rahimi2013reduction}%
  \BibitemOpen
  \bibfield  {author} {\bibinfo {author} {\bibfnamefont {S.~R.}\ \bibnamefont
  {Shamami}}\ and\ \bibinfo {author} {\bibfnamefont {A.}~\bibnamefont
  {Ghasemizad}},\ }\bibfield  {title} {\enquote {\bibinfo {title} {{Reduction
  of growth rate of Rayleigh-Taylor instability using nano-structured porous
  lining at ICF target shell}},}\ }\href {\doibase 10.1140/epjp/i2013-13141-x}
  {\bibfield  {journal} {\bibinfo  {journal} {Eur. Phys. J. Plus}\ }\textbf
  {\bibinfo {volume} {128}},\ \bibinfo {pages} {1--7} (\bibinfo {year}
  {2013})}\BibitemShut {NoStop}%
\bibitem [{\citenamefont {Ribeyre}, \citenamefont {Tikhonchuk},\ and\
  \citenamefont {Bouquet}(2004)}]{ribeyre2004compressible}%
  \BibitemOpen
  \bibfield  {author} {\bibinfo {author} {\bibfnamefont {X.}~\bibnamefont
  {Ribeyre}}, \bibinfo {author} {\bibfnamefont {V.~T.}\ \bibnamefont
  {Tikhonchuk}}, \ and\ \bibinfo {author} {\bibfnamefont {S.}~\bibnamefont
  {Bouquet}},\ }\bibfield  {title} {\enquote {\bibinfo {title} {{Compressible
  Rayleigh-Taylor instabilities in supernova remnants}},}\ }\href {\doibase
  10.1063/1.1810182} {\bibfield  {journal} {\bibinfo  {journal} {Phys. Fluids}\
  }\textbf {\bibinfo {volume} {16}},\ \bibinfo {pages} {4661--4670} (\bibinfo
  {year} {2004})}\BibitemShut {NoStop}%
\bibitem [{\citenamefont {Abarzhi}\ \emph {et~al.}(2019)\citenamefont
  {Abarzhi}, \citenamefont {Bhowmick}, \citenamefont {Naveh}, \citenamefont
  {Pandian}, \citenamefont {Swisher}, \citenamefont {Stellingwerf},\ and\
  \citenamefont {Arnett}}]{abarzhi2019supernova}%
  \BibitemOpen
  \bibfield  {author} {\bibinfo {author} {\bibfnamefont {S.~I.}\ \bibnamefont
  {Abarzhi}}, \bibinfo {author} {\bibfnamefont {A.~K.}\ \bibnamefont
  {Bhowmick}}, \bibinfo {author} {\bibfnamefont {A.}~\bibnamefont {Naveh}},
  \bibinfo {author} {\bibfnamefont {A.}~\bibnamefont {Pandian}}, \bibinfo
  {author} {\bibfnamefont {N.~C.}\ \bibnamefont {Swisher}}, \bibinfo {author}
  {\bibfnamefont {R.~F.}\ \bibnamefont {Stellingwerf}}, \ and\ \bibinfo
  {author} {\bibfnamefont {W.~D.}\ \bibnamefont {Arnett}},\ }\bibfield  {title}
  {\enquote {\bibinfo {title} {{Supernova, nuclear synthesis, fluid
  instabilities, and interfacial mixing}},}\ }\href {\doibase
  10.1073/pnas.1714502115} {\bibfield  {journal} {\bibinfo  {journal} {Proc.
  Natl. Acad. Sci. U.S.A.}\ }\textbf {\bibinfo {volume} {116}},\ \bibinfo
  {pages} {18184--18192} (\bibinfo {year} {2019})}\BibitemShut {NoStop}%
\bibitem [{\citenamefont {Selig}\ and\ \citenamefont
  {Wermund}(1966)}]{selig1966families}%
  \BibitemOpen
  \bibfield  {author} {\bibinfo {author} {\bibfnamefont {F.}~\bibnamefont
  {Selig}}\ and\ \bibinfo {author} {\bibfnamefont {E.~G.}\ \bibnamefont
  {Wermund}},\ }\bibfield  {title} {\enquote {\bibinfo {title} {{Families of
  salt domes in the Gulf coastal province}},}\ }\href {\doibase
  10.1190/1.1439806} {\bibfield  {journal} {\bibinfo  {journal} {Geophysics}\
  }\textbf {\bibinfo {volume} {31}},\ \bibinfo {pages} {726--740} (\bibinfo
  {year} {1966})}\BibitemShut {NoStop}%
\bibitem [{\citenamefont {Ghosh}\ \emph {et~al.}(2020)\citenamefont {Ghosh},
  \citenamefont {Maiti}, \citenamefont {Mandal},\ and\ \citenamefont
  {Baruah}}]{ghosh2020cold}%
  \BibitemOpen
  \bibfield  {author} {\bibinfo {author} {\bibfnamefont {D.}~\bibnamefont
  {Ghosh}}, \bibinfo {author} {\bibfnamefont {G.}~\bibnamefont {Maiti}},
  \bibinfo {author} {\bibfnamefont {N.}~\bibnamefont {Mandal}}, \ and\ \bibinfo
  {author} {\bibfnamefont {A.}~\bibnamefont {Baruah}},\ }\bibfield  {title}
  {\enquote {\bibinfo {title} {{Cold Plumes Initiated by Rayleigh-Taylor
  Instabilities in Subduction Zones, and Their Characteristic Volcanic
  Distributions: The Role of Slab Dip}},}\ }\href {\doibase
  https://doi.org/10.1029/2020JB019814} {\bibfield  {journal} {\bibinfo
  {journal} {J. Geophys. Res. Solid Earth}\ }\textbf {\bibinfo {volume}
  {125}},\ \bibinfo {pages} {e2020JB019814} (\bibinfo {year}
  {2020})}\BibitemShut {NoStop}%
\bibitem [{\citenamefont {Morgan}\ and\ \citenamefont
  {Jacobs}(2020)}]{morgan2020experiments}%
  \BibitemOpen
  \bibfield  {author} {\bibinfo {author} {\bibfnamefont {R.}~\bibnamefont
  {Morgan}}\ and\ \bibinfo {author} {\bibfnamefont {J.}~\bibnamefont
  {Jacobs}},\ }\bibfield  {title} {\enquote {\bibinfo {title} {{Experiments and
  simulations on the turbulent, rarefaction wave driven Rayleigh-Taylor
  instability}},}\ }\href {\doibase 10.1115/1.4048345} {\bibfield  {journal}
  {\bibinfo  {journal} {J. Fluids Eng.}\ }\textbf {\bibinfo {volume} {142}},\
  \bibinfo {pages} {121101} (\bibinfo {year} {2020})}\BibitemShut {NoStop}%
\bibitem [{\citenamefont {Lherm}\ \emph {et~al.}(2022)\citenamefont {Lherm},
  \citenamefont {Deguen}, \citenamefont {Alboussi{\`e}re},\ and\ \citenamefont
  {Landeau}}]{lherm2022rayleigh}%
  \BibitemOpen
  \bibfield  {author} {\bibinfo {author} {\bibfnamefont {V.}~\bibnamefont
  {Lherm}}, \bibinfo {author} {\bibfnamefont {R.}~\bibnamefont {Deguen}},
  \bibinfo {author} {\bibfnamefont {T.}~\bibnamefont {Alboussi{\`e}re}}, \ and\
  \bibinfo {author} {\bibfnamefont {M.}~\bibnamefont {Landeau}},\ }\bibfield
  {title} {\enquote {\bibinfo {title} {{ Rayleigh-Taylor instability in impact
  cratering experiments}},}\ }\href {\doibase 10.1017/jfm.2022.111} {\bibfield
  {journal} {\bibinfo  {journal} {J. Fluid Mech}\ }\textbf {\bibinfo {volume}
  {937}},\ \bibinfo {pages} {A20} (\bibinfo {year} {2022})}\BibitemShut
  {NoStop}%
\bibitem [{\citenamefont {Meshkov}\ and\ \citenamefont
  {Abarzhi}(2019)}]{meshkov2019group}%
  \BibitemOpen
  \bibfield  {author} {\bibinfo {author} {\bibfnamefont {E.~E.}\ \bibnamefont
  {Meshkov}}\ and\ \bibinfo {author} {\bibfnamefont {S.~I.}\ \bibnamefont
  {Abarzhi}},\ }\bibfield  {title} {\enquote {\bibinfo {title} {{Group theory
  and jelly's experiment of Rayleigh-Taylor instability and Rayleigh-Taylor
  interfacial mixing}},}\ }\href {\doibase 10.1088/1873-7005/ab3e83} {\bibfield
   {journal} {\bibinfo  {journal} {Fluid Dyn. Res.}\ }\textbf {\bibinfo
  {volume} {51}},\ \bibinfo {pages} {065502} (\bibinfo {year}
  {2019})}\BibitemShut {NoStop}%
\bibitem [{\citenamefont {Matsumoto}, \citenamefont {Aloy},\ and\ \citenamefont
  {Perucho}(2017)}]{matsumoto2017linear}%
  \BibitemOpen
  \bibfield  {author} {\bibinfo {author} {\bibfnamefont {J.}~\bibnamefont
  {Matsumoto}}, \bibinfo {author} {\bibfnamefont {A.~M.}\ \bibnamefont {Aloy}},
  \ and\ \bibinfo {author} {\bibfnamefont {M.}~\bibnamefont {Perucho}},\
  }\bibfield  {title} {\enquote {\bibinfo {title} {{Linear theory of the
  Rayleigh-Taylor instability at a discontinuous surface of a relativistic
  flow}},}\ }\href {\doibase 10.1093/mnras/stx2012} {\bibfield  {journal}
  {\bibinfo  {journal} {MNRAS}\ }\textbf {\bibinfo {volume} {472}},\ \bibinfo
  {pages} {1421--1431} (\bibinfo {year} {2017})}\BibitemShut {NoStop}%
\bibitem [{\citenamefont {Abarzhi}\ and\ \citenamefont
  {Williams}(2020)}]{abarzhi2020scale}%
  \BibitemOpen
  \bibfield  {author} {\bibinfo {author} {\bibfnamefont {S.~I.}\ \bibnamefont
  {Abarzhi}}\ and\ \bibinfo {author} {\bibfnamefont {K.~C.}\ \bibnamefont
  {Williams}},\ }\bibfield  {title} {\enquote {\bibinfo {title}
  {{Scale-dependent Rayleigh-Taylor dynamics with variable acceleration by
  group theory approach}},}\ }\href {\doibase 10.1063/5.0012035} {\bibfield
  {journal} {\bibinfo  {journal} {Phys. Plasmas}\ }\textbf {\bibinfo {volume}
  {27}},\ \bibinfo {pages} {072107} (\bibinfo {year} {2020})}\BibitemShut
  {NoStop}%
\bibitem [{\citenamefont {Liu}, \citenamefont {Wang},\ and\ \citenamefont
  {Ma}(2018)}]{liu2018surface}%
  \BibitemOpen
  \bibfield  {author} {\bibinfo {author} {\bibfnamefont {W.~H.}\ \bibnamefont
  {Liu}}, \bibinfo {author} {\bibfnamefont {X.}~\bibnamefont {Wang}}, \ and\
  \bibinfo {author} {\bibfnamefont {W.~F.}\ \bibnamefont {Ma}},\ }\bibfield
  {title} {\enquote {\bibinfo {title} {{Surface Tension Effect on Harmonics of
  Rayleigh-Taylor Instability}},}\ }\href {\doibase
  10.1063/1674-0068/31/cjcp1703056} {\bibfield  {journal} {\bibinfo  {journal}
  {Chin. J. Chem. Phys.}\ }\textbf {\bibinfo {volume} {31}},\ \bibinfo {pages}
  {39--44} (\bibinfo {year} {2018})}\BibitemShut {NoStop}%
\bibitem [{\citenamefont {Shin}, \citenamefont {Sohn},\ and\ \citenamefont
  {Hwang}(2022)}]{shin2022numerical}%
  \BibitemOpen
  \bibfield  {author} {\bibinfo {author} {\bibfnamefont {S.}~\bibnamefont
  {Shin}}, \bibinfo {author} {\bibfnamefont {S.~I.}\ \bibnamefont {Sohn}}, \
  and\ \bibinfo {author} {\bibfnamefont {W.~J.}\ \bibnamefont {Hwang}},\
  }\bibfield  {title} {\enquote {\bibinfo {title} {{Numerical simulation of
  single-and multi-mode Rayleigh-Taylor instability with surface tension in two
  dimensions}},}\ }\href {\doibase
  https://doi.org/10.1016/j.euromechflu.2021.10.005} {\bibfield  {journal}
  {\bibinfo  {journal} {Eur. J. Mech. B}\ }\textbf {\bibinfo {volume} {91}},\
  \bibinfo {pages} {141--151} (\bibinfo {year} {2022})}\BibitemShut {NoStop}%
\bibitem [{\citenamefont {Luo}\ and\ \citenamefont
  {Wang}(2021)}]{luo2021effects}%
  \BibitemOpen
  \bibfield  {author} {\bibinfo {author} {\bibfnamefont {T.~F.}\ \bibnamefont
  {Luo}}\ and\ \bibinfo {author} {\bibfnamefont {J.~C.}\ \bibnamefont {Wang}},\
  }\bibfield  {title} {\enquote {\bibinfo {title} {{Effects of Atwood number
  and stratification parameter on compressible multi-mode Rayleigh-Taylor
  instability}},}\ }\href {\doibase 10.1063/5.0071437} {\bibfield  {journal}
  {\bibinfo  {journal} {Phys. Fluids}\ }\textbf {\bibinfo {volume} {33}},\
  \bibinfo {pages} {115111} (\bibinfo {year} {2021})}\BibitemShut {NoStop}%
\bibitem [{\citenamefont {Liang}, \citenamefont {Hu},\ and\ \citenamefont
  {X.~F.~Huang}(2019)}]{liang2019direct}%
  \BibitemOpen
  \bibfield  {author} {\bibinfo {author} {\bibfnamefont {H.}~\bibnamefont
  {Liang}}, \bibinfo {author} {\bibfnamefont {X.~L.}\ \bibnamefont {Hu}}, \
  and\ \bibinfo {author} {\bibfnamefont {a.~J. R.~X.}\ \bibnamefont
  {X.~F.~Huang}},\ }\bibfield  {title} {\enquote {\bibinfo {title} {{Direct
  numerical simulations of multi-mode immiscible Rayleigh-Taylor instability
  with high Reynolds numbers}},}\ }\href {\doibase 10.1063/1.5127888}
  {\bibfield  {journal} {\bibinfo  {journal} {Phys. Fluids}\ }\textbf {\bibinfo
  {volume} {31}},\ \bibinfo {pages} {112104} (\bibinfo {year}
  {2019})}\BibitemShut {NoStop}%
\bibitem [{\citenamefont {Guo}\ \emph {et~al.}(2018)\citenamefont {Guo},
  \citenamefont {Chen}, \citenamefont {Li}, \citenamefont {Tian}, \citenamefont
  {Su},\ and\ \citenamefont {Qiu}}]{guo2018numerical}%
  \BibitemOpen
  \bibfield  {author} {\bibinfo {author} {\bibfnamefont {K.~L.}\ \bibnamefont
  {Guo}}, \bibinfo {author} {\bibfnamefont {R.~H.}\ \bibnamefont {Chen}},
  \bibinfo {author} {\bibfnamefont {Y.~L.}\ \bibnamefont {Li}}, \bibinfo
  {author} {\bibfnamefont {W.~X.}\ \bibnamefont {Tian}}, \bibinfo {author}
  {\bibfnamefont {G.~H.}\ \bibnamefont {Su}}, \ and\ \bibinfo {author}
  {\bibfnamefont {S.~Z.}\ \bibnamefont {Qiu}},\ }\bibfield  {title} {\enquote
  {\bibinfo {title} {{Numerical simulation of Rayleigh-Taylor Instability with
  periodic boundary condition using MPS method}},}\ }\href {\doibase
  https://doi.org/10.1016/j.pnucene.2018.08.008} {\bibfield  {journal}
  {\bibinfo  {journal} {Prog. Nucl. Energ.}\ }\textbf {\bibinfo {volume}
  {109}},\ \bibinfo {pages} {130--144} (\bibinfo {year} {2018})}\BibitemShut
  {NoStop}%
\bibitem [{\citenamefont {Youngs}(2017)}]{youngs2017rayleigh}%
  \BibitemOpen
  \bibfield  {author} {\bibinfo {author} {\bibfnamefont {D.~L.}\ \bibnamefont
  {Youngs}},\ }\bibfield  {title} {\enquote {\bibinfo {title} {{Rayleigh-Taylor
  mixing: Direct numerical simulation and implicit large eddy simulation}},}\
  }\href {\doibase 10.1088/1402-4896/aa732b} {\bibfield  {journal} {\bibinfo
  {journal} {Phys. Scr}\ }\textbf {\bibinfo {volume} {92}},\ \bibinfo {pages}
  {074006} (\bibinfo {year} {2017})}\BibitemShut {NoStop}%
\bibitem [{\citenamefont {Rahmat}\ \emph {et~al.}(2014)\citenamefont {Rahmat},
  \citenamefont {Tofighi}, \citenamefont {Shadloo},\ and\ \citenamefont
  {Yildiz}}]{rahmat2014numerical}%
  \BibitemOpen
  \bibfield  {author} {\bibinfo {author} {\bibfnamefont {A.}~\bibnamefont
  {Rahmat}}, \bibinfo {author} {\bibfnamefont {N.}~\bibnamefont {Tofighi}},
  \bibinfo {author} {\bibfnamefont {M.~S.}\ \bibnamefont {Shadloo}}, \ and\
  \bibinfo {author} {\bibfnamefont {M.}~\bibnamefont {Yildiz}},\ }\bibfield
  {title} {\enquote {\bibinfo {title} {{Numerical simulation of wall bounded
  and electrically excited Rayleigh-Taylor instability using incompressible
  smoothed particle hydrodynamics}},}\ }\href {\doibase
  https://doi.org/10.1016/j.colsurfa.2014.02.044} {\bibfield  {journal}
  {\bibinfo  {journal} {Colloids and Surfaces A}\ }\textbf {\bibinfo {volume}
  {460}},\ \bibinfo {pages} {60--70} (\bibinfo {year} {2014})}\BibitemShut
  {NoStop}%
\bibitem [{\citenamefont {Hamzehloo}, \citenamefont {Bartholomew},\ and\
  \citenamefont {Laizet}(2021)}]{hamzehloo2021direct}%
  \BibitemOpen
  \bibfield  {author} {\bibinfo {author} {\bibfnamefont {A.}~\bibnamefont
  {Hamzehloo}}, \bibinfo {author} {\bibfnamefont {P.}~\bibnamefont
  {Bartholomew}}, \ and\ \bibinfo {author} {\bibfnamefont {S.}~\bibnamefont
  {Laizet}},\ }\bibfield  {title} {\enquote {\bibinfo {title} {{Direct
  numerical simulations of incompressible Rayleigh-Taylor instabilities at low
  and medium Atwood numbers}},}\ }\href {\doibase 10.1063/5.0049867} {\bibfield
   {journal} {\bibinfo  {journal} {Phys. Fluids}\ }\textbf {\bibinfo {volume}
  {33}},\ \bibinfo {pages} {054114} (\bibinfo {year} {2021})}\BibitemShut
  {NoStop}%
\bibitem [{\citenamefont {Song}, \citenamefont {Wang},\ and\ \citenamefont
  {Wang}(2021)}]{song2021numerical}%
  \BibitemOpen
  \bibfield  {author} {\bibinfo {author} {\bibfnamefont {Y.}~\bibnamefont
  {Song}}, \bibinfo {author} {\bibfnamefont {P.}~\bibnamefont {Wang}}, \ and\
  \bibinfo {author} {\bibfnamefont {L.~L.}\ \bibnamefont {Wang}},\ }\bibfield
  {title} {\enquote {\bibinfo {title} {{Numerical investigations of
  Rayleigh-Taylor instability with a density gradient layer}},}\ }\href
  {\doibase https://doi.org/10.1016/j.compfluid.2021.104869} {\bibfield
  {journal} {\bibinfo  {journal} {Comput. Fluids}\ }\textbf {\bibinfo {volume}
  {220}},\ \bibinfo {pages} {104869} (\bibinfo {year} {2021})}\BibitemShut
  {NoStop}%
\bibitem [{\citenamefont {Shadloo}, \citenamefont {Zainali},\ and\
  \citenamefont {Laizet}(2013)}]{shadloo2013simulation}%
  \BibitemOpen
  \bibfield  {author} {\bibinfo {author} {\bibfnamefont {M.~S.}\ \bibnamefont
  {Shadloo}}, \bibinfo {author} {\bibfnamefont {A.}~\bibnamefont {Zainali}}, \
  and\ \bibinfo {author} {\bibfnamefont {S.}~\bibnamefont {Laizet}},\
  }\bibfield  {title} {\enquote {\bibinfo {title} {{Simulation of single mode
  Rayleigh-Taylor instability by SPH method}},}\ }\href {\doibase
  10.1007/s00466-012-0746-2} {\bibfield  {journal} {\bibinfo  {journal}
  {Comput. Mech}\ }\textbf {\bibinfo {volume} {51}},\ \bibinfo {pages}
  {699--715} (\bibinfo {year} {2013})}\BibitemShut {NoStop}%
\bibitem [{\citenamefont {Aslangil}, \citenamefont {Banerjee},\ and\
  \citenamefont {Lawrie}(2016)}]{aslangil2016numerical}%
  \BibitemOpen
  \bibfield  {author} {\bibinfo {author} {\bibfnamefont {D.}~\bibnamefont
  {Aslangil}}, \bibinfo {author} {\bibfnamefont {A.}~\bibnamefont {Banerjee}},
  \ and\ \bibinfo {author} {\bibfnamefont {A.~G.}\ \bibnamefont {Lawrie}},\
  }\bibfield  {title} {\enquote {\bibinfo {title} {{Numerical investigation of
  initial condition effects on Rayleigh-Taylor instability with acceleration
  reversals}},}\ }\href {\doibase 10.1103/PhysRevE.94.053114} {\bibfield
  {journal} {\bibinfo  {journal} {Phys. Rev. E.}\ }\textbf {\bibinfo {volume}
  {94}},\ \bibinfo {pages} {053114} (\bibinfo {year} {2016})}\BibitemShut
  {NoStop}%
\bibitem [{\citenamefont {Aslangil}, \citenamefont {Lawrie},\ and\
  \citenamefont {Banerjee}(2022)}]{aslangil2022effects}%
  \BibitemOpen
  \bibfield  {author} {\bibinfo {author} {\bibfnamefont {D.}~\bibnamefont
  {Aslangil}}, \bibinfo {author} {\bibfnamefont {A.~G.}\ \bibnamefont
  {Lawrie}}, \ and\ \bibinfo {author} {\bibfnamefont {A.}~\bibnamefont
  {Banerjee}},\ }\bibfield  {title} {\enquote {\bibinfo {title} {{Effects of
  variable deceleration periods on Rayleigh-Taylor instability with
  acceleration reversals}},}\ }\href {\doibase 10.1103/PhysRevE.105.065103}
  {\bibfield  {journal} {\bibinfo  {journal} {Phys. Rev. E}\ }\textbf {\bibinfo
  {volume} {105}},\ \bibinfo {pages} {065103} (\bibinfo {year}
  {2022})}\BibitemShut {NoStop}%
\bibitem [{\citenamefont {Boffetta}, \citenamefont {Magnani},\ and\
  \citenamefont {Musacchio}(2019)}]{boffetta2019suppression}%
  \BibitemOpen
  \bibfield  {author} {\bibinfo {author} {\bibfnamefont {G.}~\bibnamefont
  {Boffetta}}, \bibinfo {author} {\bibfnamefont {M.}~\bibnamefont {Magnani}}, \
  and\ \bibinfo {author} {\bibfnamefont {S.}~\bibnamefont {Musacchio}},\
  }\bibfield  {title} {\enquote {\bibinfo {title} {{Suppression of
  Rayleigh-Taylor turbulence by time-periodic acceleration}},}\ }\href
  {\doibase 10.1103/PhysRevE.99.033110} {\bibfield  {journal} {\bibinfo
  {journal} {Phys. Rev. E.}\ }\textbf {\bibinfo {volume} {99}},\ \bibinfo
  {pages} {033110} (\bibinfo {year} {2019})}\BibitemShut {NoStop}%
\bibitem [{\citenamefont {Ramaprabhu}, \citenamefont {Karkhanis},\ and\
  \citenamefont {Lawrie}(2013)}]{ramaprabhu2013rayleigh}%
  \BibitemOpen
  \bibfield  {author} {\bibinfo {author} {\bibfnamefont {P.}~\bibnamefont
  {Ramaprabhu}}, \bibinfo {author} {\bibfnamefont {V.}~\bibnamefont
  {Karkhanis}}, \ and\ \bibinfo {author} {\bibfnamefont {A.~G.}\ \bibnamefont
  {Lawrie}},\ }\bibfield  {title} {\enquote {\bibinfo {title} {{The
  Rayleigh-Taylor instability driven by an accel-decel-accel profile}},}\
  }\href {\doibase 10.1063/1.4829765} {\bibfield  {journal} {\bibinfo
  {journal} {Phys. Fluids}\ }\textbf {\bibinfo {volume} {25}},\ \bibinfo
  {pages} {115104} (\bibinfo {year} {2013})}\BibitemShut {NoStop}%
\bibitem [{\citenamefont {Ramaprabhu}\ \emph {et~al.}(2016)\citenamefont
  {Ramaprabhu}, \citenamefont {Karkhanis}, \citenamefont {Banerjee},
  \citenamefont {Varshochi}, \citenamefont {Khan},\ and\ \citenamefont
  {Lawrie}}]{ramaprabhu2016evolution}%
  \BibitemOpen
  \bibfield  {author} {\bibinfo {author} {\bibfnamefont {P.}~\bibnamefont
  {Ramaprabhu}}, \bibinfo {author} {\bibfnamefont {V.}~\bibnamefont
  {Karkhanis}}, \bibinfo {author} {\bibfnamefont {R.}~\bibnamefont {Banerjee}},
  \bibinfo {author} {\bibfnamefont {H.}~\bibnamefont {Varshochi}}, \bibinfo
  {author} {\bibfnamefont {M.}~\bibnamefont {Khan}}, \ and\ \bibinfo {author}
  {\bibfnamefont {A.~G.}\ \bibnamefont {Lawrie}},\ }\bibfield  {title}
  {\enquote {\bibinfo {title} {{Evolution of the single-mode Rayleigh-Taylor
  instability under the influence of time-dependent accelerations}},}\ }\href
  {\doibase 10.1103/PhysRevE.93.013118} {\bibfield  {journal} {\bibinfo
  {journal} {Phys. Rev. E.}\ }\textbf {\bibinfo {volume} {93}},\ \bibinfo
  {pages} {013118} (\bibinfo {year} {2016})}\BibitemShut {NoStop}%
\bibitem [{\citenamefont {Hu}, \citenamefont {Zhang},\ and\ \citenamefont
  {Tian}(2020)}]{hu2020evolution}%
  \BibitemOpen
  \bibfield  {author} {\bibinfo {author} {\bibfnamefont {Z.~X.}\ \bibnamefont
  {Hu}}, \bibinfo {author} {\bibfnamefont {Y.~S.}\ \bibnamefont {Zhang}}, \
  and\ \bibinfo {author} {\bibfnamefont {B.~L.}\ \bibnamefont {Tian}},\
  }\bibfield  {title} {\enquote {\bibinfo {title} {{Evolution of
  Rayleigh-Taylor instability under interface discontinuous acceleration
  induced by radiation}},}\ }\href {\doibase 10.1103/PhysRevE.101.043115}
  {\bibfield  {journal} {\bibinfo  {journal} {Phys. Rev. E.}\ }\textbf
  {\bibinfo {volume} {101}},\ \bibinfo {pages} {043115} (\bibinfo {year}
  {2020})}\BibitemShut {NoStop}%
\bibitem [{\citenamefont {Livescu}, \citenamefont {Wei},\ and\ \citenamefont
  {Brady}(2021)}]{livescu2021rayleigh}%
  \BibitemOpen
  \bibfield  {author} {\bibinfo {author} {\bibfnamefont {D.}~\bibnamefont
  {Livescu}}, \bibinfo {author} {\bibfnamefont {T.}~\bibnamefont {Wei}}, \ and\
  \bibinfo {author} {\bibfnamefont {P.~T.}\ \bibnamefont {Brady}},\ }\bibfield
  {title} {\enquote {\bibinfo {title} {{Rayleigh-Taylor instability with
  gravity reversal}},}\ }\href {\doibase
  https://doi.org/10.1016/j.physd.2020.132832} {\bibfield  {journal} {\bibinfo
  {journal} {Phys. D: Nonlinear Phenom.}\ }\textbf {\bibinfo {volume} {417}},\
  \bibinfo {pages} {132832} (\bibinfo {year} {2021})}\BibitemShut {NoStop}%
\bibitem [{\citenamefont {Banerjee}(2023)}]{banerjee2023ablative}%
  \BibitemOpen
  \bibfield  {author} {\bibinfo {author} {\bibfnamefont {R.}~\bibnamefont
  {Banerjee}},\ }\bibfield  {title} {\enquote {\bibinfo {title} {{Ablative
  Rayleigh-Taylor instability driven by time-varying acceleration}},}\ }\href
  {\doibase 10.1007/s12648-023-02755-3} {\bibfield  {journal} {\bibinfo
  {journal} {Indian J. Phys.}\ }\textbf {\bibinfo {volume} {97}},\ \bibinfo
  {pages} {4365–4371} (\bibinfo {year} {2023})}\BibitemShut {NoStop}%
\bibitem [{\citenamefont {Wang}, \citenamefont {Wei},\ and\ \citenamefont
  {Qian}(2020)}]{Wang2020}%
  \BibitemOpen
  \bibfield  {author} {\bibinfo {author} {\bibfnamefont {Z.~D.}\ \bibnamefont
  {Wang}}, \bibinfo {author} {\bibfnamefont {Y.~K.}\ \bibnamefont {Wei}}, \
  and\ \bibinfo {author} {\bibfnamefont {Y.~H.}\ \bibnamefont {Qian}},\
  }\bibfield  {title} {\enquote {\bibinfo {title} {{A bounce back-immersed
  boundary-lattice Boltzmann model for curved boundary}},}\ }\href {\doibase
  https://doi.org/10.1016/j.apm.2020.01.012} {\bibfield  {journal} {\bibinfo
  {journal} {Appl. Math. Model.}\ }\textbf {\bibinfo {volume} {81}},\ \bibinfo
  {pages} {428--440} (\bibinfo {year} {2020})}\BibitemShut {NoStop}%
\bibitem [{\citenamefont {Wei}\ \emph {et~al.}(2018)\citenamefont {Wei},
  \citenamefont {Yang}, \citenamefont {Dou}, \citenamefont {Lin}, \citenamefont
  {Wang},\ and\ \citenamefont {Qian}}]{wei2018novel}%
  \BibitemOpen
  \bibfield  {author} {\bibinfo {author} {\bibfnamefont {Y.~K.}\ \bibnamefont
  {Wei}}, \bibinfo {author} {\bibfnamefont {H.}~\bibnamefont {Yang}}, \bibinfo
  {author} {\bibfnamefont {H.~S.}\ \bibnamefont {Dou}}, \bibinfo {author}
  {\bibfnamefont {Z.}~\bibnamefont {Lin}}, \bibinfo {author} {\bibfnamefont
  {Z.~D.}\ \bibnamefont {Wang}}, \ and\ \bibinfo {author} {\bibfnamefont
  {Y.~H.}\ \bibnamefont {Qian}},\ }\bibfield  {title} {\enquote {\bibinfo
  {title} {{A novel two-dimensional coupled lattice Boltzmann model for thermal
  incompressible flows}},}\ }\href {\doibase
  https://doi.org/10.1016/j.amc.2018.07.047} {\bibfield  {journal} {\bibinfo
  {journal} {Appl. Math. Comput.}\ }\textbf {\bibinfo {volume} {339}},\
  \bibinfo {pages} {556--567} (\bibinfo {year} {2018})}\BibitemShut {NoStop}%
\bibitem [{\citenamefont {Wei}\ \emph {et~al.}(2022)\citenamefont {Wei},
  \citenamefont {Li}, \citenamefont {Wang}, \citenamefont {Yang}, \citenamefont
  {Zhu}, \citenamefont {Qian},\ and\ \citenamefont {Luo}}]{wei2022small}%
  \BibitemOpen
  \bibfield  {author} {\bibinfo {author} {\bibfnamefont {Y.~K.}\ \bibnamefont
  {Wei}}, \bibinfo {author} {\bibfnamefont {Y.~M.}\ \bibnamefont {Li}},
  \bibinfo {author} {\bibfnamefont {Z.~D.}\ \bibnamefont {Wang}}, \bibinfo
  {author} {\bibfnamefont {H.}~\bibnamefont {Yang}}, \bibinfo {author}
  {\bibfnamefont {Z.~C.}\ \bibnamefont {Zhu}}, \bibinfo {author} {\bibfnamefont
  {Y.~H.}\ \bibnamefont {Qian}}, \ and\ \bibinfo {author} {\bibfnamefont
  {K.~H.}\ \bibnamefont {Luo}},\ }\bibfield  {title} {\enquote {\bibinfo
  {title} {{Small-scale fluctuation and scaling law of mixing in
  three-dimensional rotating turbulent Rayleigh-Taylor instability}},}\ }\href
  {\doibase 10.1103/PhysRevE.105.015103} {\bibfield  {journal} {\bibinfo
  {journal} {Phys. Rev. E}\ }\textbf {\bibinfo {volume} {105}},\ \bibinfo
  {pages} {015103} (\bibinfo {year} {2022})}\BibitemShut {NoStop}%
\bibitem [{\citenamefont {Chai}\ \emph {et~al.}(2008)\citenamefont {Chai},
  \citenamefont {Guo}, \citenamefont {Zheng},\ and\ \citenamefont
  {Shi}}]{chai2008lattice}%
  \BibitemOpen
  \bibfield  {author} {\bibinfo {author} {\bibfnamefont {Z.~H.}\ \bibnamefont
  {Chai}}, \bibinfo {author} {\bibfnamefont {Z.~L.}\ \bibnamefont {Guo}},
  \bibinfo {author} {\bibfnamefont {L.}~\bibnamefont {Zheng}}, \ and\ \bibinfo
  {author} {\bibfnamefont {B.~C.}\ \bibnamefont {Shi}},\ }\bibfield  {title}
  {\enquote {\bibinfo {title} {{Lattice Boltzmann simulation of surface
  roughness effect on gaseous flow in a microchannel}},}\ }\href {\doibase
  10.1063/1.2949273} {\bibfield  {journal} {\bibinfo  {journal} {J. Appl.
  Phys}\ }\textbf {\bibinfo {volume} {104}},\ \bibinfo {pages} {014902}
  (\bibinfo {year} {2008})}\BibitemShut {NoStop}%
\bibitem [{\citenamefont {Chai}\ and\ \citenamefont {Shi}(2020)}]{chai2020}%
  \BibitemOpen
  \bibfield  {author} {\bibinfo {author} {\bibfnamefont {Z.~H.}\ \bibnamefont
  {Chai}}\ and\ \bibinfo {author} {\bibfnamefont {B.~C.}\ \bibnamefont {Shi}},\
  }\bibfield  {title} {\enquote {\bibinfo {title} {{Multiple-relaxation-time
  lattice Boltzmann method for the Navier-Stokes and nonlinear
  convection-diffusion equations: Modeling, analysis, and elements}},}\ }\href
  {\doibase 10.1103/PhysRevE.102.023306} {\bibfield  {journal} {\bibinfo
  {journal} {Phys. Rev. E}\ }\textbf {\bibinfo {volume} {102}},\ \bibinfo
  {pages} {023306} (\bibinfo {year} {2020})}\BibitemShut {NoStop}%
\bibitem [{\citenamefont {Xu}\ \emph {et~al.}(2012)\citenamefont {Xu},
  \citenamefont {Zhang}, \citenamefont {Gan},\ and\ \citenamefont
  {Yu}}]{xu2012lattice}%
  \BibitemOpen
  \bibfield  {author} {\bibinfo {author} {\bibfnamefont {A.~G.}\ \bibnamefont
  {Xu}}, \bibinfo {author} {\bibfnamefont {G.~C.}\ \bibnamefont {Zhang}},
  \bibinfo {author} {\bibfnamefont {Y.~B.}\ \bibnamefont {Gan}}, \ and\
  \bibinfo {author} {\bibfnamefont {X.~J.}\ \bibnamefont {Yu}},\ }\bibfield
  {title} {\enquote {\bibinfo {title} {{Lattice Boltzmann modeling and
  simulation of compressible flows}},}\ }\href {\doibase
  10.1007/s11467-012-0269-5} {\bibfield  {journal} {\bibinfo  {journal} {Front.
  Phys.}\ }\textbf {\bibinfo {volume} {7}},\ \bibinfo {pages} {582--600}
  (\bibinfo {year} {2012})}\BibitemShut {NoStop}%
\bibitem [{\citenamefont {Xu}, \citenamefont {Zhang},\ and\ \citenamefont
  {Gan}(2024)}]{Xu2024}%
  \BibitemOpen
  \bibfield  {author} {\bibinfo {author} {\bibfnamefont {A.~G.}\ \bibnamefont
  {Xu}}, \bibinfo {author} {\bibfnamefont {D.~J.}\ \bibnamefont {Zhang}}, \
  and\ \bibinfo {author} {\bibfnamefont {Y.~B.}\ \bibnamefont {Gan}},\
  }\bibfield  {title} {\enquote {\bibinfo {title} {{Advances in the kinetics of
  heat and mass transfer in near-continuous complex flows}},}\ }\href {\doibase
  10.1007/s11467-023-1353-8} {\bibfield  {journal} {\bibinfo  {journal} {Front.
  Phys.}\ }\textbf {\bibinfo {volume} {19}},\ \bibinfo {pages} {42500}
  (\bibinfo {year} {2024})}\BibitemShut {NoStop}%
\bibitem [{\citenamefont {Gan}\ \emph {et~al.}(2018)\citenamefont {Gan},
  \citenamefont {Xu}, \citenamefont {Zhang}, \citenamefont {Zhang},\ and\
  \citenamefont {Succi}}]{gan2018discrete}%
  \BibitemOpen
  \bibfield  {author} {\bibinfo {author} {\bibfnamefont {Y.~B.}\ \bibnamefont
  {Gan}}, \bibinfo {author} {\bibfnamefont {A.~G.}\ \bibnamefont {Xu}},
  \bibinfo {author} {\bibfnamefont {G.~C.}\ \bibnamefont {Zhang}}, \bibinfo
  {author} {\bibfnamefont {Y.~D.}\ \bibnamefont {Zhang}}, \ and\ \bibinfo
  {author} {\bibfnamefont {S.}~\bibnamefont {Succi}},\ }\bibfield  {title}
  {\enquote {\bibinfo {title} {Discrete boltzmann trans-scale modeling of
  high-speed compressible flows},}\ }\href {\doibase
  10.1103/PhysRevE.97.053312} {\bibfield  {journal} {\bibinfo  {journal}
  {Physical Review E}\ }\textbf {\bibinfo {volume} {97}},\ \bibinfo {pages}
  {053312} (\bibinfo {year} {2018})}\BibitemShut {NoStop}%
\bibitem [{\citenamefont {Gan}\ \emph {et~al.}(2022)\citenamefont {Gan},
  \citenamefont {Xu}, \citenamefont {Lai}, \citenamefont {Li}, \citenamefont
  {Sun},\ and\ \citenamefont {Succi}}]{gan2022discrete}%
  \BibitemOpen
  \bibfield  {author} {\bibinfo {author} {\bibfnamefont {Y.~B.}\ \bibnamefont
  {Gan}}, \bibinfo {author} {\bibfnamefont {A.~G.}\ \bibnamefont {Xu}},
  \bibinfo {author} {\bibfnamefont {H.~L.}\ \bibnamefont {Lai}}, \bibinfo
  {author} {\bibfnamefont {W.}~\bibnamefont {Li}}, \bibinfo {author}
  {\bibfnamefont {G.~L.}\ \bibnamefont {Sun}}, \ and\ \bibinfo {author}
  {\bibfnamefont {S.}~\bibnamefont {Succi}},\ }\bibfield  {title} {\enquote
  {\bibinfo {title} {{Discrete Boltzmann multi-scale modelling of
  non-equilibrium multiphase flows}},}\ }\href {\doibase 10.1017/jfm.2022.844}
  {\bibfield  {journal} {\bibinfo  {journal} {J. Fluid Mech.}\ }\textbf
  {\bibinfo {volume} {951}},\ \bibinfo {pages} {A8} (\bibinfo {year}
  {2022})}\BibitemShut {NoStop}%
\bibitem [{\citenamefont {Lin}, \citenamefont {Su},\ and\ \citenamefont
  {Zhang}(2020)}]{lin2020hydrodynamic}%
  \BibitemOpen
  \bibfield  {author} {\bibinfo {author} {\bibfnamefont {C.~D.}\ \bibnamefont
  {Lin}}, \bibinfo {author} {\bibfnamefont {X.~L.}\ \bibnamefont {Su}}, \ and\
  \bibinfo {author} {\bibfnamefont {Y.~D.}\ \bibnamefont {Zhang}},\ }\bibfield
  {title} {\enquote {\bibinfo {title} {{Hydrodynamic and thermodynamic
  nonequilibrium effects around shock waves: based on a discrete Boltzmann
  method}},}\ }\href {\doibase 10.3390/e22121397} {\bibfield  {journal}
  {\bibinfo  {journal} {Entropy}\ }\textbf {\bibinfo {volume} {22}},\ \bibinfo
  {pages} {1397} (\bibinfo {year} {2020})}\BibitemShut {NoStop}%
\bibitem [{\citenamefont {Liu}\ \emph {et~al.}(2023)\citenamefont {Liu},
  \citenamefont {Song}, \citenamefont {Xu}, \citenamefont {Zhang},\ and\
  \citenamefont {Xie}}]{liu2023discrete}%
  \BibitemOpen
  \bibfield  {author} {\bibinfo {author} {\bibfnamefont {Z.~P.}\ \bibnamefont
  {Liu}}, \bibinfo {author} {\bibfnamefont {J.~H.}\ \bibnamefont {Song}},
  \bibinfo {author} {\bibfnamefont {A.~G.}\ \bibnamefont {Xu}}, \bibinfo
  {author} {\bibfnamefont {Y.~D.}\ \bibnamefont {Zhang}}, \ and\ \bibinfo
  {author} {\bibfnamefont {K.}~\bibnamefont {Xie}},\ }\bibfield  {title}
  {\enquote {\bibinfo {title} {{Discrete Boltzmann modeling of plasma shock
  wave}},}\ }\href {\doibase 10.1177/09544062221075943} {\bibfield  {journal}
  {\bibinfo  {journal} {Proc. Inst. Mech. Eng. Part C J. Mech. Eng. Sci.}\
  }\textbf {\bibinfo {volume} {237}},\ \bibinfo {pages} {2532--2548} (\bibinfo
  {year} {2023})}\BibitemShut {NoStop}%
\bibitem [{\citenamefont {Zhang}\ \emph {et~al.}(2023)\citenamefont {Zhang},
  \citenamefont {Xu}, \citenamefont {Song}, \citenamefont {Gan}, \citenamefont
  {Zhang},\ and\ \citenamefont {Li}}]{zhang2023specific}%
  \BibitemOpen
  \bibfield  {author} {\bibinfo {author} {\bibfnamefont {D.~J.}\ \bibnamefont
  {Zhang}}, \bibinfo {author} {\bibfnamefont {A.~G.}\ \bibnamefont {Xu}},
  \bibinfo {author} {\bibfnamefont {J.~H.}\ \bibnamefont {Song}}, \bibinfo
  {author} {\bibfnamefont {Y.~B.}\ \bibnamefont {Gan}}, \bibinfo {author}
  {\bibfnamefont {Y.~D.}\ \bibnamefont {Zhang}}, \ and\ \bibinfo {author}
  {\bibfnamefont {Y.~J.}\ \bibnamefont {Li}},\ }\bibfield  {title} {\enquote
  {\bibinfo {title} {{Specific-heat ratio effects on the interaction between
  shock wave and heavy-cylindrical bubble: based on discrete Boltzmann
  method}},}\ }\href {\doibase https://doi.org/10.1016/j.compfluid.2023.106021}
  {\bibfield  {journal} {\bibinfo  {journal} {Comput. Fluids}\ }\textbf
  {\bibinfo {volume} {265}},\ \bibinfo {pages} {106021} (\bibinfo {year}
  {2023})}\BibitemShut {NoStop}%
\bibitem [{\citenamefont {Gan}\ \emph {et~al.}(2015)\citenamefont {Gan},
  \citenamefont {Xu}, \citenamefont {Zhang},\ and\ \citenamefont
  {Succi}}]{gan2015discrete}%
  \BibitemOpen
  \bibfield  {author} {\bibinfo {author} {\bibfnamefont {Y.~B.}\ \bibnamefont
  {Gan}}, \bibinfo {author} {\bibfnamefont {A.~G.}\ \bibnamefont {Xu}},
  \bibinfo {author} {\bibfnamefont {G.~C.}\ \bibnamefont {Zhang}}, \ and\
  \bibinfo {author} {\bibfnamefont {S.}~\bibnamefont {Succi}},\ }\bibfield
  {title} {\enquote {\bibinfo {title} {{Discrete Boltzmann modeling of
  multiphase flows: Hydrodynamic and thermodynamic non-equilibrium effects}},}\
  }\href {\doibase 10.1039/C5SM01125F} {\bibfield  {journal} {\bibinfo
  {journal} {Soft Matter}\ }\textbf {\bibinfo {volume} {11}},\ \bibinfo {pages}
  {5336--5345} (\bibinfo {year} {2015})}\BibitemShut {NoStop}%
\bibitem [{\citenamefont {Gan}\ \emph {et~al.}(2012)\citenamefont {Gan},
  \citenamefont {Xu}, \citenamefont {Zhang},\ and\ \citenamefont
  {Li}}]{gan2012physical}%
  \BibitemOpen
  \bibfield  {author} {\bibinfo {author} {\bibfnamefont {Y.~B.}\ \bibnamefont
  {Gan}}, \bibinfo {author} {\bibfnamefont {A.~G.}\ \bibnamefont {Xu}},
  \bibinfo {author} {\bibfnamefont {G.~C.}\ \bibnamefont {Zhang}}, \ and\
  \bibinfo {author} {\bibfnamefont {Y.~J.}\ \bibnamefont {Li}},\ }\bibfield
  {title} {\enquote {\bibinfo {title} {{Physical modeling of multiphase flow
  via lattice Boltzmann method: Numerical effects, equation of state and
  boundary conditions}},}\ }\href {\doibase 10.1007/s11467-012-0245-0}
  {\bibfield  {journal} {\bibinfo  {journal} {Front. Phys.}\ }\textbf {\bibinfo
  {volume} {7}},\ \bibinfo {pages} {481--490} (\bibinfo {year}
  {2012})}\BibitemShut {NoStop}%
\bibitem [{\citenamefont {Wang}\ \emph {et~al.}(2023)\citenamefont {Wang},
  \citenamefont {Lin}, \citenamefont {Yan}, \citenamefont {Su},\ and\
  \citenamefont {Yang}}]{wang2023high}%
  \BibitemOpen
  \bibfield  {author} {\bibinfo {author} {\bibfnamefont {S.~E.}\ \bibnamefont
  {Wang}}, \bibinfo {author} {\bibfnamefont {C.~D.}\ \bibnamefont {Lin}},
  \bibinfo {author} {\bibfnamefont {W.~W.}\ \bibnamefont {Yan}}, \bibinfo
  {author} {\bibfnamefont {X.~L.}\ \bibnamefont {Su}}, \ and\ \bibinfo {author}
  {\bibfnamefont {L.~C.}\ \bibnamefont {Yang}},\ }\bibfield  {title} {\enquote
  {\bibinfo {title} {{High-order modeling of multiphase flows: Based on
  discrete Boltzmann method}},}\ }\href {\doibase
  https://doi.org/10.1016/j.compfluid.2023.106009} {\bibfield  {journal}
  {\bibinfo  {journal} {Comput. Fluids}\ }\textbf {\bibinfo {volume} {265}},\
  \bibinfo {pages} {106009} (\bibinfo {year} {2023})}\BibitemShut {NoStop}%
\bibitem [{\citenamefont {Sun}\ \emph {et~al.}(2024)\citenamefont {Sun},
  \citenamefont {Gan}, \citenamefont {Xu},\ and\ \citenamefont
  {Shi}}]{sun2024droplet}%
  \BibitemOpen
  \bibfield  {author} {\bibinfo {author} {\bibfnamefont {G.~L.}\ \bibnamefont
  {Sun}}, \bibinfo {author} {\bibfnamefont {Y.~B.}\ \bibnamefont {Gan}},
  \bibinfo {author} {\bibfnamefont {A.~G.}\ \bibnamefont {Xu}}, \ and\ \bibinfo
  {author} {\bibfnamefont {Q.~F.}\ \bibnamefont {Shi}},\ }\bibfield  {title}
  {\enquote {\bibinfo {title} {Droplet coalescence kinetics: Thermodynamic
  non-equilibrium effects and entropy production mechanism},}\ }\href {\doibase
  10.1063/5.0187058} {\bibfield  {journal} {\bibinfo  {journal} {Physics of
  Fluids}\ }\textbf {\bibinfo {volume} {36}} (\bibinfo {year} {2024}),\
  10.1063/5.0187058}\BibitemShut {NoStop}%
\bibitem [{\citenamefont {Sun}\ \emph {et~al.}(2022)\citenamefont {Sun},
  \citenamefont {Gan}, \citenamefont {Xu}, \citenamefont {Zhang},\ and\
  \citenamefont {Shi}}]{sun2022thermodynamic}%
  \BibitemOpen
  \bibfield  {author} {\bibinfo {author} {\bibfnamefont {G.~L.}\ \bibnamefont
  {Sun}}, \bibinfo {author} {\bibfnamefont {Y.~B.}\ \bibnamefont {Gan}},
  \bibinfo {author} {\bibfnamefont {A.~G.}\ \bibnamefont {Xu}}, \bibinfo
  {author} {\bibfnamefont {Y.~D.}\ \bibnamefont {Zhang}}, \ and\ \bibinfo
  {author} {\bibfnamefont {Q.~F.}\ \bibnamefont {Shi}},\ }\bibfield  {title}
  {\enquote {\bibinfo {title} {Thermodynamic nonequilibrium effects in bubble
  coalescence: A discrete boltzmann study},}\ }\href {\doibase
  10.1103/PhysRevE.106.035101} {\bibfield  {journal} {\bibinfo  {journal}
  {Physical Review E}\ }\textbf {\bibinfo {volume} {106}},\ \bibinfo {pages}
  {035101} (\bibinfo {year} {2022})}\BibitemShut {NoStop}%
\bibitem [{\citenamefont {Lin}\ and\ \citenamefont {Luo}(2018)}]{lin2018mrt}%
  \BibitemOpen
  \bibfield  {author} {\bibinfo {author} {\bibfnamefont {C.~D.}\ \bibnamefont
  {Lin}}\ and\ \bibinfo {author} {\bibfnamefont {K.~H.}\ \bibnamefont {Luo}},\
  }\bibfield  {title} {\enquote {\bibinfo {title} {{MRT discrete Boltzmann
  method for compressible exothermic reactive flows}},}\ }\href {\doibase
  https://doi.org/10.1016/j.compfluid.2018.02.012} {\bibfield  {journal}
  {\bibinfo  {journal} {Comput. Fluids}\ }\textbf {\bibinfo {volume} {166}},\
  \bibinfo {pages} {176--183} (\bibinfo {year} {2018})}\BibitemShut {NoStop}%
\bibitem [{\citenamefont {Su}\ and\ \citenamefont
  {Lin}(2022)}]{su2022nonequilibrium}%
  \BibitemOpen
  \bibfield  {author} {\bibinfo {author} {\bibfnamefont {X.~L.}\ \bibnamefont
  {Su}}\ and\ \bibinfo {author} {\bibfnamefont {C.~D.}\ \bibnamefont {Lin}},\
  }\bibfield  {title} {\enquote {\bibinfo {title} {{Nonequilibrium effects of
  reactive flow based on gas kinetic theory}},}\ }\href {\doibase
  10.1088/1572-9494/ac53a0} {\bibfield  {journal} {\bibinfo  {journal} {Commum.
  Theor. Phys.}\ }\textbf {\bibinfo {volume} {74}},\ \bibinfo {pages} {035604}
  (\bibinfo {year} {2022})}\BibitemShut {NoStop}%
\bibitem [{\citenamefont {Yan}\ \emph {et~al.}(2013)\citenamefont {Yan},
  \citenamefont {Xu}, \citenamefont {Zhang}, \citenamefont {Ying},\ and\
  \citenamefont {Li}}]{yan2013lattice}%
  \BibitemOpen
  \bibfield  {author} {\bibinfo {author} {\bibfnamefont {B.}~\bibnamefont
  {Yan}}, \bibinfo {author} {\bibfnamefont {A.~G.}\ \bibnamefont {Xu}},
  \bibinfo {author} {\bibfnamefont {G.~C.}\ \bibnamefont {Zhang}}, \bibinfo
  {author} {\bibfnamefont {Y.~J.}\ \bibnamefont {Ying}}, \ and\ \bibinfo
  {author} {\bibfnamefont {H.}~\bibnamefont {Li}},\ }\bibfield  {title}
  {\enquote {\bibinfo {title} {{Lattice Boltzmann model for combustion and
  detonation}},}\ }\href {https://doi.org/10.1007/s11467-013-0286-z} {\bibfield
   {journal} {\bibinfo  {journal} {Front. Phys.}\ }\textbf {\bibinfo {volume}
  {8}},\ \bibinfo {pages} {94--110} (\bibinfo {year} {2013})}\BibitemShut
  {NoStop}%
\bibitem [{\citenamefont {Zhang}\ \emph {et~al.}(2016)\citenamefont {Zhang},
  \citenamefont {Xu}, \citenamefont {Zhang}, \citenamefont {Zhu},\ and\
  \citenamefont {Lin}}]{zhang2016kinetic}%
  \BibitemOpen
  \bibfield  {author} {\bibinfo {author} {\bibfnamefont {Y.~D.}\ \bibnamefont
  {Zhang}}, \bibinfo {author} {\bibfnamefont {A.~G.}\ \bibnamefont {Xu}},
  \bibinfo {author} {\bibfnamefont {G.~C.}\ \bibnamefont {Zhang}}, \bibinfo
  {author} {\bibfnamefont {C.~M.}\ \bibnamefont {Zhu}}, \ and\ \bibinfo
  {author} {\bibfnamefont {C.~D.}\ \bibnamefont {Lin}},\ }\bibfield  {title}
  {\enquote {\bibinfo {title} {{Kinetic modeling of detonation and effects of
  negative temperature coefficient}},}\ }\href {\doibase
  https://doi.org/10.1016/j.combustflame.2016.04.003} {\bibfield  {journal}
  {\bibinfo  {journal} {Combust. Flame}\ }\textbf {\bibinfo {volume} {173}},\
  \bibinfo {pages} {483--492} (\bibinfo {year} {2016})}\BibitemShut {NoStop}%
\bibitem [{\citenamefont {Yu}, \citenamefont {Lin},\ and\ \citenamefont
  {Luo}(2022)}]{ji2022three}%
  \BibitemOpen
  \bibfield  {author} {\bibinfo {author} {\bibfnamefont {J.}~\bibnamefont
  {Yu}}, \bibinfo {author} {\bibfnamefont {C.~D.}\ \bibnamefont {Lin}}, \ and\
  \bibinfo {author} {\bibfnamefont {K.~H.}\ \bibnamefont {Luo}},\ }\bibfield
  {title} {\enquote {\bibinfo {title} {{A three-dimensional discrete Boltzmann
  model for steady and unsteady detonation}},}\ }\href {\doibase
  https://doi.org/10.1016/j.jcp.2022.111002} {\bibfield  {journal} {\bibinfo
  {journal} {J. Comput. Phys.}\ }\textbf {\bibinfo {volume} {455}},\ \bibinfo
  {pages} {111002} (\bibinfo {year} {2022})}\BibitemShut {NoStop}%
\bibitem [{\citenamefont {Lin}\ and\ \citenamefont
  {Luo}(2019)}]{lin2019discrete}%
  \BibitemOpen
  \bibfield  {author} {\bibinfo {author} {\bibfnamefont {C.~D.}\ \bibnamefont
  {Lin}}\ and\ \bibinfo {author} {\bibfnamefont {K.~H.}\ \bibnamefont {Luo}},\
  }\bibfield  {title} {\enquote {\bibinfo {title} {{Discrete Boltzmann modeling
  of unsteady reactive flows with nonequilibrium effects}},}\ }\href {\doibase
  10.1103/PhysRevE.99.012142} {\bibfield  {journal} {\bibinfo  {journal} {Phys.
  Rev. E}\ }\textbf {\bibinfo {volume} {99}},\ \bibinfo {pages} {012142}
  (\bibinfo {year} {2019})}\BibitemShut {NoStop}%
\bibitem [{\citenamefont {Yu}, \citenamefont {Lin},\ and\ \citenamefont
  {Luo}(2021)}]{ji2021three}%
  \BibitemOpen
  \bibfield  {author} {\bibinfo {author} {\bibfnamefont {J.}~\bibnamefont
  {Yu}}, \bibinfo {author} {\bibfnamefont {C.~D.}\ \bibnamefont {Lin}}, \ and\
  \bibinfo {author} {\bibfnamefont {K.~H.}\ \bibnamefont {Luo}},\ }\bibfield
  {title} {\enquote {\bibinfo {title} {{Three-dimensional
  multiple-relaxation-time discrete Boltzmann model of compressible reactive
  flows with nonequilibrium effects}},}\ }\href {\doibase 10.1063/5.0047480}
  {\bibfield  {journal} {\bibinfo  {journal} {AIP Adv.}\ }\textbf {\bibinfo
  {volume} {11}},\ \bibinfo {pages} {045217} (\bibinfo {year}
  {2021})}\BibitemShut {NoStop}%
\bibitem [{\citenamefont {Shan}\ \emph {et~al.}(2023)\citenamefont {Shan},
  \citenamefont {Xu}, \citenamefont {Wang},\ and\ \citenamefont
  {Zhang}}]{shan2023}%
  \BibitemOpen
  \bibfield  {author} {\bibinfo {author} {\bibfnamefont {Y.~M.}\ \bibnamefont
  {Shan}}, \bibinfo {author} {\bibfnamefont {A.~G.}\ \bibnamefont {Xu}},
  \bibinfo {author} {\bibfnamefont {L.~F.}\ \bibnamefont {Wang}}, \ and\
  \bibinfo {author} {\bibfnamefont {Y.~D.}\ \bibnamefont {Zhang}},\ }\bibfield
  {title} {\enquote {\bibinfo {title} {{Nonequilibrium kinetics effects in
  Richtmyer-Meshkov instability and reshock processes}},}\ }\href {\doibase
  10.1088/1572-9494/acf305} {\bibfield  {journal} {\bibinfo  {journal} {Commun.
  Theor. Phys.}\ }\textbf {\bibinfo {volume} {75}},\ \bibinfo {pages} {115601}
  (\bibinfo {year} {2023})}\BibitemShut {NoStop}%
\bibitem [{\citenamefont {Song}\ \emph {et~al.}(2024)\citenamefont {Song},
  \citenamefont {Xu}, \citenamefont {Miao}, \citenamefont {Chen}, \citenamefont
  {Liu}, \citenamefont {Wang}, \citenamefont {Wang},\ and\ \citenamefont
  {Hou}}]{song2024plasma}%
  \BibitemOpen
  \bibfield  {author} {\bibinfo {author} {\bibfnamefont {J.~H.}\ \bibnamefont
  {Song}}, \bibinfo {author} {\bibfnamefont {A.~G.}\ \bibnamefont {Xu}},
  \bibinfo {author} {\bibfnamefont {L.}~\bibnamefont {Miao}}, \bibinfo {author}
  {\bibfnamefont {F.}~\bibnamefont {Chen}}, \bibinfo {author} {\bibfnamefont
  {Z.~P.}\ \bibnamefont {Liu}}, \bibinfo {author} {\bibfnamefont {L.~F.}\
  \bibnamefont {Wang}}, \bibinfo {author} {\bibfnamefont {N.~F.}\ \bibnamefont
  {Wang}}, \ and\ \bibinfo {author} {\bibfnamefont {X.}~\bibnamefont {Hou}},\
  }\bibfield  {title} {\enquote {\bibinfo {title} {{Plasma kinetics: Discrete
  Boltzmann modeling and Richtmyer-Meshkov instability}},}\ }\href {\doibase
  10.1063/5.0180246} {\bibfield  {journal} {\bibinfo  {journal} {Phys. Fluids}\
  }\textbf {\bibinfo {volume} {36}},\ \bibinfo {pages} {016107} (\bibinfo
  {year} {2024})}\BibitemShut {NoStop}%
\bibitem [{\citenamefont {Yang}\ \emph {et~al.}(2023)\citenamefont {Yang},
  \citenamefont {Lin}, \citenamefont {Li},\ and\ \citenamefont
  {Lai}}]{yang2023influence}%
  \BibitemOpen
  \bibfield  {author} {\bibinfo {author} {\bibfnamefont {T.}~\bibnamefont
  {Yang}}, \bibinfo {author} {\bibfnamefont {C.~D.}\ \bibnamefont {Lin}},
  \bibinfo {author} {\bibfnamefont {D.~M.}\ \bibnamefont {Li}}, \ and\ \bibinfo
  {author} {\bibfnamefont {H.~L.}\ \bibnamefont {Lai}},\ }\bibfield  {title}
  {\enquote {\bibinfo {title} {{Influence of Density Ratios on
  Richtmyer-Meshkov Instability with Non-Equilibrium Effects in the Reshock
  Process}},}\ }\href {\doibase 10.3390/inventions8060157} {\bibfield
  {journal} {\bibinfo  {journal} {Inventions}\ }\textbf {\bibinfo {volume}
  {8}},\ \bibinfo {pages} {157} (\bibinfo {year} {2023})}\BibitemShut {NoStop}%
\bibitem [{\citenamefont {Gan}\ \emph {et~al.}(2019)\citenamefont {Gan},
  \citenamefont {Xu}, \citenamefont {Zhang}, \citenamefont {Lin}, \citenamefont
  {Lai},\ and\ \citenamefont {Liu}}]{gan2019nonequilibrium}%
  \BibitemOpen
  \bibfield  {author} {\bibinfo {author} {\bibfnamefont {Y.~B.}\ \bibnamefont
  {Gan}}, \bibinfo {author} {\bibfnamefont {A.~G.}\ \bibnamefont {Xu}},
  \bibinfo {author} {\bibfnamefont {G.~C.}\ \bibnamefont {Zhang}}, \bibinfo
  {author} {\bibfnamefont {C.~D.}\ \bibnamefont {Lin}}, \bibinfo {author}
  {\bibfnamefont {H.~L.}\ \bibnamefont {Lai}}, \ and\ \bibinfo {author}
  {\bibfnamefont {Z.~P.}\ \bibnamefont {Liu}},\ }\bibfield  {title} {\enquote
  {\bibinfo {title} {{Nonequilibrium and morphological characterizations of
  Kelvin-Helmholtz instability in compressible flows}},}\ }\href {\doibase
  10.1007/s11467-019-0885-4} {\bibfield  {journal} {\bibinfo  {journal} {Front.
  Phys.}\ }\textbf {\bibinfo {volume} {14}},\ \bibinfo {pages} {1--17}
  (\bibinfo {year} {2019})}\BibitemShut {NoStop}%
\bibitem [{\citenamefont {Li}\ \emph {et~al.}(2022{\natexlab{a}})\citenamefont
  {Li}, \citenamefont {Lai}, \citenamefont {Lin},\ and\ \citenamefont
  {Li}}]{Li2022FOP}%
  \BibitemOpen
  \bibfield  {author} {\bibinfo {author} {\bibfnamefont {Y.~F.}\ \bibnamefont
  {Li}}, \bibinfo {author} {\bibfnamefont {H.~L.}\ \bibnamefont {Lai}},
  \bibinfo {author} {\bibfnamefont {C.~D.}\ \bibnamefont {Lin}}, \ and\
  \bibinfo {author} {\bibfnamefont {D.~M.}\ \bibnamefont {Li}},\ }\bibfield
  {title} {\enquote {\bibinfo {title} {{Influence of the tangential velocity on
  the compressible Kelvin-Helmholtz instability with nonequilibrium
  effects}},}\ }\href {\doibase 10.1007/s11467-022-1200-3} {\bibfield
  {journal} {\bibinfo  {journal} {Front. Phys.}\ }\textbf {\bibinfo {volume}
  {17}},\ \bibinfo {pages} {63500} (\bibinfo {year}
  {2022}{\natexlab{a}})}\BibitemShut {NoStop}%
\bibitem [{\citenamefont {Lai}\ \emph {et~al.}(2016)\citenamefont {Lai},
  \citenamefont {Xu}, \citenamefont {Zhang}, \citenamefont {Gan}, \citenamefont
  {Ying},\ and\ \citenamefont {Succi}}]{Lai2016}%
  \BibitemOpen
  \bibfield  {author} {\bibinfo {author} {\bibfnamefont {H.~L.}\ \bibnamefont
  {Lai}}, \bibinfo {author} {\bibfnamefont {A.~G.}\ \bibnamefont {Xu}},
  \bibinfo {author} {\bibfnamefont {G.~C.}\ \bibnamefont {Zhang}}, \bibinfo
  {author} {\bibfnamefont {Y.~B.}\ \bibnamefont {Gan}}, \bibinfo {author}
  {\bibfnamefont {Y.~J.}\ \bibnamefont {Ying}}, \ and\ \bibinfo {author}
  {\bibfnamefont {S.}~\bibnamefont {Succi}},\ }\bibfield  {title} {\enquote
  {\bibinfo {title} {Nonequilibrium thermohydrodynamic effects on the
  {Rayleigh-Taylor} instability in compressible flows},}\ }\href {\doibase
  10.1103/PhysRevE.94.023106} {\bibfield  {journal} {\bibinfo  {journal} {Phys.
  Rev. E}\ }\textbf {\bibinfo {volume} {94}},\ \bibinfo {pages} {023106}
  (\bibinfo {year} {2016})}\BibitemShut {NoStop}%
\bibitem [{\citenamefont {Chen}\ \emph
  {et~al.}(2022{\natexlab{a}})\citenamefont {Chen}, \citenamefont {Xu},
  \citenamefont {Chen}, \citenamefont {Zhang},\ and\ \citenamefont
  {Chen}}]{Chen2022discrete}%
  \BibitemOpen
  \bibfield  {author} {\bibinfo {author} {\bibfnamefont {J.}~\bibnamefont
  {Chen}}, \bibinfo {author} {\bibfnamefont {A.~G.}\ \bibnamefont {Xu}},
  \bibinfo {author} {\bibfnamefont {D.~W.}\ \bibnamefont {Chen}}, \bibinfo
  {author} {\bibfnamefont {Y.~D.}\ \bibnamefont {Zhang}}, \ and\ \bibinfo
  {author} {\bibfnamefont {Z.~H.}\ \bibnamefont {Chen}},\ }\bibfield  {title}
  {\enquote {\bibinfo {title} {{Discrete Boltzmann modeling of Rayleigh-Taylor
  instability: Effects of interfacial tension, viscosity, and heat
  conductivity}},}\ }\href {\doibase 10.1103/PhysRevE.106.015102} {\bibfield
  {journal} {\bibinfo  {journal} {Phys. Rev. E}\ }\textbf {\bibinfo {volume}
  {106}},\ \bibinfo {pages} {015102} (\bibinfo {year}
  {2022}{\natexlab{a}})}\BibitemShut {NoStop}%
\bibitem [{\citenamefont {Li}\ \emph {et~al.}(2022{\natexlab{b}})\citenamefont
  {Li}, \citenamefont {Xu}, \citenamefont {Zhang},\ and\ \citenamefont
  {Shan}}]{li2022rayleigh}%
  \BibitemOpen
  \bibfield  {author} {\bibinfo {author} {\bibfnamefont {H.~W.}\ \bibnamefont
  {Li}}, \bibinfo {author} {\bibfnamefont {A.~G.}\ \bibnamefont {Xu}}, \bibinfo
  {author} {\bibfnamefont {G.}~\bibnamefont {Zhang}}, \ and\ \bibinfo {author}
  {\bibfnamefont {Y.~M.}\ \bibnamefont {Shan}},\ }\bibfield  {title} {\enquote
  {\bibinfo {title} {Rayleigh-taylor instability under multi-mode perturbation:
  Discrete boltzmann modeling with tracers},}\ }\href {\doibase
  10.1088/1572-9494/ac85d9} {\bibfield  {journal} {\bibinfo  {journal} {Commum.
  Theor. Phys.}\ }\textbf {\bibinfo {volume} {74}},\ \bibinfo {pages} {115601}
  (\bibinfo {year} {2022}{\natexlab{b}})}\BibitemShut {NoStop}%
\bibitem [{\citenamefont {Chen}\ \emph
  {et~al.}(2022{\natexlab{b}})\citenamefont {Chen}, \citenamefont {Lai},
  \citenamefont {Lin},\ and\ \citenamefont {Li}}]{Chen2022}%
  \BibitemOpen
  \bibfield  {author} {\bibinfo {author} {\bibfnamefont {L.}~\bibnamefont
  {Chen}}, \bibinfo {author} {\bibfnamefont {H.~L.}\ \bibnamefont {Lai}},
  \bibinfo {author} {\bibfnamefont {C.~D.}\ \bibnamefont {Lin}}, \ and\
  \bibinfo {author} {\bibfnamefont {D.~M.}\ \bibnamefont {Li}},\ }\bibfield
  {title} {\enquote {\bibinfo {title} {{Numerical study of multimode
  Rayleigh-Taylor instability by using the discrete Boltzmann method}},}\
  }\href {\doibase {}} {\bibfield  {journal} {\bibinfo  {journal} {Acta
  Aerodyn. Sin.}\ }\textbf {\bibinfo {volume} {40}},\ \bibinfo {pages}
  {140--150} (\bibinfo {year} {2022}{\natexlab{b}})}\BibitemShut {NoStop}%
\bibitem [{\citenamefont {Ye}\ \emph {et~al.}(2020)\citenamefont {Ye},
  \citenamefont {Lai}, \citenamefont {Li}, \citenamefont {Gan}, \citenamefont
  {Lin}, \citenamefont {Chen},\ and\ \citenamefont {Xu}}]{Ye2020}%
  \BibitemOpen
  \bibfield  {author} {\bibinfo {author} {\bibfnamefont {H.~Y.}\ \bibnamefont
  {Ye}}, \bibinfo {author} {\bibfnamefont {H.~L.}\ \bibnamefont {Lai}},
  \bibinfo {author} {\bibfnamefont {D.~M.}\ \bibnamefont {Li}}, \bibinfo
  {author} {\bibfnamefont {Y.~B.}\ \bibnamefont {Gan}}, \bibinfo {author}
  {\bibfnamefont {C.~D.}\ \bibnamefont {Lin}}, \bibinfo {author} {\bibfnamefont
  {L.}~\bibnamefont {Chen}}, \ and\ \bibinfo {author} {\bibfnamefont {A.~G.}\
  \bibnamefont {Xu}},\ }\bibfield  {title} {\enquote {\bibinfo {title}
  {{Knudsen number effects on two-dimensional Rayleigh-Taylor instability in
  compressible fluid: Based on a discrete Boltzmann method}},}\ }\href
  {\doibase 10.3390/e22050500} {\bibfield  {journal} {\bibinfo  {journal}
  {Entropy}\ }\textbf {\bibinfo {volume} {22}},\ \bibinfo {pages} {500}
  (\bibinfo {year} {2020})}\BibitemShut {NoStop}%
\bibitem [{\citenamefont {Chen}\ \emph {et~al.}(2021)\citenamefont {Chen},
  \citenamefont {Lai}, \citenamefont {Lin},\ and\ \citenamefont
  {Li}}]{Chen2021}%
  \BibitemOpen
  \bibfield  {author} {\bibinfo {author} {\bibfnamefont {L.}~\bibnamefont
  {Chen}}, \bibinfo {author} {\bibfnamefont {H.~L.}\ \bibnamefont {Lai}},
  \bibinfo {author} {\bibfnamefont {C.~D.}\ \bibnamefont {Lin}}, \ and\
  \bibinfo {author} {\bibfnamefont {D.~M.}\ \bibnamefont {Li}},\ }\bibfield
  {title} {\enquote {\bibinfo {title} {{Specific heat ratio effects of
  compressible Rayleigh-Taylor instability studied by discrete Boltzmann
  method}},}\ }\href {\doibase 10.1007/s11467-021-1096-3} {\bibfield  {journal}
  {\bibinfo  {journal} {Front. Phys.}\ }\textbf {\bibinfo {volume} {16}},\
  \bibinfo {pages} {52500} (\bibinfo {year} {2021})}\BibitemShut {NoStop}%
\bibitem [{\citenamefont {Li}\ \emph {et~al.}(2018)\citenamefont {Li},
  \citenamefont {Lai}, \citenamefont {Xu}, \citenamefont {Zhang}, \citenamefont
  {Lin},\ and\ \citenamefont {Gan}}]{Li2018}%
  \BibitemOpen
  \bibfield  {author} {\bibinfo {author} {\bibfnamefont {D.~M.}\ \bibnamefont
  {Li}}, \bibinfo {author} {\bibfnamefont {H.~L.}\ \bibnamefont {Lai}},
  \bibinfo {author} {\bibfnamefont {A.~G.}\ \bibnamefont {Xu}}, \bibinfo
  {author} {\bibfnamefont {G.~C.}\ \bibnamefont {Zhang}}, \bibinfo {author}
  {\bibfnamefont {C.~D.}\ \bibnamefont {Lin}}, \ and\ \bibinfo {author}
  {\bibfnamefont {Y.~B.}\ \bibnamefont {Gan}},\ }\bibfield  {title} {\enquote
  {\bibinfo {title} {{Discrete Boltzmann simulation of Rayleigh-Taylor
  instability in compressible flows}},}\ }\href {\doibase {}} {\bibfield
  {journal} {\bibinfo  {journal} {Acta Physica Sinica}\ }\textbf {\bibinfo
  {volume} {67}},\ \bibinfo {pages} {080501} (\bibinfo {year}
  {2018})}\BibitemShut {NoStop}%
\bibitem [{\citenamefont {Chen}, \citenamefont {Xu},\ and\ \citenamefont
  {Zhang}(2016)}]{Chen2016Prandtl}%
  \BibitemOpen
  \bibfield  {author} {\bibinfo {author} {\bibfnamefont {F.}~\bibnamefont
  {Chen}}, \bibinfo {author} {\bibfnamefont {A.~G.}\ \bibnamefont {Xu}}, \ and\
  \bibinfo {author} {\bibfnamefont {G.~C.}\ \bibnamefont {Zhang}},\ }\bibfield
  {title} {\enquote {\bibinfo {title} {{Viscosity, heat conductivity, and
  Prandtl number effects in the Rayleigh-Taylor instability}},}\ }\href
  {\doibase 10.1007/s11467-016-0603-4} {\bibfield  {journal} {\bibinfo
  {journal} {Front. Phys.}\ }\textbf {\bibinfo {volume} {11}},\ \bibinfo
  {pages} {114703} (\bibinfo {year} {2016})}\BibitemShut {NoStop}%
\bibitem [{\citenamefont {Lai}\ \emph {et~al.}(2023)\citenamefont {Lai},
  \citenamefont {Lin}, \citenamefont {Gan}, \citenamefont {Li},\ and\
  \citenamefont {Chen}}]{lai2023influences}%
  \BibitemOpen
  \bibfield  {author} {\bibinfo {author} {\bibfnamefont {H.~L.}\ \bibnamefont
  {Lai}}, \bibinfo {author} {\bibfnamefont {C.~D.}\ \bibnamefont {Lin}},
  \bibinfo {author} {\bibfnamefont {Y.~B.}\ \bibnamefont {Gan}}, \bibinfo
  {author} {\bibfnamefont {D.~M.}\ \bibnamefont {Li}}, \ and\ \bibinfo {author}
  {\bibfnamefont {L.}~\bibnamefont {Chen}},\ }\bibfield  {title} {\enquote
  {\bibinfo {title} {{The influences of acceleration on compressible
  Rayleigh-Taylor instability with non-equilibrium effects}},}\ }\href
  {\doibase https://doi.org/10.1016/j.compfluid.2023.106037} {\bibfield
  {journal} {\bibinfo  {journal} {Comput. Fluids}\ }\textbf {\bibinfo {volume}
  {266}},\ \bibinfo {pages} {106037} (\bibinfo {year} {2023})}\BibitemShut
  {NoStop}%
\bibitem [{\citenamefont {Lin}\ \emph {et~al.}(2017)\citenamefont {Lin},
  \citenamefont {Xu}, \citenamefont {Zhang}, \citenamefont {Luo},\ and\
  \citenamefont {Li}}]{lin2017discrete}%
  \BibitemOpen
  \bibfield  {author} {\bibinfo {author} {\bibfnamefont {C.~D.}\ \bibnamefont
  {Lin}}, \bibinfo {author} {\bibfnamefont {A.~G.}\ \bibnamefont {Xu}},
  \bibinfo {author} {\bibfnamefont {G.~C.}\ \bibnamefont {Zhang}}, \bibinfo
  {author} {\bibfnamefont {K.~H.}\ \bibnamefont {Luo}}, \ and\ \bibinfo
  {author} {\bibfnamefont {Y.~J.}\ \bibnamefont {Li}},\ }\bibfield  {title}
  {\enquote {\bibinfo {title} {{Discrete Boltzmann modeling of Rayleigh-Taylor
  instability in two-component compressible flows}},}\ }\href {\doibase
  10.1103/PhysRevE.96.053305} {\bibfield  {journal} {\bibinfo  {journal} {Phys.
  Rev. E}\ }\textbf {\bibinfo {volume} {96}},\ \bibinfo {pages} {053305}
  (\bibinfo {year} {2017})}\BibitemShut {NoStop}%
\bibitem [{\citenamefont {Lin}\ \emph {et~al.}(2019)\citenamefont {Lin},
  \citenamefont {Luo}, \citenamefont {Gan},\ and\ \citenamefont
  {Liu}}]{lin2019kinetic}%
  \BibitemOpen
  \bibfield  {author} {\bibinfo {author} {\bibfnamefont {C.~D.}\ \bibnamefont
  {Lin}}, \bibinfo {author} {\bibfnamefont {K.~H.}\ \bibnamefont {Luo}},
  \bibinfo {author} {\bibfnamefont {Y.~B.}\ \bibnamefont {Gan}}, \ and\
  \bibinfo {author} {\bibfnamefont {Z.~P.}\ \bibnamefont {Liu}},\ }\bibfield
  {title} {\enquote {\bibinfo {title} {{Kinetic simulation of nonequilibrium
  Kelvin-Helmholtz instability}},}\ }\href {\doibase
  10.1088/0253-6102/71/1/132} {\bibfield  {journal} {\bibinfo  {journal}
  {Commun. Theor. Phys.}\ }\textbf {\bibinfo {volume} {71}},\ \bibinfo {pages}
  {132--142} (\bibinfo {year} {2019})}\BibitemShut {NoStop}%
\bibitem [{\citenamefont {Zhang}\ \emph {et~al.}(2020)\citenamefont {Zhang},
  \citenamefont {Xu}, \citenamefont {Zhang},\ and\ \citenamefont
  {Li}}]{Zhang2020Two}%
  \BibitemOpen
  \bibfield  {author} {\bibinfo {author} {\bibfnamefont {D.~J.}\ \bibnamefont
  {Zhang}}, \bibinfo {author} {\bibfnamefont {A.~G.}\ \bibnamefont {Xu}},
  \bibinfo {author} {\bibfnamefont {Y.~D.}\ \bibnamefont {Zhang}}, \ and\
  \bibinfo {author} {\bibfnamefont {Y.~J.}\ \bibnamefont {Li}},\ }\bibfield
  {title} {\enquote {\bibinfo {title} {{Two-fluid discrete Boltzmann model for
  compressible flows: Based on ellipsoidal statistical
  Bhatnagar-Gross-Krook}},}\ }\href {\doibase 10.1063/5.0017673} {\bibfield
  {journal} {\bibinfo  {journal} {Phys. Fluids}\ }\textbf {\bibinfo {volume}
  {32}},\ \bibinfo {pages} {126110} (\bibinfo {year} {2020})}\BibitemShut
  {NoStop}%
\bibitem [{\citenamefont {Chen}\ \emph {et~al.}(2023)\citenamefont {Chen},
  \citenamefont {Lai}, \citenamefont {Lin},\ and\ \citenamefont
  {Li}}]{chen2023effects}%
  \BibitemOpen
  \bibfield  {author} {\bibinfo {author} {\bibfnamefont {B.~L.}\ \bibnamefont
  {Chen}}, \bibinfo {author} {\bibfnamefont {H.~L.}\ \bibnamefont {Lai}},
  \bibinfo {author} {\bibfnamefont {C.~D.}\ \bibnamefont {Lin}}, \ and\
  \bibinfo {author} {\bibfnamefont {D.~M.}\ \bibnamefont {Li}},\ }\bibfield
  {title} {\enquote {\bibinfo {title} {{ Effects of Inclined Interface Angle on
  Compressible Rayleigh-Taylor Instability: A Numerical Study Based on the
  Discrete Boltzmann Method}},}\ }\href {\doibase 10.3390/e25121623} {\bibfield
   {journal} {\bibinfo  {journal} {Entropy}\ }\textbf {\bibinfo {volume}
  {25}},\ \bibinfo {pages} {1623} (\bibinfo {year} {2023})}\BibitemShut
  {NoStop}%
\bibitem [{\citenamefont {Lin}, \citenamefont {Luo},\ and\ \citenamefont
  {Lai}(2024)}]{Lin2024CTP}%
  \BibitemOpen
  \bibfield  {author} {\bibinfo {author} {\bibfnamefont {C.}~\bibnamefont
  {Lin}}, \bibinfo {author} {\bibfnamefont {K.~H.}\ \bibnamefont {Luo}}, \ and\
  \bibinfo {author} {\bibfnamefont {H.}~\bibnamefont {Lai}},\ }\bibfield
  {title} {\enquote {\bibinfo {title} {{Discrete Boltzmann model with split
  collision for nonequilibrium reactive flows}},}\ }\href {\doibase
  10.1088/1572-9494/ad4a36} {\bibfield  {journal} {\bibinfo  {journal}
  {Communications in Theoretical Physics}\ }\textbf {\bibinfo {volume} {76}},\
  \bibinfo {pages} {085602} (\bibinfo {year} {2024})}\BibitemShut {NoStop}%
\bibitem [{\citenamefont {Lin}\ \emph {et~al.}(2021)\citenamefont {Lin},
  \citenamefont {Luo}, \citenamefont {Xu}, \citenamefont {Gan},\ and\
  \citenamefont {Lai}}]{Lin2021PRE}%
  \BibitemOpen
  \bibfield  {author} {\bibinfo {author} {\bibfnamefont {C.~D.}\ \bibnamefont
  {Lin}}, \bibinfo {author} {\bibfnamefont {K.~H.}\ \bibnamefont {Luo}},
  \bibinfo {author} {\bibfnamefont {A.~G.}\ \bibnamefont {Xu}}, \bibinfo
  {author} {\bibfnamefont {Y.~B.}\ \bibnamefont {Gan}}, \ and\ \bibinfo
  {author} {\bibfnamefont {H.~L.}\ \bibnamefont {Lai}},\ }\bibfield  {title}
  {\enquote {\bibinfo {title} {{Multiple-relaxation-time discrete Boltzmann
  modeling of multicomponent mixture with nonequilibrium effects}},}\ }\href
  {\doibase 10.1103/PhysRevE.103.013305} {\bibfield  {journal} {\bibinfo
  {journal} {Phys. Rev. E}\ }\textbf {\bibinfo {volume} {103}},\ \bibinfo
  {pages} {013305} (\bibinfo {year} {2021})}\BibitemShut {NoStop}%
\bibitem [{\citenamefont {Lin}\ \emph {et~al.}(2016)\citenamefont {Lin},
  \citenamefont {Xu}, \citenamefont {Zhang},\ and\ \citenamefont
  {Li}}]{Lin2016CNF}%
  \BibitemOpen
  \bibfield  {author} {\bibinfo {author} {\bibfnamefont {C.~D.}\ \bibnamefont
  {Lin}}, \bibinfo {author} {\bibfnamefont {A.~G.}\ \bibnamefont {Xu}},
  \bibinfo {author} {\bibfnamefont {G.~C.}\ \bibnamefont {Zhang}}, \ and\
  \bibinfo {author} {\bibfnamefont {Y.~J.}\ \bibnamefont {Li}},\ }\bibfield
  {title} {\enquote {\bibinfo {title} {{Double-distribution-function discrete
  Boltzmann model for combustion}},}\ }\href {\doibase
  https://doi.org/10.1016/j.combustflame.2015.11.010} {\bibfield  {journal}
  {\bibinfo  {journal} {Combust. Flame}\ }\textbf {\bibinfo {volume} {164}},\
  \bibinfo {pages} {137--51} (\bibinfo {year} {2016})}\BibitemShut {NoStop}%
\end{thebibliography}%

\end{document}